\documentclass[oneside,11pt]{article}

\usepackage[utf8]{inputenc}
\usepackage[T1]{fontenc}

\usepackage{amsthm}
\usepackage{amsmath}
\usepackage{amsfonts}
\usepackage{amssymb}

\usepackage{graphicx} 
\usepackage{float}
\usepackage{booktabs}
\usepackage{multirow}
\usepackage{rotating}
\usepackage{array}

\usepackage{breakcites}

\usepackage{enumitem}

\usepackage[top=1.5cm,bottom=2cm,right=2.5cm,left=2.5cm]{geometry}
\usepackage{titling}

\usepackage{natbib}

\usepackage{ifplatform}

\ifwindows
\fi

\ifwindows
	\usepackage{bbm}
\else
	\iflinux
		\usepackage{bbm}
	\else
		\usepackage{bbm}
	\fi
\fi

\newcommand{\lp}{\left(}
\newcommand{\rp}{\right)}
\newcommand{\lc}{\left[}
\newcommand{\rc}{\right]}
\newcommand{\lb}{\left\{}

\newcommand{\ri}{\right.}
\newcommand{\R}{\mathbb{R}}

\newcommand{\bx}{\mathbf{x}}

\newcommand{\by}{\mathbf{y}}
\newcommand{\bX}{\mathbf{X}}

\newcommand{\bmu}{\boldsymbol\mu}

\newcommand{\bb}{\mathbf{b}}

\newcommand{\bxi}{\boldsymbol\xi}

\newcommand{\btheta}{\boldsymbol\theta}
\newcommand{\bTheta}{\boldsymbol\Theta}
\newcommand{\Ical}{\mathcal{I}}
\newcommand{\bSigma}{\boldsymbol\Sigma}
\newcommand{\bO}{\mathbf{0}_q}
\newcommand{\bOd}{\mathbf{0}_{q-1}}

\newcommand{\bHcal}{\boldsymbol{\mathcal{H}}}

\newcommand{\bnab}{\boldsymbol\nabla}
\newcommand{\bB}{\mathbf{B}}

\newcommand{\bI}{\mathbf{I}}

\newcommand{\lrp}[1]{\left(#1\right)}
\newcommand{\lrc}[1]{\left[#1\right]}
\newcommand{\lrb}[1]{\left\{#1\right\}}

\newcommand{\E}[1]{\mathbb{E}\lc #1\rc}

\newcommand{\norm}[1]{\left|\left| #1\right|\right|}

\newcommand{\Om}[1]{\Omega_{#1}}
\newcommand{\om}[1]{\omega_{#1}}
\newcommand{\Iq}[2]{\int_{\Omega_{q}} #1\,\omega_{q}(d #2)}

\DeclareFontFamily{OT1}{pzc}{}
\DeclareFontShape{OT1}{pzc}{m}{it}{<-> s * [1.10] pzcmi7t}{}
\DeclareMathAlphabet{\mathpzc}{OT1}{pzc}{m}{it}
\newcommand{\order}[1]{\mathpzc{o}\lp#1\rp}
\newcommand{\Order}[1]{\mathcal{O}\lp#1\rp}

\usepackage[pdftex,final,bookmarksnumbered,bookmarksopen=false,breaklinks,colorlinks]{hyperref}
\hypersetup{
	final,
    bookmarksnumbered=true,
    bookmarksopen=true,
    bookmarksopenlevel=0,
    unicode=false,          
    pdftoolbar=true,        
    pdfmenubar=true,        
    pdffitwindow=false,     
    pdftitle={Exact risk improvement of bandwidth selectors for kernel density estimation with directional data},    
	pdfdisplaydoctitle=true, 
	pdftoolbar=true,
	pdfmenubar=true,
	pdflang={English},
    pdfauthor={Eduardo Garcia-Portugues},     
    pdfsubject={arXiv paper},   
    pdfcreator={Eduardo Garcia-Portugues},   
    pdfproducer={Eduardo Garcia-Portugues}, 
    pdfkeywords={Bandwidth selection} {Directional data} {Mixtures} {Kernel density estimator} {Von Mises}, 
    pdfnewwindow=true,      
    breaklinks=true,
    hidelinks,       
    linkcolor=black,          
    citecolor=black,        
    filecolor=magenta,      
    urlcolor=cyan           
}

\newtheorem{coro}{Corollary}
\newtheorem{rem}{Remark}
\newtheorem{prop}{Proposition}

\newtheorem{algo}{Algorithm}

\allowdisplaybreaks

\begin{document}

\title{Exact risk improvement of bandwidth selectors for kernel\\ density estimation with directional data}
\setlength{\droptitle}{-1cm}
\predate{}%
\postdate{}%
\author{Eduardo Garc\'ia-Portugu\'es$^{1,2}$}

\date{}

\footnotetext[1]{
Department of Statistics and Operations Research, University of Santiago de Compostela (Spain).}
\footnotetext[2]{Corresponding author. e-mail: \href{mailto:eduardo.garcia@usc.es}{eduardo.garcia@usc.es}.}

\maketitle


\begin{abstract}
New bandwidth selectors for kernel density estimation with directional data are presented in this work. These selectors are based on asymptotic and exact error expressions for the kernel density estimator combined with mixtures of von Mises distributions. The performance of the proposed selectors is investigated in a simulation study and compared with other existing rules for a large variety of directional scenarios, sample sizes and dimensions. The selector based on the exact error expression turns out to have the best behaviour of the studied selectors for almost all the situations. This selector is illustrated with real data for the circular and spherical cases.
\end{abstract}
\begin{flushleft}
\end{flushleft}

\section{Introduction}
\label{kdebwd:sec:introduction}

Bandwidth selection is a key issue in kernel density estimation that has deserved considerable attention during the last decades. The problem of selecting the most suitable bandwidth for the nonparametric kernel density estimator introduced by \cite{Rosenblatt1956} and \cite{Parzen1962} is the main topic of the reviews of \cite{Cao1994}, \cite{Jones1996} and \cite{Chiu1996}, among others. Comprehensive references on kernel smoothing and bandwidth selection include the books by \cite{Silverman1986}, \cite{Scott1992} and \cite{Wand1995}. Bandwidth selection is still an active research field in density estimation, with some recent contributions like \cite{Horova2013} and \cite{Chacon2013} in the last years.\\

Kernel density estimation has been also adapted to directional data, that is, data in the unit hypersphere of dimension $q$. Due to the particular nature of directional data (periodicity for $q=1$ and manifold structure for any $q$), the usual multivariate techniques are not appropriate and specific methodology that accounts for their characteristics has to be considered. The classical references for the theory of directional statistics are the complete review of \cite{Jupp1989} and the book by \cite{Mardia2000}. The kernel density estimation with directional data was firstly proposed by \cite{Hall1987}, studying the properties of two types of kernel density estimators and providing cross-validatory bandwidth selectors. Almost simultaneously, \cite{Bai1988} provided a similar definition of kernel estimator, establishing its pointwise and $\mathcal{L}_1$ consistency. Some of the results by \cite{Hall1987} were extended by \cite{Klemela2000}, who studied the estimation of the Laplacian of the density and other types of derivatives. Whereas the framework for all these references is the general $q$-sphere, which comprises as particular case the circle ($q=1$), there exists a remarkable collection of works devoted to kernel density estimation and bandwidth selection for the circular scenario. Specifically, \cite{Taylor2008} presented the first plug-in bandwidth selector in this context and \cite{Oliveira2012} derived a selector based on mixtures and on the results of \cite{DiMarzio2009} for the circular Asymptotic Mean Integrated Squared Error (AMISE). Recently, \cite{DiMarzio2011} proposed a product kernel density estimator on the $q$-dimensional torus and cross-validatory bandwidth selection methods for that situation. Another nonparametric approximation for density estimation with circular data was given in \cite{Fern'andez-Dur'an2004} and \cite{Fernandez-Duran2010}. In the general setting of spherical random fields \cite{ClaudioDurastanti2013} derived an estimation method based on a needlet basis representation.\\ 

Directional data arise in many applied fields. For the circular case ($q=1$) a typical example is wind direction, studied among others in \cite{Jammalamadaka2006}, \cite{Fern'andez-Dur'an2007} and \cite{Garcia-Portugues:so2}. The spherical case ($q=2$) poses challenging applications in astronomy, for example in the study of stars position in the celestial sphere or in the study of the cosmic microwave background radiation \citep{Cabella2009}. Finally, a novel field where directional data is present for large $q$ is text mining  \citep{Banerjee2005}, where documents are usually codified as high dimensional unit vectors. For all these situations, a reliable method for choosing the bandwidth parameter seems necessary to trust the density estimate. \\

The aim of this work is to introduce new bandwidth selectors for the kernel density estimator for directional data. The first one is a rule of thumb which assumes that the underlying density is a von Mises and it is intended to be the directional analogue of the rule of thumb proposed by \cite{Silverman1986} for data in the real line. This selector uses the AMISE expression that can be seen, among others, in \cite{Garcia-Portugues:dirlin}. The novelty of the selector is that it is more general and robust than the previous proposal by \cite{Taylor2008}, although both rules exhibit an unsatisfactory behaviour when the reference density spreads off from the von Mises. To overcome this problem, two new selectors based on the use of mixtures of von Mises for the reference density are proposed. One of them uses the aforementioned AMISE expression, whereas the other one uses the exact MISE computation for mixtures of von Mises densities given in \cite{Garcia-Portugues:dirlin}. Both of them use the Expectation-Maximization algorithm of \cite{Banerjee2005} to fit the mixtures and, to select the number of components, the BIC criteria is employed. These selectors based on mixtures are inspired by the earlier ideas of \cite{Cwik1997}, for the multivariate setting, and \cite{Oliveira2012} for the circular scenario.\\

This paper is organized as follows. Section \ref{kdebwd:sec:kdedir} presents some background on kernel density estimation for directional data and the available bandwidth selectors. The rule of thumb selector is introduced in Section \ref{kdebwd:sec:ruleofthumb} and the two selectors based on mixtures of von Mises are presented in Section \ref{kdebwd:sec:mixtures}. Section \ref{kdebwd:sec:comparative} contains a simulation study comparing the proposed selectors with the ones available in the literature. Finally, Section \ref{kdebwd:sec:data} illustrates a real data application and some conclusions are given in Section \ref{kdebwd:sec:conclusions}. Supplementary materials with proofs, simulated models and extended tables are given in the appendix.

\section{Kernel density estimation with directional data}
\label{kdebwd:sec:kdedir}

Denote by $\bX$ a directional random variable with density $f$. The support of such variable is the $q$-dimensional sphere, namely $\Om{q}=\big\{\bx\in\R^{q+1}:x^2_1+\cdots+x^2_{q+1}=1\big\}$, endowed with the Lebesgue measure in $\Om{q}$, that will be denoted by $\om{q}$. Then, a directional density is a nonnegative function that satisfies $\Iq{f(\bx)}{\bx}=1$. Also, when there is no possible confusion, the area of $\Om{q}$ will be denoted by
\begin{align*}
\om{q}=\om{q}\lrp{\Om{q}}=\frac{2\pi^\frac{q+1}{2}}{\Gamma\lrp{\frac{q+1}{2}}},\quad q\geq 1, 
\end{align*}
where $\Gamma$ represents the Gamma function defined as $\Gamma(p)=\int_0^\infty x^{p-1}e^{-x}\,dx$, $p>-1$.\\

Among the directional distributions, the von Mises--Fisher distribution (see \cite{Watson1983}) is perhaps the most widely used. The von Mises density, denoted by $\mathrm{vM}(\bmu,\kappa)$, is given by
\begin{align}
f_{\mathrm{vM}}(\bx;\bmu,\kappa)=C_q(\kappa) \exp{\lrb{\kappa\bx^T\bmu}},\quad C_q(\kappa)=\frac{\kappa^{\frac{q-1}{2}}}{(2\pi)^{\frac{q+1}{2}}\mathcal{I}_{\frac{q-1}{2}}(\kappa)},\label{kdebwd:dir:cq}
\end{align}
where $\bmu\in\Omega_q$ is the directional mean, $\kappa\geq0$ the concentration parameter around the mean, $^T$ stands for the transpose operator and $\mathcal{I}_p$ is the modified Bessel function of order $\nu$,
\begin{align*}
\mathcal{I}_\nu(z)=\frac{\lrp{\frac{z}{2}}^\nu}{\pi^{1/2}\Gamma\lrp{\nu+\frac{1}{2}}}\int_{-1}^1 (1-t^2)^{\nu-\frac{1}{2}}e^{zt}\,dt.
\end{align*}
This distribution is the main reference for directional models and, in that sense, plays the role of the normal distribution for directional data (is also a multivariate normal $\mathcal{N}(\bmu,\kappa^{-1}\bI_{q+1})$ conditioned on $\Om{q}$; see \cite{Mardia2000}). A particular case of this density sets $\kappa=0$, which corresponds to the uniform density that assigns probability $\om{q}^{-1}$ to any direction in $\Om{q}$.\\

Given a random sample $\bX_1,\ldots,\bX_n$ from the directional random variable $\bX$, the proposal of \cite{Bai1988} for the directional kernel density estimator at a point $\bx\in \Omega_q$ is 
\begin{align}
\hat f_h(\bx)=\frac{c_{h,q}(L)}{n}\sum_{i=1}^n L\lrp{\frac{1-\bx^T\bX_i}{h^2}},\label{kdebwd:kernel_directional}
\end{align}
where $L$ is a directional kernel (a rapidly decaying function with nonnegative values and defined in $[0,\infty)$), $h>0$ is the bandwidth parameter and $c_{h,q}(L)$ is a normalizing constant. This constant is needed in order to ensure that the estimator is indeed a density and satisfies that
\begin{align*}
c_{h,q}(L)^{-1}=\Iq{L\lrp{\frac{1-\bx^T\by}{h^2}}}{\bx}=\Order{h^q}.
\end{align*}
As usual in kernel smoothing, the selection of the bandwidth is a crucial step that affects notably the final estimation: large values of $h$ result in a uniform density in the sphere, whereas small values of $h$ provide an undersmoothed estimator with high concentrations around the sample observations. On the other hand, the choice of the kernel is not seen as important for practical purposes and the most common choice is the so called von Mises kernel $L(r)=e^{-r}$. Its name is due to the fact that the kernel estimator can be viewed as a mixture of von Mises--Fisher densities as follows:
\begin{align*}
\hat f_h(\bx)=\frac{1}{n}\sum_{i=1}^n f_{\mathrm{vM}}\lrp{\bx;\bX_i,1/h^2},%
\end{align*}
where, for each von Mises component, the mean value is the $i$-th observation $\bX_i$ and the common concentration parameter is given by $1/h^{2}$.\\

The classical error measurement in kernel density estimation is the $\mathcal{L}_2$ distance between the estimator $\hat f_h$ and the target density $f$, the so called Integrated Squared Error (ISE). As this is a random quantity depending on the sample, its expected value, the Mean Integrated Squared Error (MISE), is usually considered:
\begin{align*}
\mathrm{MISE}(h)=\E{\mathrm{ISE}\big[\hat f_h\big]}=\E{\Iq{\lrp{\hat f_h(\bx)-f(\bx)}^2}{\bx}},
\end{align*}
which depends on the bandwidth $h$, the kernel $L$, the sample size $n$ and the target density $f$. Whereas the two last elements are fixed when estimating a density from a random sample, the bandwidth has to be chosen (also the kernel, although this does not present a big impact in the performance of the estimator). Then, a possibility is to search for the bandwidth that minimizes the MISE:
\begin{align*}
h_{\mathrm{MISE}}=\arg\min_{h>0}\mathrm{MISE}(h).
\end{align*}
To derive an easier form for the MISE that allows to obtain $h_{\mathrm{MISE}}$, the following conditions on the elements of the estimator (\ref{kdebwd:kernel_directional}) are required:
\begin{enumerate}[label=\textbf{D\arabic{*}}.,ref=\textbf{D\arabic{*}}]
\item Extend $f$ from $\Omega_q$ to $\R^{q+1}\backslash\lrb{\mathbf{0}}$ by $f(\bx)\equiv f\lrp{\bx/\norm{\bx}}$ for all $\bx\in\R^{q+1}\backslash\lrb{\mathbf{0}}$, where $\norm{\cdot}$ denotes the Euclidean norm. Assume that the gradient vector $\bnab f(\bx)$ and the Hessian matrix $\bHcal f(\bx)$ exist and are continuous. \label{kdebwd:cond:d1}
 \item Assume that $L:[0,\infty)\rightarrow[0,\infty)$ is a bounded and integrable function such that
$0<\int_0^\infty L^k(r) r^{\frac{q}{2}-1}\,dr<\infty$, $\forall q\geq1$, for $k=1,2$.\label{kdebwd:cond:d2}
\item Assume that $h=h_n$ is a positive sequence such that $h_n\rightarrow0$ and $n h_n^q\rightarrow\infty$ as $n\rightarrow\infty$.\label{kdebwd:cond:d3}
\end{enumerate}
The following result, available from \cite{Garcia-Portugues:dirlin}, provides the MISE expansion for the estimator (\ref{kdebwd:kernel_directional}). It is worth mentioning that, under similar conditions, \cite{Hall1987} and \cite{Klemela2000} also derived analogous expressions. 
\begin{prop}[\cite{Garcia-Portugues:dirlin}]
\label{kdebwd:dir:prop:3}
Under conditions \ref{kdebwd:cond:d1}--\ref{kdebwd:cond:d3}, the MISE for the directional kernel density estimator (\ref{kdebwd:kernel_directional}) is given by
\begin{align*}
\mathrm{MISE}(h)=&b_q(L)^2R(\Psi(f,\cdot))h^4+\frac{c_{h,q}(L)}{n}d_q(L)+\order{h^4+(nh^q)^{-1}},
\end{align*}
where $R(\Psi(f,\cdot))=\int_{\Omega_{q}}\Psi(f,\bx)^2\,\omega_q(d\bx)$, $b_q(L)=\frac{\int_0^\infty L(r) r^{\frac{q}{2}}\,dr}{\int_0^\infty L(r) r^{\frac{q}{2}-1}\,dr}$, $d_q(L)=\frac{\int_0^\infty L^2(r) r^{\frac{q}{2}-1}\,dr}{\int_0^\infty L(r) r^{\frac{q}{2}-1}\,dr}$ and
\begin{align}
\Psi(f,\bx)=&-\bx^T\bnab f(\bx)+q^{-1}\lrp{\nabla^2f(\bx)-\bx^T\bHcal f(\bx)\bx}.\label{kdebwd:Psi_dir}
\end{align}
\end{prop}
This results leads to the decomposition $\mathrm{MISE}(h)=\mathrm{AMISE}(h)+\order{h^4+(nh^q)^{-1}}$, where AMISE stands for the Asymptotic MISE. It is possible to derive an optimal bandwidth for the AMISE in this sense, $h_{\mathrm{AMISE}}=\arg\min_{h>0}\mathrm{AMISE}(h)$, that will be close to $h_{\mathrm{MISE}}$ when $h^4+(nh^q)^{-1}$ is small enough.
\begin{coro}[\cite{Garcia-Portugues:dirlin}]
\label{kdebwd:dir:cor:1}
The AMISE optimal bandwidth for the directional kernel density estimator (\ref{kdebwd:kernel_directional}) is given by
\begin{align}
h_{\mathrm{AMISE}}=&\lrc{\frac{qd_q(L)}{4b_q(L)^2\lambda_q(L) R(\Psi(f,\cdot)) n}}^{\frac{1}{4+q}},\label{kdebwd:dir:cor:1:1}
\end{align}
where $\lambda_q(L)=2^{\frac{q}{2}-1}\om{q-1}\int_0^{\infty} L(r) r^{\frac{q}{2}-1}\,dr$.
\end{coro}
Unfortunately, expression (\ref{kdebwd:dir:cor:1:1}) can not be used in practise since it depends on the curvature term $R(\Psi(f,\cdot))$ of the unknown density $f$.

\subsection{Available bandwidth selectors}
\label{kdebwd:subsec:bwsels}

The first proposals for data-driven bandwidth selection with directional data are from \cite{Hall1987}, who provide cross-validatory selectors. Specifically, Least Squares Cross-Validation (LSCV) and Likelihood Cross-Validation (LCV) selectors are introduced, arising as the minimizers of the cross-validated estimates of the squared error loss and the Kullback--Leibler loss, respectively. The selectors have the following expressions:
\begin{align*}
h_\mathrm{LSCV}&=\arg\max_{h>0} \mathrm{CV}_2(h),\quad \mathrm{CV}_2(h)=2n^{-1}\sum_{i=1}^n \hat f^{-i}_h(\bX_i)-\Iq{\hat f_h(\bx)^2}{\bx},\\
h_\mathrm{LCV}&=\arg\max_{h>0} \mathrm{CV}_{\mathrm{KL}}(h),\quad \mathrm{CV}_{\mathrm{KL}}(h)=\sum_{i=1}^n \log \hat f^{-i}_h(\bX_i),
\end{align*}
where $\hat f^{-i}_h$ represents the kernel estimator computed without the $i$-th observation. See Remark \ref{kdebwd:rem:3} for an efficient computation of $h_\mathrm{LSCV}$. \\

Recently, \cite{Taylor2008} proposed a plug-in selector for the case of circular data ($q=1$) for the estimator with the von Mises kernel. The selector of \cite{Taylor2008} uses from the beginning the assumption that the reference density is a von Mises to construct the AMISE. This contrasts with the classic rule of thumb selector of \cite{Silverman1986}, which supposes at the end (\textit{i.e.}, after deriving the AMISE expression) that the reference density is a normal. The bandwidth parameter is chosen by first obtaining an estimation $\hat\kappa$ of the concentration parameter $\kappa$ in the reference density (for example, by maximum likelihood) and using the formula
\begin{align*}
h_{\mathrm{TAY}}=\lrc{\frac{4\pi^\frac{1}{2}\mathcal{I}_0(\hat\kappa)^2}{3\hat{\kappa}^2\Ical_2(2\hat\kappa)n}}^\frac{1}{5}.
\end{align*}
Note that the parametrization of \cite{Taylor2008} has been adapted to the context of the estimator (\ref{kdebwd:kernel_directional}) by denoting by $h$ the inverse of the squared concentration parameter employed in his paper.\\

More recently, \cite{Oliveira2012} proposed a selector that improves the performance of \cite{Taylor2008} allowing for more flexibility in the reference density, considering a mixture of von Mises. This selector is also devoted to the circular case and is mainly based on two elements. First, the AMISE expansion that \cite{DiMarzio2009} derived for the circular kernel density estimator by the use of Fourier expansions of the circular kernels. This expression has the following form when the kernel is a circular von Mises (the estimator is equivalent to consider $L(r)=e^{-r}$, $q=1$ and $h$ as the inverse of the squared concentration parameter in (\ref{kdebwd:kernel_directional})):
\begin{align}
\mathrm{AMISE}(h)=\frac{1}{16}\lrc{1-\frac{\Ical_2\big(h^{-2}\big)}{\Ical_0\lrp{h^{-2}}}}^2\int_0^{2\pi} f''(\theta)^2\,d\theta+\frac{\Ical_0\big(2h^{-2}\big)}{2n\pi\Ical_0\lrp{h^{-2}}^2}. \label{kdebwd:dimarzio}
\end{align}
The second element is the Expectation-Maximization (EM) algorithm of \cite{Banerjee2005} for fitting mixtures of directional von Mises. The selector, that will be denoted by $h_{\mathrm{OLI}}$, proceeds as follows:
\begin{enumerate}[label=\textit{\roman{*}}.]
\item Use the EM algorithm to fit mixtures from a determined range of components.
\item Choose the fitted mixture with minimum AIC.
\item Compute the curvature term in (\ref{kdebwd:dimarzio}) using the fitted mixture and seek for the $h$ that minimizes this expression, that will be $h_{\mathrm{OLI}}$.
\end{enumerate}

\section{A new rule of thumb selector}
\label{kdebwd:sec:ruleofthumb}

Using the properties of the von Mises density it is possible to derive a directional analogue to the rule of thumb of \cite{Silverman1986}, which is the optimal AMISE bandwidth for normal reference density and normal kernel. The rule is resumed in the following result.

\begin{prop}[Rule of thumb]
\label{kdebwd:prop:rot}
The curvature term for a von Mises density $\mathrm{vM}(\bmu,\kappa)$ is
\begin{align*}
R(\Psi(f_{\mathrm{vM}}(\cdot;\bmu,\kappa),\cdot))&=\frac{\kappa^{\frac{q+1}{2}}}{2^{q+2}\pi^\frac{q+1}{2}\mathcal{I}_{\frac{q-1}{2}}(\kappa)^2q}\lrc{2q\mathcal{I}_{\frac{q+1}{2}}(2\kappa)+(2+q)\kappa\mathcal{I}_{\frac{q+3}{2}}(2\kappa)}.
\end{align*}
If $\hat\kappa$ is a suitable estimator for $\kappa$, then the rule of thumb selector for the kernel estimator (\ref{kdebwd:kernel_directional}) with a directional kernel $L$ is
\begin{align*}
h_{\mathrm{ROT}}=&\lrc{\frac{q^2d_q(L)2^{q+2}\pi^\frac{q+1}{2}\mathcal{I}_{\frac{q-1}{2}}(\hat\kappa)^2}{\hat\kappa^{\frac{q+1}{2}}4b_q(L)^2\lambda_q(L) \lrp{2q\mathcal{I}_{\frac{q+1}{2}}(2\hat\kappa)+(2+q)\hat\kappa\mathcal{I}_{\frac{q+3}{2}}(2\hat\kappa)} n}}^{\frac{1}{4+q}}.
\end{align*}
If $L$ is the von Mises kernel, then:
\begin{align}
h_{\mathrm{ROT}}=\lb\begin{array}{ll}
\displaystyle\lrc{\frac{4\pi^\frac{1}{2}\mathcal{I}_0(\hat\kappa)^2}{\hat\kappa\lrc{2\Ical_1(2\hat\kappa)+3\hat\kappa\Ical_2(2\hat\kappa)}n}}^\frac{1}{5}, &q=1,\\
\displaystyle\lrc{\frac{8\sinh^2(\hat\kappa)}{\hat\kappa\lrc{(1+4\hat\kappa^2)\sinh(2\hat\kappa)-2\hat\kappa\cosh(2\hat\kappa)}n}}^\frac{1}{6}, &q=2,\\
\displaystyle\lrc{\frac{4\pi^\frac{1}{2}\Ical_{\frac{q-1}{2}}(\hat\kappa)^2}{\hat\kappa^{\frac{q+1}{2}}\lrc{2q\Ical_{\frac{q+1}{2}}(2\hat\kappa)+(2+q)\hat\kappa\Ical_{\frac{q+3}{2}}(2\hat\kappa)}n}}^\frac{1}{4+q}, &q\geq3.
\end{array}\ri\label{kdebwd:rot}
\end{align}
The parameter $\kappa$ can be estimated by maximum likelihood.
\end{prop}

In view of the expression for $h_{\mathrm{ROT}}$ in (\ref{kdebwd:rot}), it is interesting to compare it with $h_\mathrm{TAY}$ when $q=1$. As it can be seen, both selectors coincide except for one difference: the term $2\Ical_1(2\hat\kappa)$ in the sum in the denominator of $h_{\mathrm{ROT}}$. This ``extra term'' can be explained by examining the way that both selectors are derived. Whereas the selector $h_{\mathrm{ROT}}$ derives the bandwidth supposing that the reference density is a von Mises when the AMISE is already derived in a general way, the selector $h_\mathrm{TAY}$ uses the von Mises assumption to compute it. Therefore, it is expected that the selector $h_\mathrm{ROT}$  will be more robust against deviations from the von Mises density.\\

\begin{figure}[h!]
\centering
\includegraphics[scale=0.45]{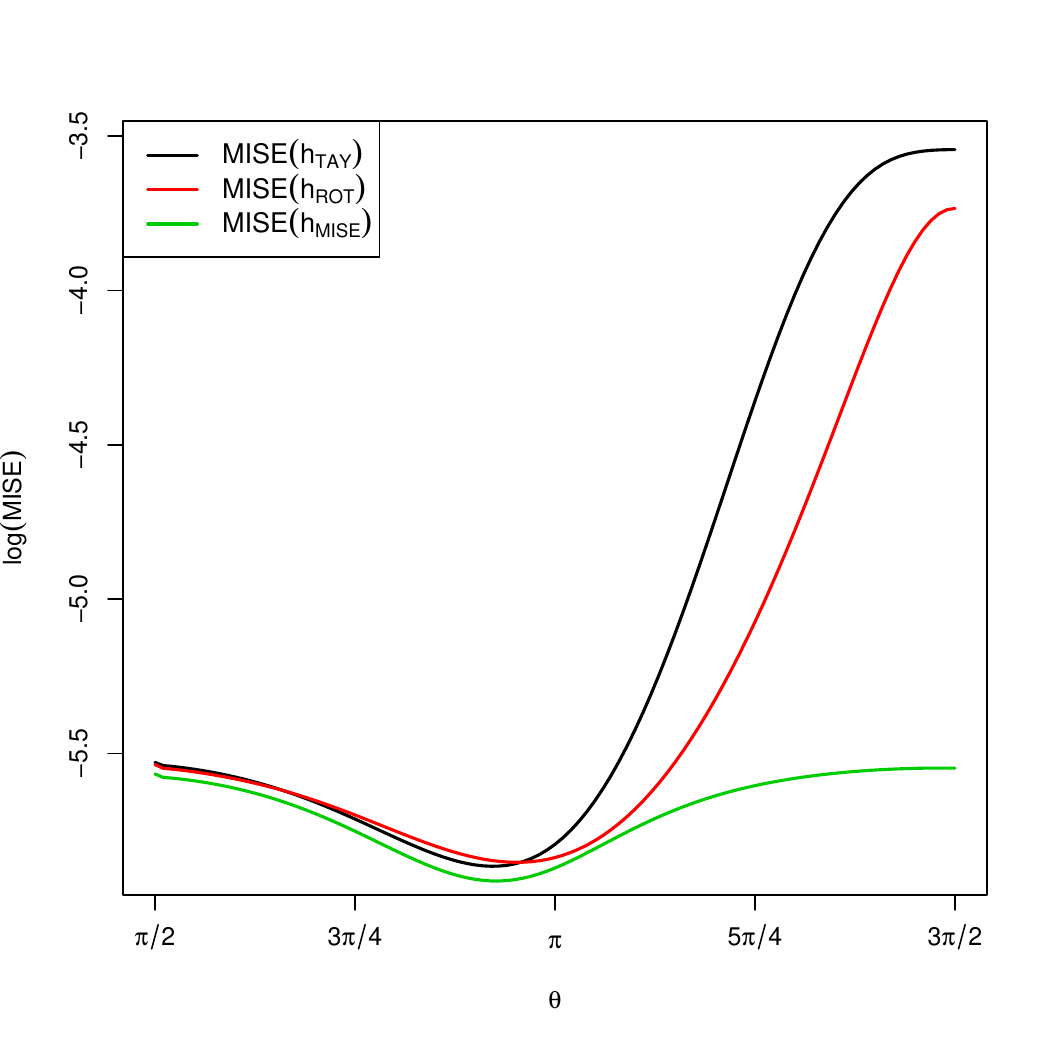}
\includegraphics[scale=0.45]{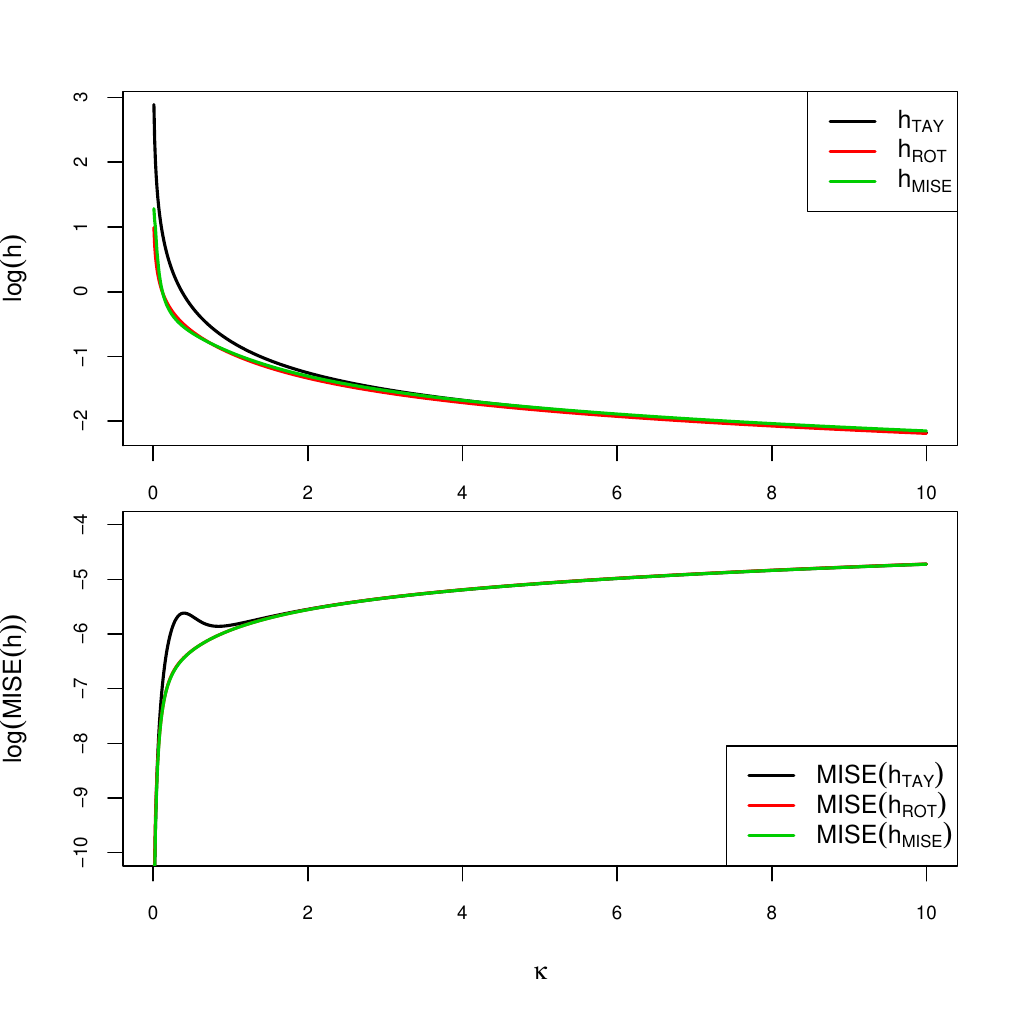}
\caption{\small The effect of the ``extra term'' in $h_\mathrm{ROT}$. Left plot: logarithm of the curves of $\mathrm{MISE}(h_\mathrm{TAY})$, $\mathrm{MISE}(h_\mathrm{ROT})$ and $\mathrm{MISE}(h_\mathrm{MISE})$ for sample size $n=250$. The curves are computed by $1000$ Monte Carlo samples and $h_\mathrm{MISE}$ is obtained exactly. The abscissae axis represents the variation of the parameter $\theta\in\lrc{\frac{\pi}{2},\frac{3\pi}{2}}$, which indexes the reference density $\frac{1}{2}\mathrm{vM}\lrp{(0,1),2}+\frac{1}{2}\mathrm{vM}\lrp{(\cos(\theta),\sin(\theta)),2}$. Right plot: logarithm of $h_\mathrm{TAY}$, $h_\mathrm{ROT}$, $h_\mathrm{MISE}$ and their corresponding MISE for different values of $\kappa$, with $n=250$.\label{kdebwd:fig:vs}}
\end{figure}

Figure \ref{kdebwd:fig:vs} collects two graphs exposing these comments, that are also corroborated in Section \ref{kdebwd:sec:comparative}. The left plot shows the MISE for $h_{\mathrm{TAY}}$ and $h_{\mathrm{ROT}}$ for the density $\frac{1}{2}\mathrm{vM}\lrp{(0,1),2}+\frac{1}{2}\mathrm{vM}\lrp{(\cos(\theta),\sin(\theta)),2}$, where $\theta\in\big[\frac{\pi}{2},\frac{3\pi}{2}\big]$. This model represents two equally concentrated von Mises densities that spread off from being the same to being antipodal. As it can be seen, the $h_\mathrm{ROT}$ selector is slightly more accurate when the von Mises model holds ($\theta=\frac{\pi}{2}$) and when the deviation is large ($\theta\in\big[\pi,\frac{3\pi}{2}\big]$). When $\theta\in\big[\frac{\pi}{2},\pi\big]$, both selectors perform similar. This graph also illustrates the main problem of these selectors: the von Mises density is not flexible enough to capture densities with multimodality and it approximates them by the flat uniform density.\\

When the density is a $\mathrm{vM}(\bmu,\kappa)$, the right plot of Figure \ref{kdebwd:fig:vs} shows the output of $h_\mathrm{TAY}$, $h_\mathrm{ROT}$, $h_\mathrm{MISE}$ and their corresponding errors with respect to $\kappa$. The effect of the ``extra term'' is visible for low values of $\kappa$, where $\mathrm{MISE}(h_\mathrm{TAY})$ presents a local maxima. This corresponds with higher values of $h_\mathrm{TAY}$ with respect to $h_\mathrm{ROT}$ and $h_\mathrm{MISE}$, which means that the former produces oversmoothed estimations of the density (\textit{i.e.} tend to the uniform case faster). Despite the worse behaviour of $h_\mathrm{TAY}$, when the concentration parameter increases the effect of the ``extra term'' is mitigated and both selectors are almost the same. 

\section{Selectors based on mixtures}
\label{kdebwd:sec:mixtures}

The results of the previous section show that, although the rule of thumb presents a significant improvement with respect to the \cite{Taylor2008} selector in terms of generality and robustness, it also shares the same drawbacks when the underlying density is not the von Mises model (see Figure \ref{kdebwd:fig:vs}). To overcome these problems, two alternatives for improving $h_\mathrm{ROT}$ will be considered.\\

The first one is related with improving the reference density to plug-in into the curvature term. The von Mises density has been proved to be not flexible enough to estimate properly the curvature term in (\ref{kdebwd:dir:cor:1:1}). This is specially visible when the underlying model is a mixture of antipodal von Mises, but the estimated curvature term is close to zero (the curvature of a uniform density). A modification in this  direction is to consider a suitable mixture of von Mises for the reference density, that will be able to capture the curvature of rather complex underlying densities. This idea was employed first by \cite{Cwik1997} considering mixtures of multivariate normals and by \cite{Oliveira2012} in the circular setting.\\

The second improvement is concerned with the error criterion for the choice of the bandwidth. Until now, the error criterion considered was the AMISE, which is the usual in the literature of kernel smoothing. However, as \cite{Marron1992} showed for the linear case and \cite{Garcia-Portugues:dirlin} did for the directional situation, the AMISE and MISE may differ significantly for moderate and even large sample sizes, with a potential significative misfit between $h_\mathrm{AMISE}$ and $h_\mathrm{MISE}$. Then, a substantial decreasing of the error of the estimator (\ref{kdebwd:kernel_directional}) is likely to happen if the bandwidth is obtained from the exact MISE, instead of the asymptotic version. Obviously, the problem of this new approach is how to compute exactly the MISE, but this can be done if the reference density is a mixture of von Mises. \\

The previous two considerations, improve the reference density and the error criterion, will lead to the bandwidth selectors of Asymptotic MIxtures (AMI), denoted by $h_\mathrm{AMI}$, and Exact MIxtures (EMI), denoted by  $h_\mathrm{EMI}$. Before explaining in detail the two proposed selectors, it is required to introduce some notation on mixtures of von Mises.\\

\pagebreak

An $M$-mixture of von Mises densities with means $\bmu_j$, concentration parameters $\kappa_j$ and weights $p_j$,\nopagebreak[4] with $j=1,\ldots,M$, is denoted by
\begin{align}
f_M(\bx)=\sum_{j=1}^M p_j f_{\mathrm{vM}}(\bx;\bmu_j,\kappa_j),\quad \sum_{j=1}^M p_j=1,\quad p_j\geq 0.\label{kdebwd:mise:mvm}
\end{align}
When dealing with mixtures, the tuning parameter is the number of components, $M$, which can be estimated from the sample. The notation $f_{\widehat M}$ will be employed to represent the mixture of $\widehat{M}$ components where the parameters are estimated and $\widehat{M}$ is obtained from the sample. The details of this fitting are explained later in Algorithm \ref{kdebwd:algo:nm}.\\

Then, the AMI selector follows from modifying the rule of thumb selector to allow fitted mixtures of von Mises. It is stated in the next procedure.
\begin{algo}[AMI selector]
\label{kdebwd:algo:ami}
Let $\bX_1,\ldots,\bX_n$ be a random sample of a directional variable\nolinebreak[4] $\bX$.
\begin{enumerate}[label=\textit{\roman{*}}.]
\item Compute a suitable estimation $f_{\widehat M}$ using Algorithm \ref{kdebwd:algo:nm}.
\item For a directional kernel $L$, set
\begin{align*}
h_{\mathrm{AMI}}=\lrc{\frac{qd_q(L)}{4b_q(L)^2\lambda_q(L) R\Big(\Psi\big(f_{\widehat M},\cdot\big)\Big) n}}^{\frac{1}{4+q}}
\end{align*}
and for the von Mises kernel,
\begin{align*}
h_{\mathrm{AMI}}=\lrc{q2^q\pi^{\frac{q}{2}} R\Big(\Psi\big(f_{\widehat M},\cdot\big)\Big) n}^{-\frac{1}{4+q}}.
\end{align*}
\end{enumerate}
\end{algo}

\begin{rem}
\label{kdebwd:rem:1}
Unfortunately, the curvature term $R\big(\Psi\big(f_{\widehat M},\cdot\big)\big)$ does not admit a simple closed expression, unless for the case where $\widehat M=1$, \textit{i.e.}, when $h_\mathrm{AMI}$ is equivalent to $h_\mathrm{ROT}$. This is due to the cross-product terms between the derivatives of the mixtures that appear in the integrand. However, this issue can be bypassed by using either numerical integration in $q$-spherical coordinates or Monte Carlo integration to compute $R\big(\Psi\big(f_{\widehat M},\cdot\big)\big)$ for any $\widehat M$.
\end{rem}

The EMI selector relies on the exact expression of the MISE for densities of the type (\ref{kdebwd:mise:mvm}), that will be denoted by
\begin{align*}
\mathrm{MISE}_M(h)=\E{\Iq{\lrp{\hat f_h(\bx)-f_M(\bx)}^2}{\bx}}.
\end{align*}
Similarly to what \cite{Marron1992} did for the linear case, \cite{Garcia-Portugues:dirlin} derived the closed expression of $\mathrm{MISE}_M(h)$ when the directional kernel is the von Mises one. The calculations are based on the convolution properties of the von Mises, which unfortunately are not so straightforward as the ones for the normal, resulting in more complex expressions.

\begin{prop}[\cite{Garcia-Portugues:dirlin}]
\label{kdebwd:mise:th:1}
Let $f_M$ be the density of an $M$-mixture of directional von Mises (\ref{kdebwd:mise:mvm}). The exact MISE of the directional kernel estimator (\ref{kdebwd:kernel_directional}) with von Mises kernel and obtained from a random sample of size $n$ is
\begin{align}
\mathrm{MISE}_M(h)=\lrp{D_q(h)n}^{-1}+\mathbf{p}^T\lrc{(1-n^{-1})\mathbf{\Psi_2}(h)-2\mathbf{\Psi_1}(h)+\mathbf{\Psi_0}(h)}\mathbf{p}, \label{kdebwd:mise:mvm:1}
\end{align}
where $\mathbf{p}=\lrp{p_1,\ldots,p_M}^T$ and $D_q(h)=C_q\lrp{1/h^2}^2C_q\lrp{2/h^2}^{-1}$. The matrices $\mathbf{\Psi_a}(h)$, $a=0,1,2$ have entries:
\begin{align*}
\mathbf{\Psi_0}(h)=&\lrp{\frac{C_q(\kappa_i)C_q(\kappa_j)}{C_q\big(||\kappa_i\bmu_i+\kappa_j\bmu_j||\big)}}_{ij},\,\mathbf{\Psi_1}(h)= \lrp{\Iq{\frac{C_q\lrp{1/h^2}C_q(\kappa_i)C_q(\kappa_j)}{C_q\lrp{\norm {\bx/h^2+\kappa_i\bmu_i}}}e^{\kappa_j\bx^T\bmu_j}}{\bx}}_{ij},\\
\mathbf{\Psi_2}(h)=& \lrp{\Iq{\frac{C_q\lrp{1/h^2}^2C_q(\kappa_i)C_q(\kappa_j)}{C_q\lrp{||\bx/h^2+\kappa_i\bmu_i||}C_q\big(||\bx/h^2+\kappa_j\bmu_j||\big)}}{\bx}}_{ij},
\end{align*}
where $C_q$ is defined in equation (\ref{kdebwd:dir:cq}).
\end{prop}

\begin{rem}
\label{kdebwd:rem:2}
A more efficient way to implement (\ref{kdebwd:mise:mvm:1}), specially for large sample sizes and higher dimensions, is the following expression:
\begin{align*}
\mathrm{MISE}_M(h)&=\lrp{D_q(h)n}^{-1}+\Iq{\lrb{\lrp{\E{\hat f_h(\bx)}-f_M(\bx)}^2-\E{\hat f_h(\bx)}^2}}{\bx},
\end{align*}
where the integral is either evaluated numerically using $q$-spherical coordinates or Monte Carlo integration and $\E{\hat f_h(\bx)}$ is computed using 
\begin{align*}
\E{\hat f_h(\bx)}&=\sum_{j=1}^M p_j \frac{C_q(\kappa_j)C_q\lrp{1/h^2}}{C_q\big(||\bx/h^2+\kappa_j\bmu_j||\big)}.
\end{align*}
\end{rem}

\begin{rem}
\label{kdebwd:rem:3}
By the use of similar techniques, when the kernel is von Mises, the LSCV selector admits an easier expression for the CV$_2$ loss that avoids the calculation of the integral of $\hat f_h^{-i}$:
\begin{align*}
\mathrm{CV}_2(h)=\frac{2C_q\lrp{1/h^2}}{n}\sum_{i=1}^n\sum_{j>i}^n\lrc{ \frac{2}{n-1}e^{\bX_i^T\bX_j/h^2}-\frac{C_q\lrp{1/h^2}}{n C_q\lrp{\norm{\bX_i+\bX_j}/h^2}}}-\frac{C_q(1/h^2)^2}{nC_q(2/h^2)}.
\end{align*}
\end{rem}

Based on the previous result, the philosophy of the EMI selector is the following: using a suitable pilot parametric estimation of the unknown density (given by Algorithm \ref{kdebwd:algo:nm}), build the exact MISE and obtain the bandwidth that minimizes it. This is summarized in the following procedure.

\begin{algo}[EMI selector]
\label{kdebwd:algo:emi}
Consider the von Mises kernel and let $\bX_1,\ldots,\bX_n$ be a random sample of a directional variable $\bX$.
\begin{enumerate}[label=\textit{\roman{*}}.]
\item Compute a suitable estimation $f_{\widehat M}$ using Algorithm \ref{kdebwd:algo:nm}.
\item Obtain $h_{\mathrm{EMI}}=\arg\min_{h>0}\mathrm{MISE}_{\widehat M}(h)$.
\end{enumerate}
\end{algo}

\subsection{Mixtures fitting and selection of the number of components}
\label{kdebwd:subsec:fitting}

The EM algorithm of \cite{Banerjee2005}, implemented in the \texttt{R} package \texttt{movMF} (see \cite{Hornik2014}), provides a complete solution to the problem of estimation of the parameters in a mixture of directional von Mises of dimension $q$. However, the issue of selecting the number of components of the mixture in an automatic and optimal way is still an open problem. \\

The propose considered in this work is an heuristic approach based on the Bayesian Information Criterion (BIC), defined as $\text{BIC}=-2\mathit{l}+k\log n$, where $\mathit{l}$ is the log-likelihood of the model and $k$ is the number of parameters. The procedure looks for the fitted mixture with a number of components $M$ that minimizes the BIC. This problem can be summarized as the global minimization of a function (BIC) defined on the naturals (number of components).  \\

The heuristic procedure starts by fitting mixtures from $M=1$ to $M=M_B$, computing their BIC and providing $\widehat M$, the number of components with minimum BIC. Then, in order to ensure that $\widehat M$ is a global minimum and not a local one, $M_N$ neighbours next to $\widehat M$ are explored (\textit{i.e.} fit mixture, compute BIC and update $\widehat M$), if they were not previously explored. This procedure continues until $\widehat M$ has at least $M_N$ neighbours at each side with larger BICs. A reasonable compromise for $M_B$ and $M_N$, checked by simulations, is to set $M_B=\lfloor\log n\rfloor$ and $M_N=3$. In order to avoid spurious solutions, fitted mixtures with any $\kappa_j>250$ are removed. The procedure is detailed as follows.

\begin{algo}[Mixture estimation with data-driven selection of the number of components]
\label{kdebwd:algo:nm}
Let $\bX_1,\ldots,\bX_n$ be a random sample of a directional variable $\bX$ with density $f$.
\begin{enumerate}[label=\textit{\roman{*}}., ref=\textit{\roman{*}}]
\item Set $M_B=\lfloor\log n\rfloor$ and $M_N$ as the user supplies, usually $M_N=3$. \label{kdebwd:algo:nm:1}
\item For $M$ varying from $1$ to $M_B$, \label{kdebwd:algo:nm:2}
\begin{enumerate}
\item estimate the $M$-mixture with the EM algorithm of \cite{Banerjee2005} and
\item compute the BIC of the fitted mixture.
\end{enumerate}
\item Set $\widehat M$ as the number of components of the mixture with lower BIC.  \label{kdebwd:algo:nm:3}
\item If $M_B-M_N<\widehat M$, set $M_B=M_B+1$ and turn back to step \ref{kdebwd:algo:nm:2}. Otherwise, end with the final estimation $f_{\widehat M}$. \label{kdebwd:algo:nm:4}
\end{enumerate}
\end{algo}
Other informative criteria, such as the Akaike Information Criterion (AIC) and its corrected version, AICc, were checked in the simulation study together with BIC. The BIC turned out to be the best choice to use with the AMI and EMI selectors, as it yielded the minimum errors.

\section{Comparative study}
\label{kdebwd:sec:comparative}

Along this section, the three new bandwidth selectors will be compared with the already proposed selectors described in Subsection \ref{kdebwd:subsec:bwsels}. A collection of directional models, with their corresponding simulation schemes, are considered. Subsection \ref{kdebwd:subsec:dirmods} is devoted to comment the directional models used in the simulation study (all of them are defined for any arbitrary dimension $q$, not just for the circular or spherical case). These models are also described in the appendix. \\

For each of the different combinations of dimension, sample size and model, the MISE of each selector was estimated empirically by $1000$ Monte Carlo samples, with the same seed for the different selectors. This is used in the computation of $\mathrm{MISE}(h_\mathrm{MISE})$, where $h_\mathrm{MISE}$ is obtained as a numerical minimization of the estimated MISE. The calculus of the ISE was done by: Simpson quadrature rule with $2000$ discretization points for $q=1$; \cite{Lebedev} rule  with $5810$ nodes for $q=2$ and Monte Carlo integration with $10000$ sampling points for $q>2$ (same seed for all the integrations). Finally, the kernel considered in the study is the von Mises.%

\subsection{Directional models}
\label{kdebwd:subsec:dirmods}

The first models considered are the uniform density in $\Om{q}$ and the von Mises density given in (\ref{kdebwd:dir:cq}). The analogous of the von Mises for axial data (\textit{i.e.}, directional data where $f(\bx)=f(-\bx)$) is the Watson distribution $\mathrm{W}(\bmu,\kappa)$ \citep{Mardia2000}:
\begin{align*}
f_{\mathrm{W}}(\bx;\bmu,\kappa)=M_q(\kappa)\exp\lrb{\kappa (\bx^T\bmu)^2},
\end{align*}
where $M_q(\kappa)=\big(\om{q-1}\int_{-1}^1 e^{\kappa t^2}(1-t^2)^{\frac{q}{2}-1}\,dt\big)^{-1}$. This density has two antipodal modes: $\bmu$ and $-\bmu$, both of them with concentration parameter $\kappa\geq0$. A further extension of this density is the called Small Circle distribution $\mathrm{SC}(\bmu,\tau,\nu)$ \citep{Bingham1978}:
\begin{align*}
f_{\mathrm{SC}}(\bx;\bmu,\tau,\nu)=A_q(\tau,\nu)\exp\lrb{-\tau (\bx^T\bmu-\nu)^2},
\end{align*}
where $A_q(\tau,\nu)=\big(\om{q-1}\int_{-1}^1 e^{-\tau (t-\nu)^2}(1-t^2)^{\frac{q}{2}-1}\,dt\big)^{-1}$, $\nu\in(-1,1)$ and $\tau\in\R$. For the case $\tau\geq0$, this density has a kind of modal strip along the $(q-1)$-sphere $\big\{\bx\in\Om{q}:\bx^T\bmu=\nu\big\}$.\\

A common feature of all these densities is that they are rotationally symmetric, that is, their contourlines are $(q-1)$-spheres orthogonal to a particular direction. This characteristic can be exploited by means of the so called tangent-normal decomposition (see \cite{Mardia2000}), that leads to the change of variables
\begin{align}
\lb\begin{array}{l}
\bx=t\bmu+(1-t^2)^\frac{1}{2}\bB_{\bmu}\bxi,\\
\om{q}(d\bx)=(1-t^2)^{\frac{q}{2}-1}\,dt\,\om{q-1}(d\bxi),
\end{array}\ri\label{kdebwd:change}
\end{align}
where  $\bmu\in\Om{q}$ is a fixed vector, $t=\bmu^T\bx$ (measures the distance of $\bx$ from $\bmu$), $\bxi\in\Om{q-1}$ and $\bB_{\bmu} = (\bb_1,\ldots, \bb_q)_{(q+1)\times q}$ is the semi-orthonormal matrix ($\bB_{\bmu}^T\bB_{\bmu}=\bI_q$ and $\bB_{\bmu}\bB_{\bmu}^T=\bI_{q+1}-\bmu\bmu^T$, with $\bI_q$ the $q$-identity matrix) resulting from the completion of $\bmu$ to the orthonormal basis $\lrb{\bmu,\bb_1,\ldots,\bb_q}$. The family of rotationally symmetric densities can be parametrized as
\begin{align}
f_{g_{\btheta},\bmu}(\bx)=g_{\btheta}(\bmu^T\bx),\label{kdebwd:rotsym}
\end{align}
where $g_{\btheta}$ is a function depending on a vector parameter $\btheta\in\bTheta\subset\R^p$ and such that $\om{q-1} g_{\btheta}(t) (1-t^2)^{\frac{q}{2}-1}$ is a density in $(-1,1)$, for all $\btheta\in\bTheta$. Using this property, it is easy to simulate from (\ref{kdebwd:rotsym}).

\begin{algo}[Sampling from a rotationally symmetric density]
\label{kdebwd:algo:rotsym}
Let be the rotationally symmetric density (\ref{kdebwd:rotsym}) and consider the notation of (\ref{kdebwd:change}).
\begin{enumerate}[label=\textit{\roman{*}}., ref=\textit{\roman{*}}]
\item Sample $T$ from the density $\om{q-1} g_{\btheta}(t) (1-t^2)^{\frac{q}{2}-1}$.\label{kdebwd:algo:rotsym:1}
\item Sample $\bxi$ from a uniform in $\Om{q-1}$ $(\Om{0}=\lrb{-1,1})$.\label{kdebwd:algo:rotsym:2}
\item $T\bmu+(1-T^2)^\frac{1}{2}\bB_{\bmu}\bxi$ is a sample from $f_{g_{\btheta},\bmu}$.\label{kdebwd:algo:rotsym:3}
\end{enumerate}
\end{algo}

\begin{rem}
Step \ref{kdebwd:algo:rotsym:1} can always be performed using the inversion method \citep{Johnson1987}. This approach can be computationally expensive: it involves solving the root of the distribution function, which is computed from an integral evaluated numerically if no closed expression is available. A reasonable solution to this (for a fixed choice of $g_{\btheta}$ and $\bmu$) is to evaluate once the quantile function in a dense grid (for example, $2000$ points equispaced in $(0,1)$), save the grid and use it to interpolate using cubic splines the new evaluations, which is computationally fast.
\end{rem}

Extending these ideas for rotationally symmetric models, two new directional densities are proposed. The first one is the Directional Cauchy density $\mathrm{DC}(\bmu,\kappa)$, defined as an analogy with the usual Cauchy distribution as
\begin{align*}
f_{\mathrm{DC}}(\bx;\bmu,\kappa)=\frac{1}{D_q(\kappa)(1+2\kappa(1-\bx^T\bmu))},\quad D_q(\kappa)=\left\{\begin{array}{ll}
2\pi\lrp{1+4\kappa}^{-1/2},&q=1,\\
\pi\log(1+4\kappa)\kappa^{-1},&q=2,\\
\om{q-1}\int_{-1}^1 \frac{(1-t^2)^{\frac{q}{2}-1}}{1+2\kappa(1-t)}\,dt,&q>2,\\
\end{array}\ri
\end{align*}
where $\bmu$ is the mode direction and $\kappa\geq0$ the concentration parameter around it ($\kappa=0$ gives the uniform density). This density shares also some of the characteristics of the usual Cauchy distribution: high concentration around a peaked mode and a power decay of the density. The other proposed density is the Skew Normal Directional density $\mathrm{SND}(\bmu,m,\sigma,\lambda)$,
\begin{align*}
f_{\mathrm{SND}}(\bx;\bmu,m,\sigma,\lambda)=S_q(m,\sigma,\lambda)g_{m,\sigma,\lambda}(\bmu^T\bx),\, S_q(m,\sigma,\lambda)\!=\!\lrp{\om{q-1}\!\!\int_{-1}^1 g_{m,\sigma,\lambda}(t)(1-t^2)^{\frac{q}{2}-1}\,dt}^{\!\!-1}\!\!\!\!,
\end{align*}
where $g_{m,\sigma,\lambda}$ is the skew normal density of \cite{Azzalini1985} with location $m$, scale $\sigma$ and shape $\lambda$ that is truncated to the interval $(-1,1)$. The density is inspired by the wrapped skew normal distribution of \cite{Pewsey2006}, although it is based on the rotationally symmetry rather than in wrapping techniques. A particular form of this density is an homogeneous ``cap'' in a neighbourhood of $\bmu$ that decreases very fast outside of it. \\

Non rotationally symmetric densities can be created by mixtures of rotationally symmetric. However, it is interesting to introduce a purely non rotationally symmetric density: the Projected Normal distribution of \cite{Pukkila1988}. Denoted by $\mathrm{PN}(\bmu,\bSigma)$, the corresponding density\nolinebreak[4] is
\begin{align*}
f_{\mathrm{PN}}(\bx;\bmu,\bSigma)=\lrp{2\pi}^{-\frac{p}{2}}|\bSigma|^{-\frac{1}{2}}\mathbf{Q}_3^{-\frac{p}{2}} I_p\lrp{\mathbf{Q}_2\mathbf{Q}^{-\frac{1}{2}}_3}\exp\lrb{-2^{-1}\lrp{\mathbf{Q}_1-\mathbf{Q}^2_2\mathbf{Q}^{-1}_3}},
\end{align*}
where $\mathbf{Q}_1=\bx^T\bSigma^{-1}\bx$, $\mathbf{Q}_2=\bmu^T\bSigma^{-1}\bx$, $\mathbf{Q}_3=\bmu^T\bSigma^{-1}\bmu$ and $I_p(\alpha)=\int_0^\infty t^{p-1}\exp\lrb{-2^{-1}(t-\alpha)^2}dt$.

Sampling from this distribution is extremely easy: just sample $\bX\sim \mathcal{N}\lrp{\bmu,\bSigma}$ and then project $\bX$ to $\Om{q}$ by $\bX/\norm{\bX}$.\\

The whole collection of models, with $20$ densities in total, are detailed in Table \ref{kdebwd:tab:models} in Appendix \ref{kdebwd:ap:models}. Figures \ref{kdebwd:fig:circ} and \ref{kdebwd:fig:sph} show the plots of these densities for the circular and spherical cases.

\subsection{Circular case}
\label{kdebwd:subsec:circular}

\begin{figure}[H]
	\centering
	\vspace{-0.3cm}
	\includegraphics[width=0.225\textwidth]{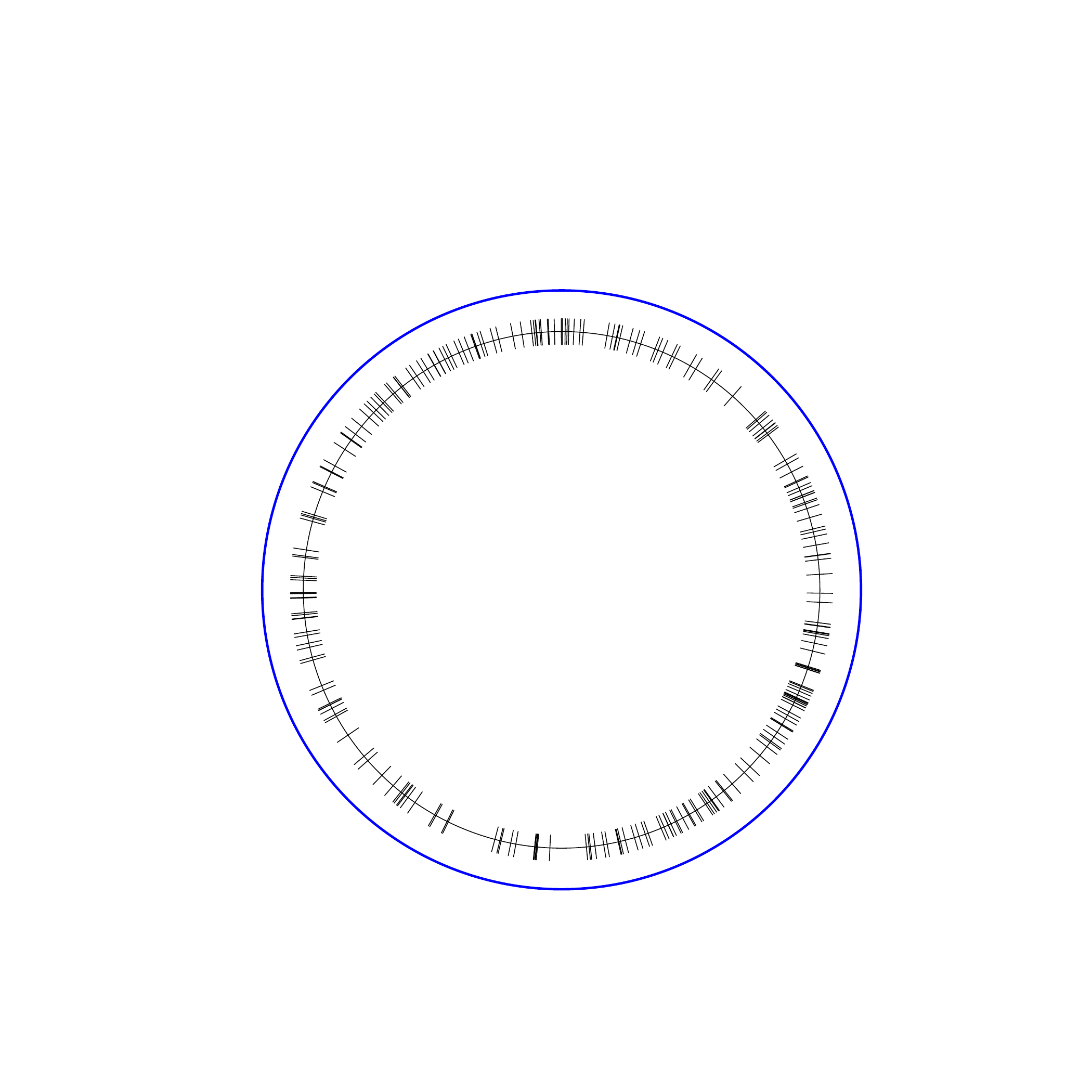}
	\includegraphics[width=0.225\textwidth]{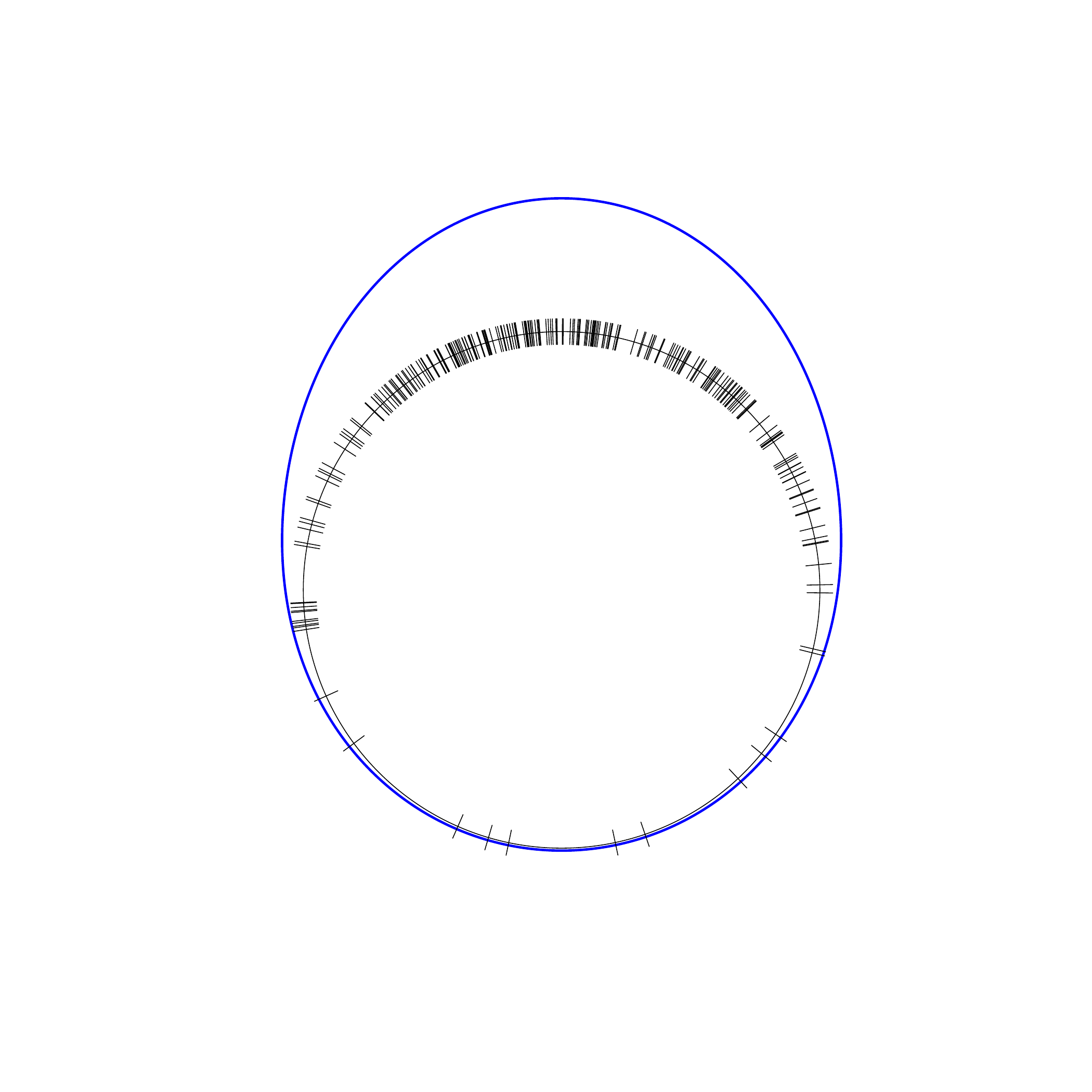}
	\includegraphics[width=0.225\textwidth]{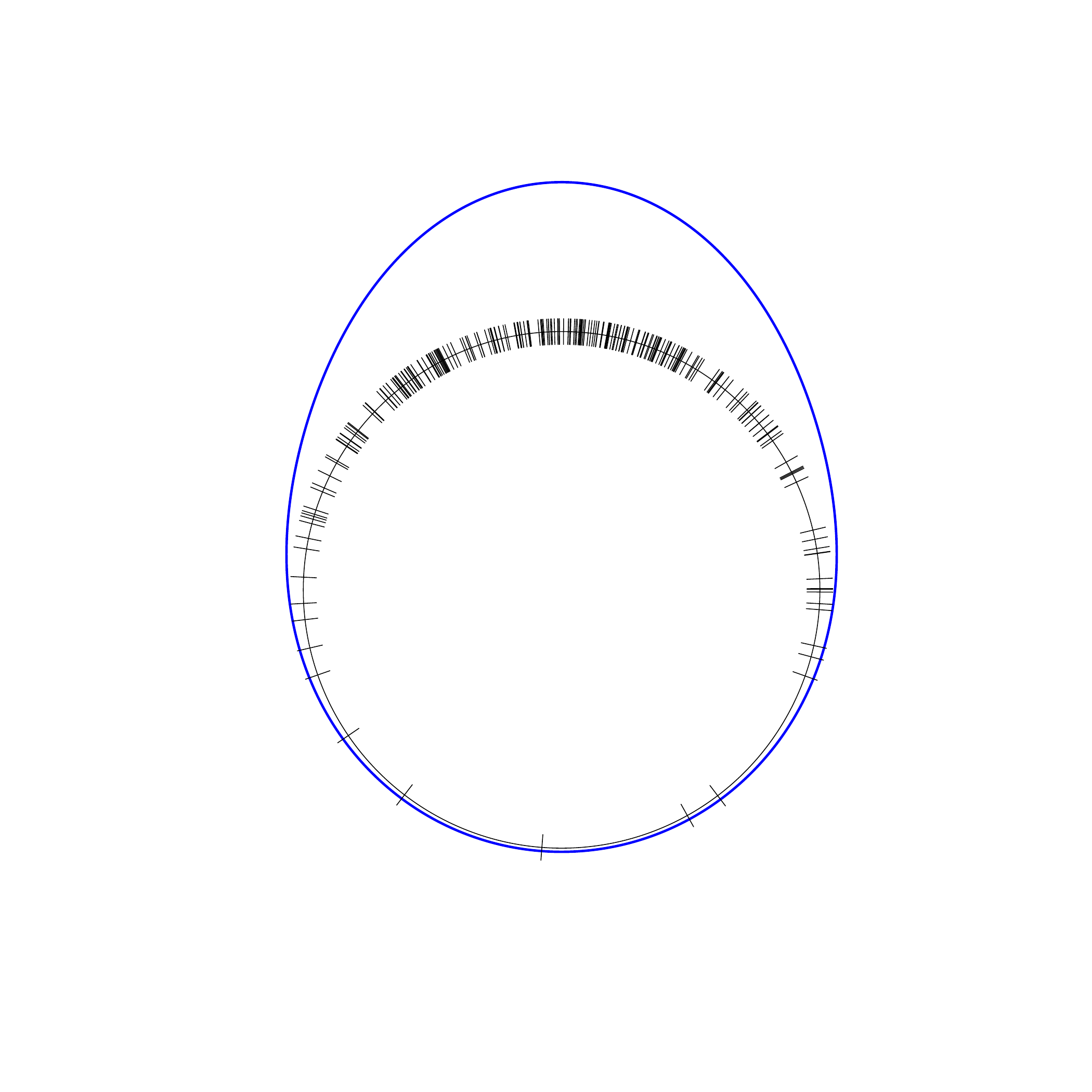}
	\includegraphics[width=0.225\textwidth]{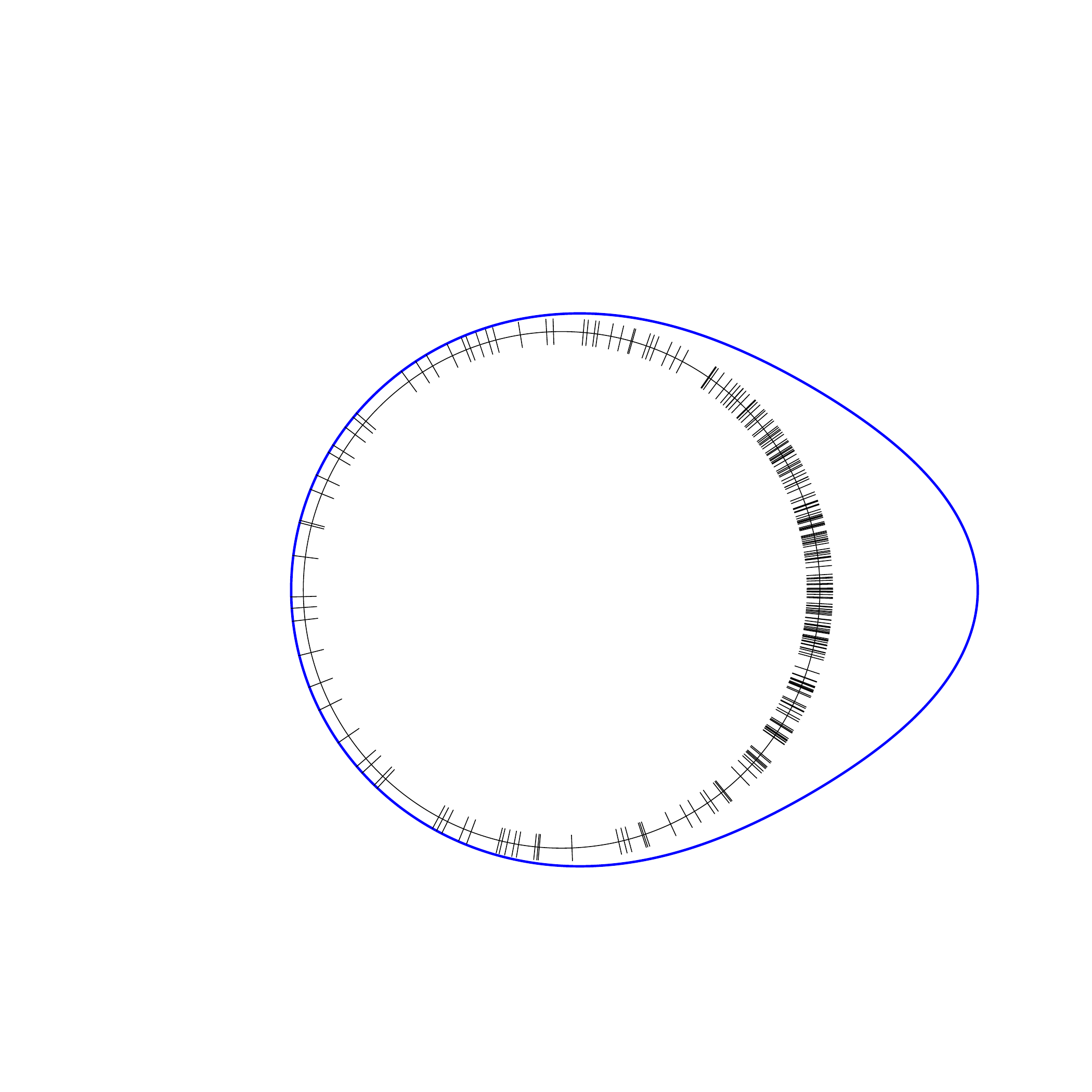}\\
	\includegraphics[width=0.225\textwidth]{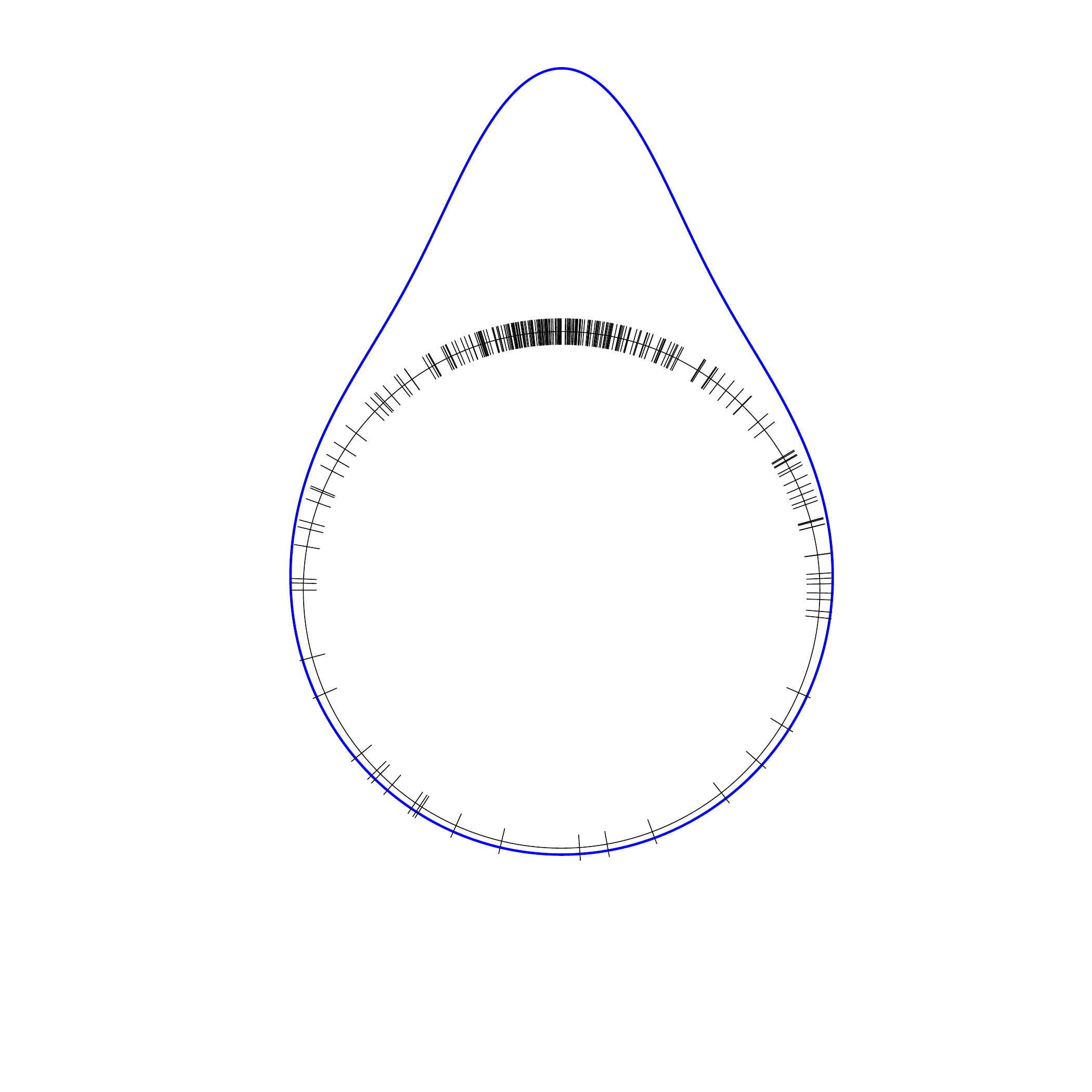}
	\includegraphics[width=0.225\textwidth]{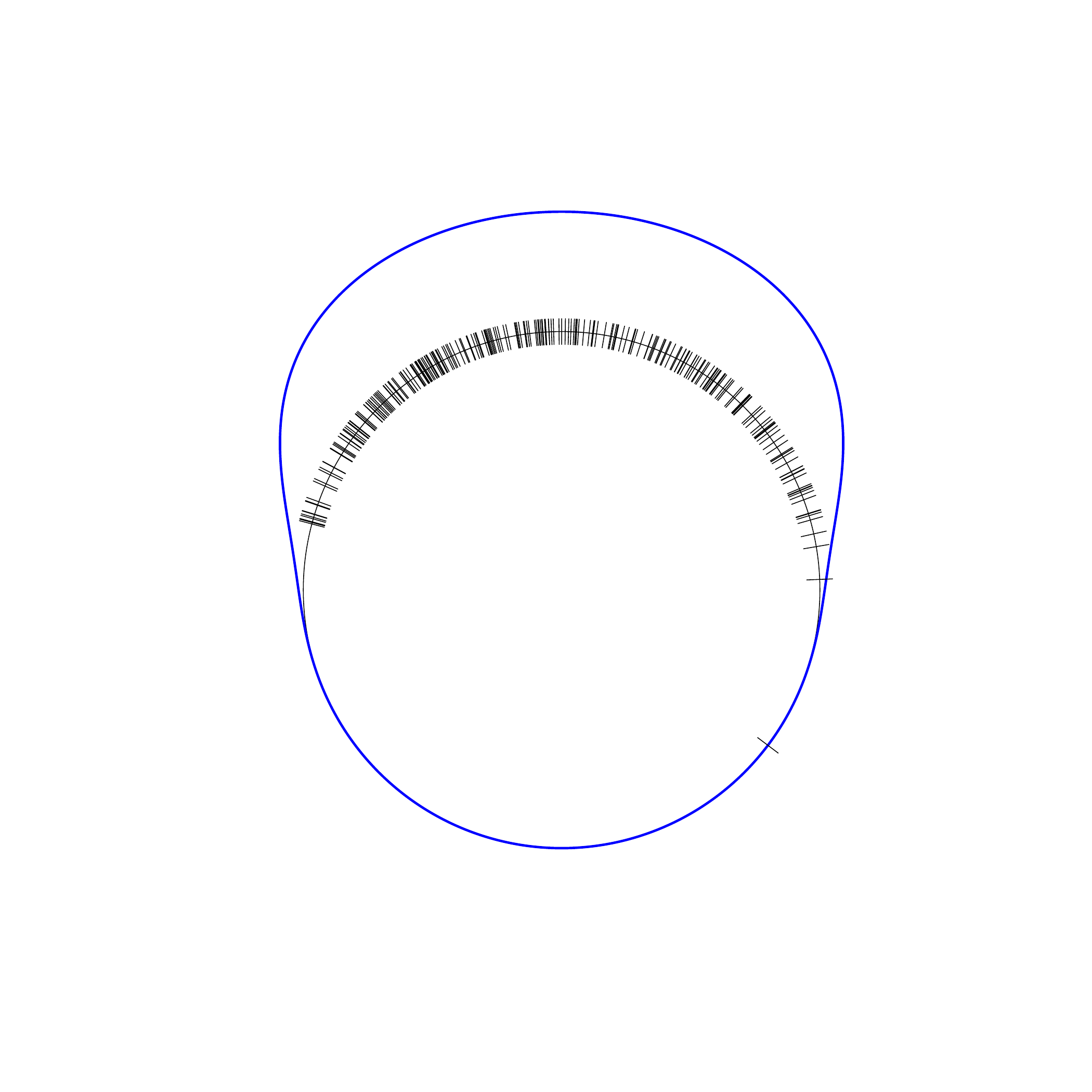}
	\includegraphics[width=0.225\textwidth]{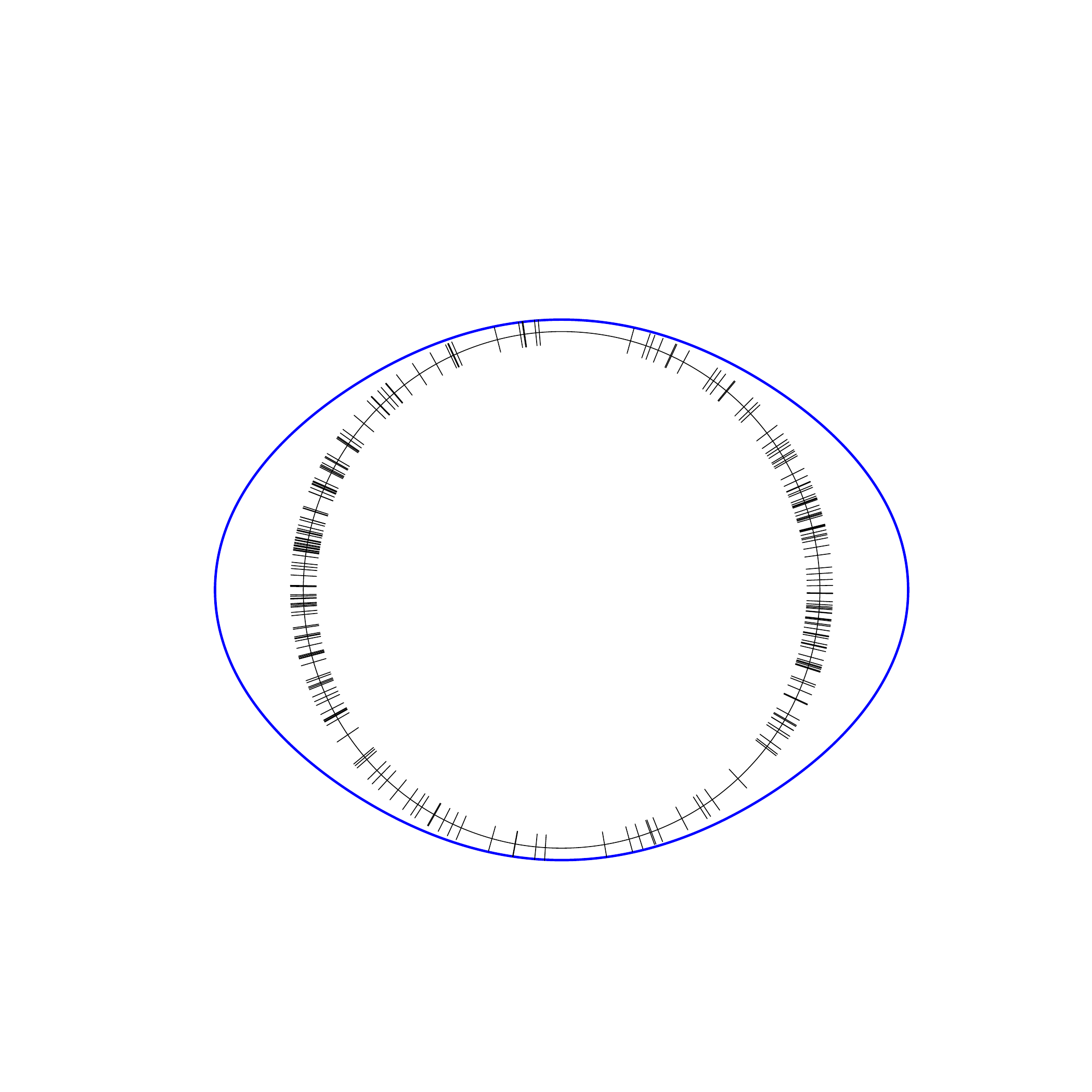}
	\includegraphics[width=0.225\textwidth]{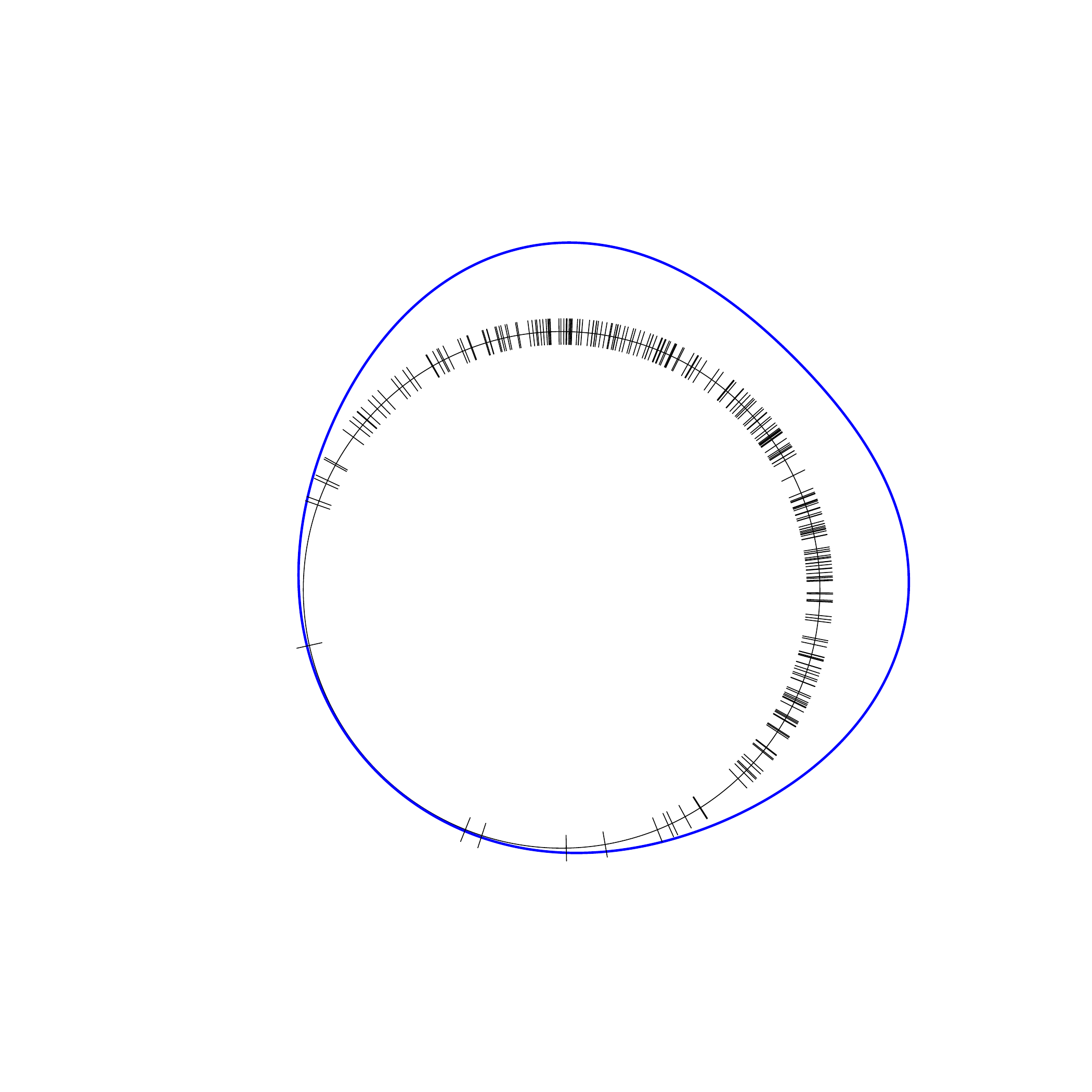}\\
	\includegraphics[width=0.225\textwidth]{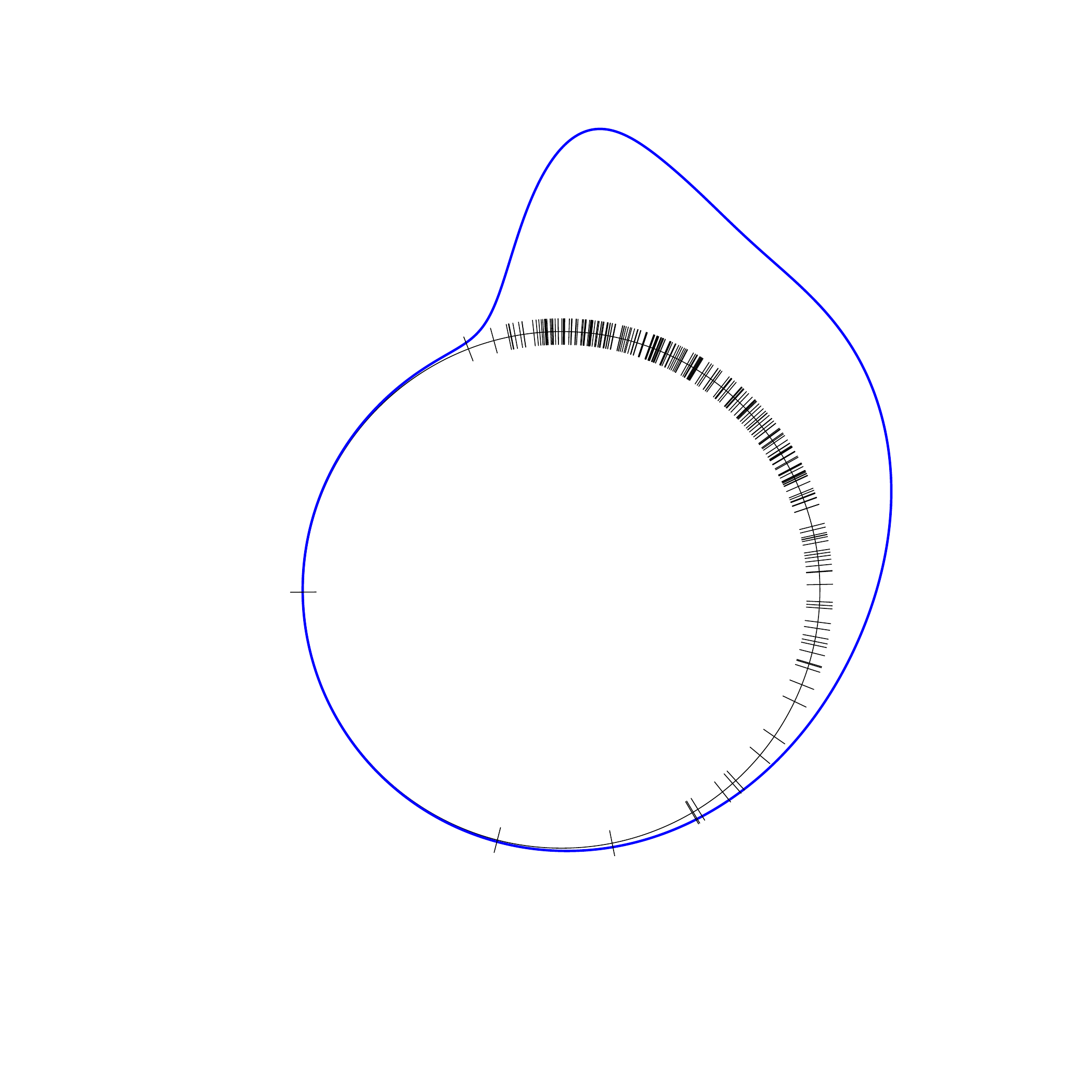}
	\includegraphics[width=0.225\textwidth]{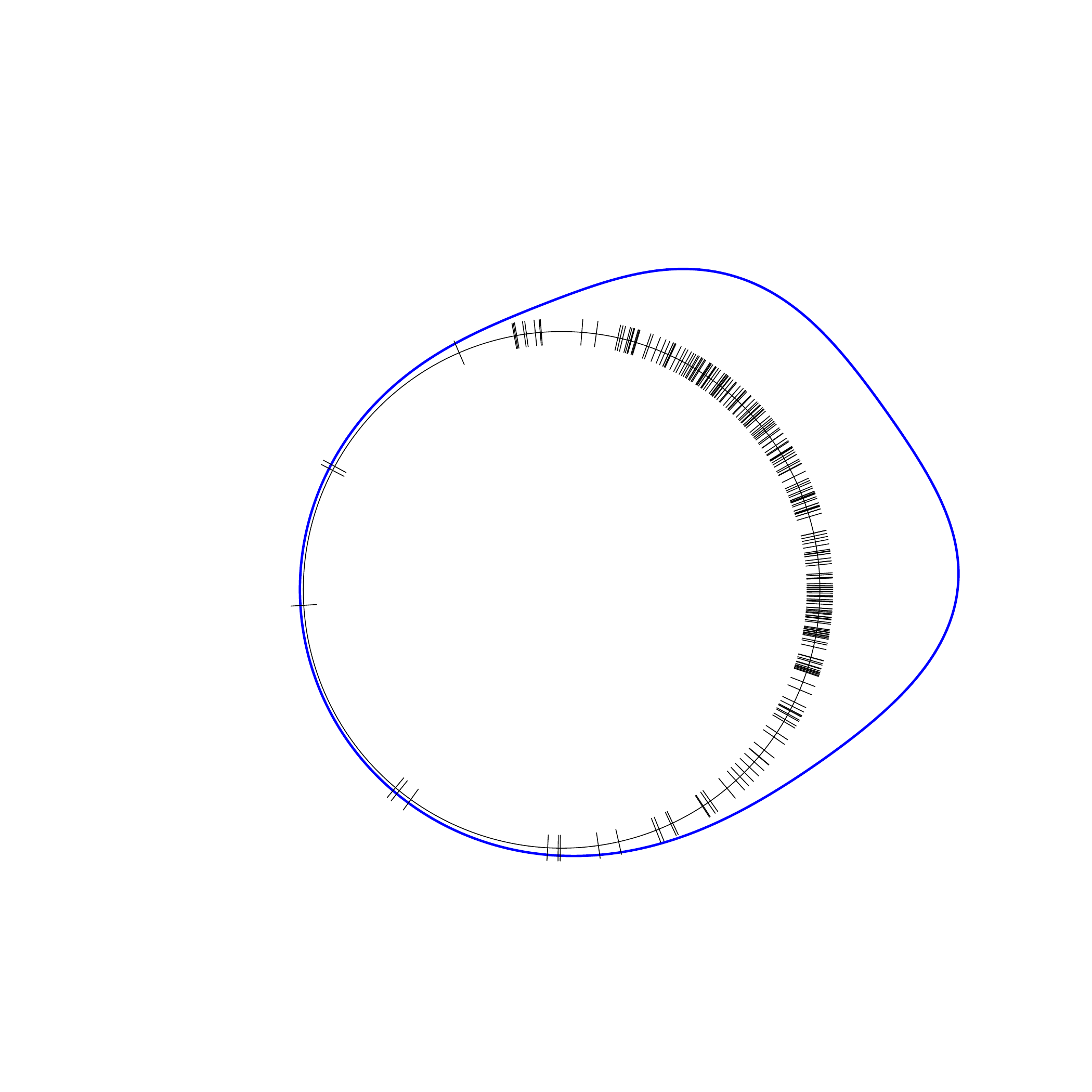}
	\includegraphics[width=0.225\textwidth]{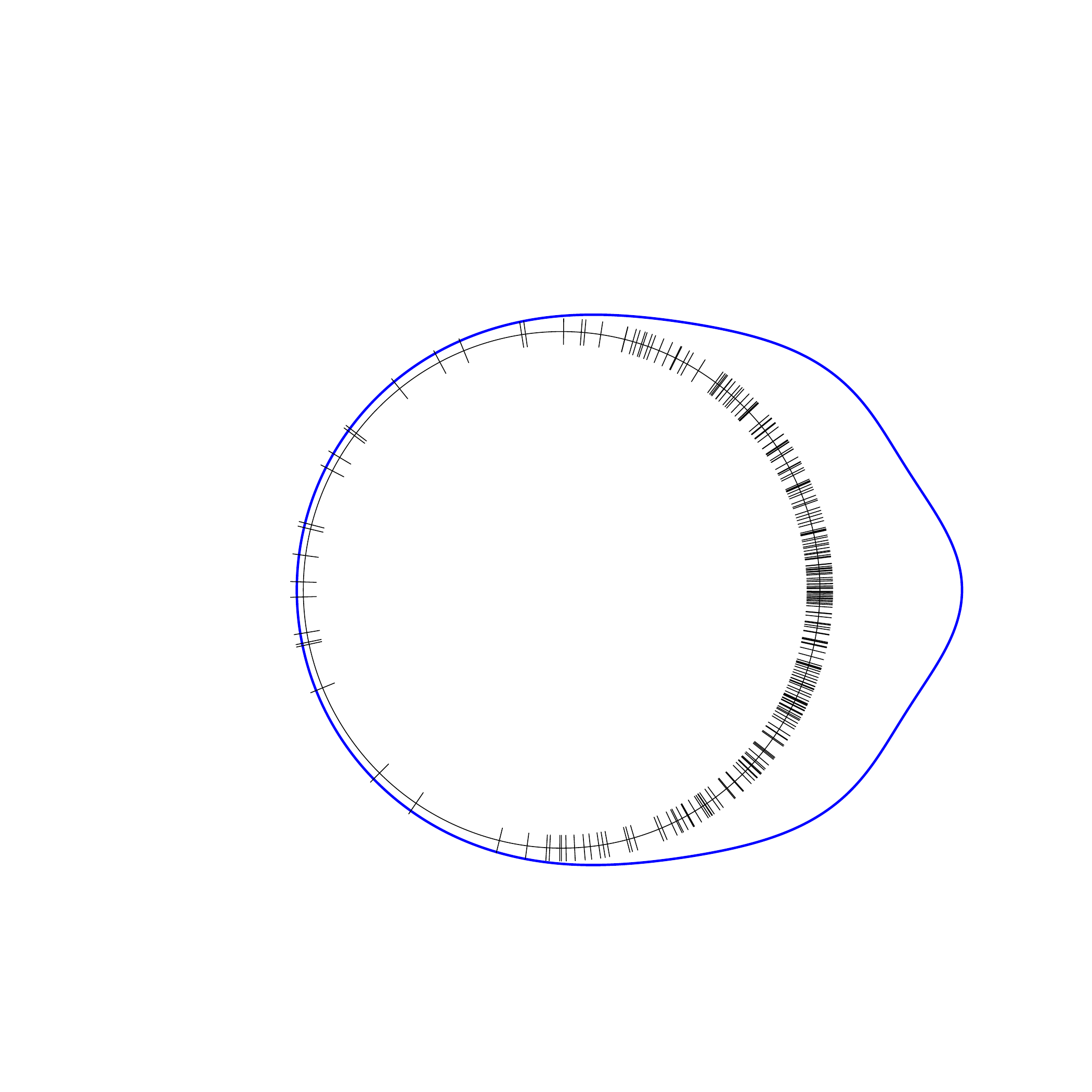}
	\includegraphics[width=0.225\textwidth]{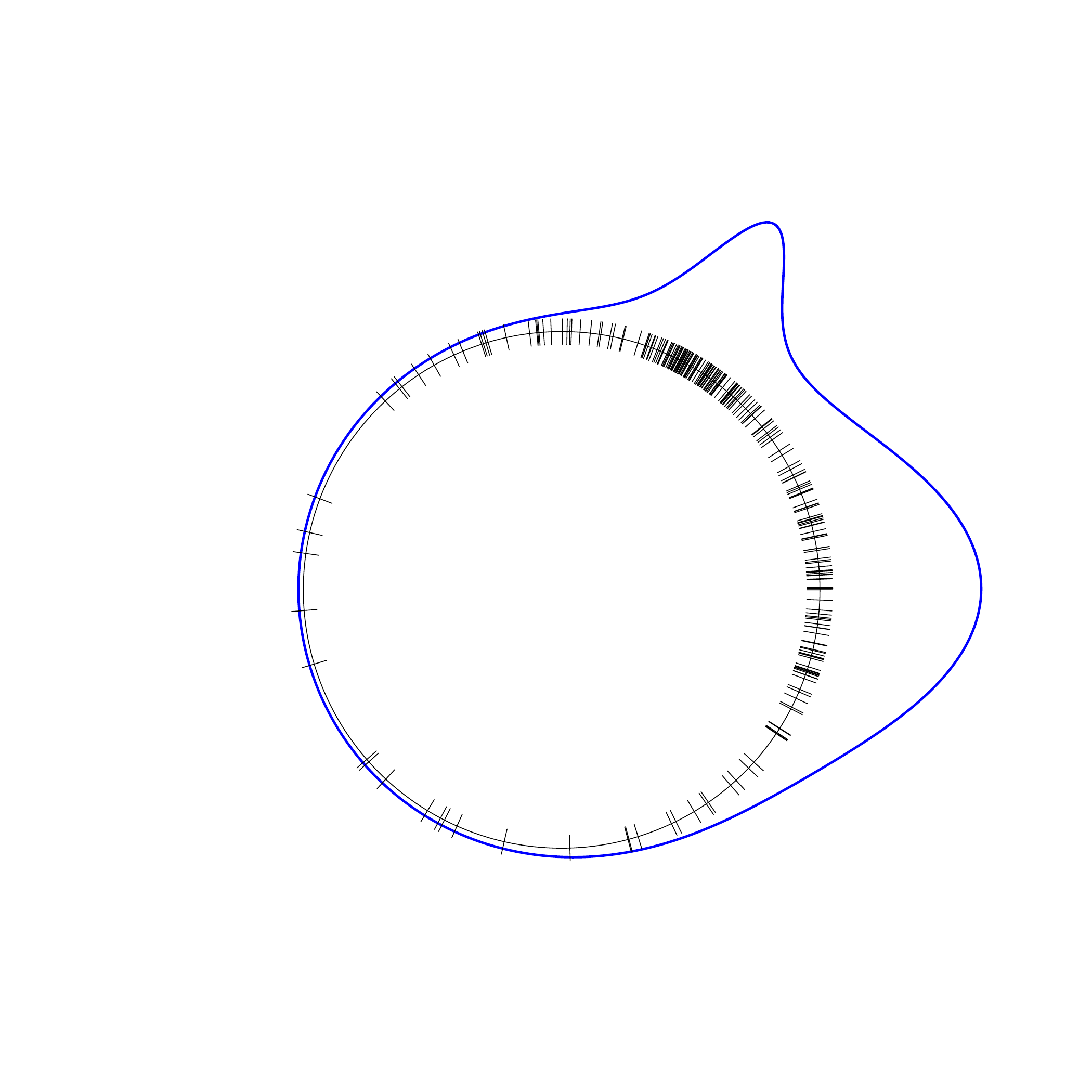}\\
	\includegraphics[width=0.225\textwidth]{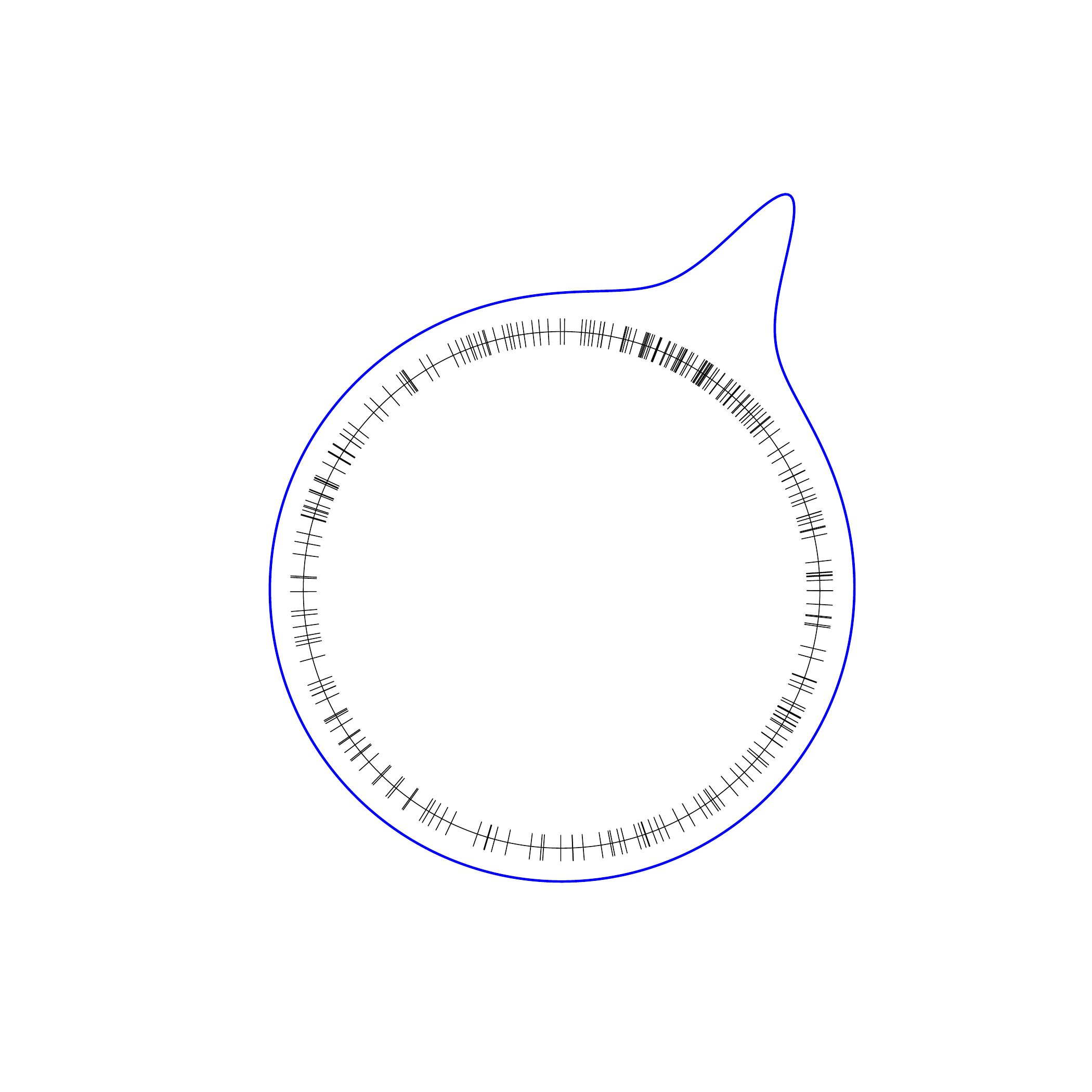}
	\includegraphics[width=0.225\textwidth]{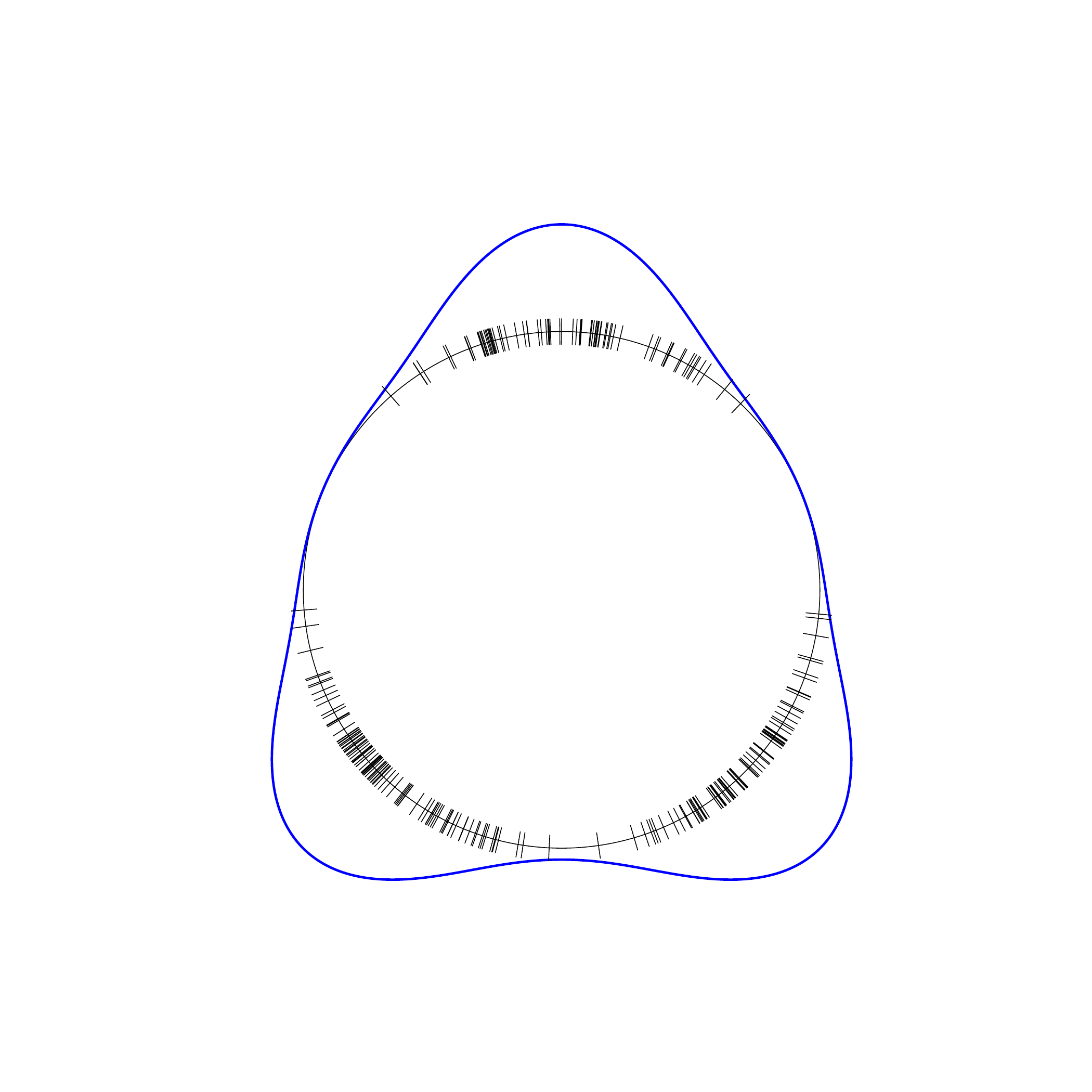}
	\includegraphics[width=0.225\textwidth]{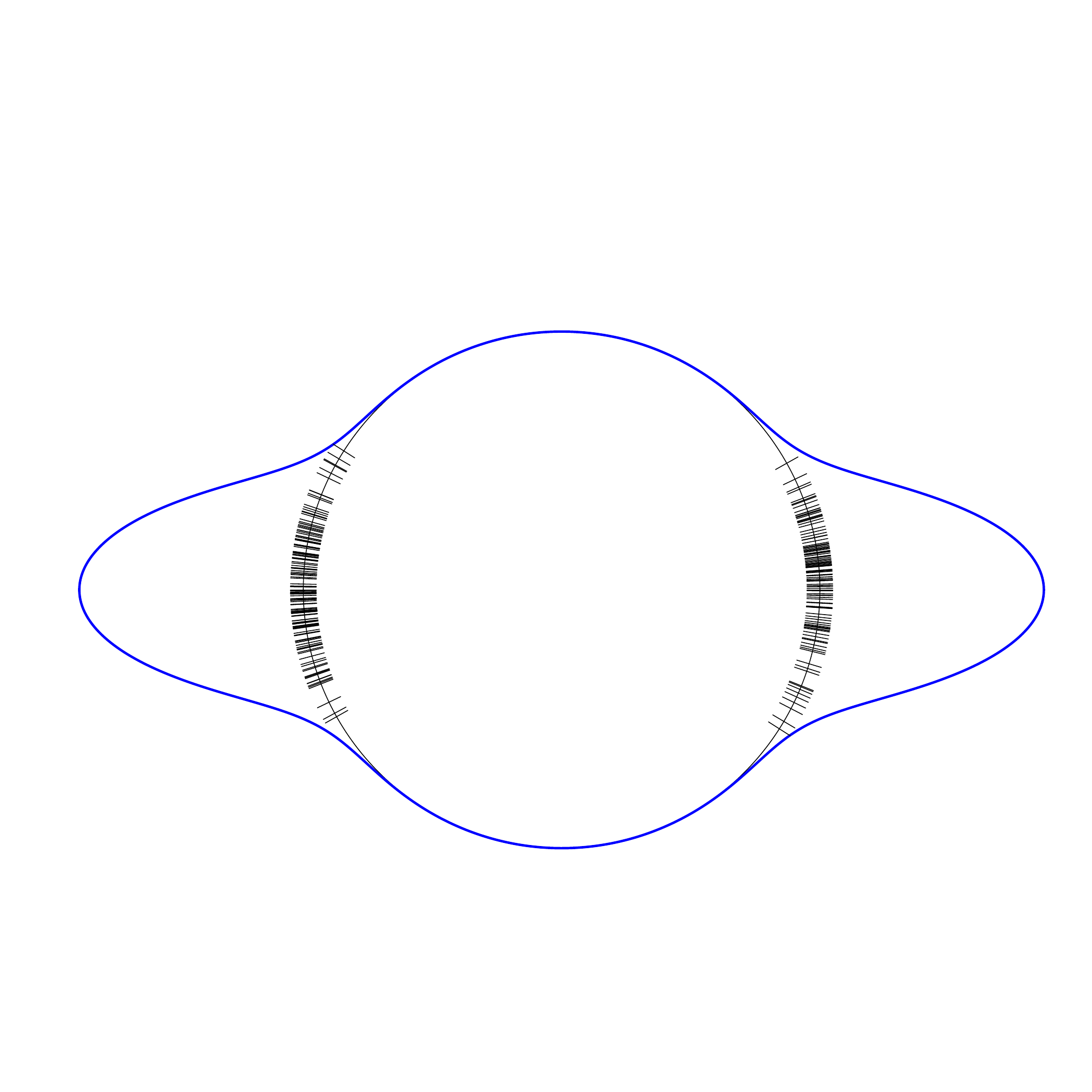}
	\includegraphics[width=0.225\textwidth]{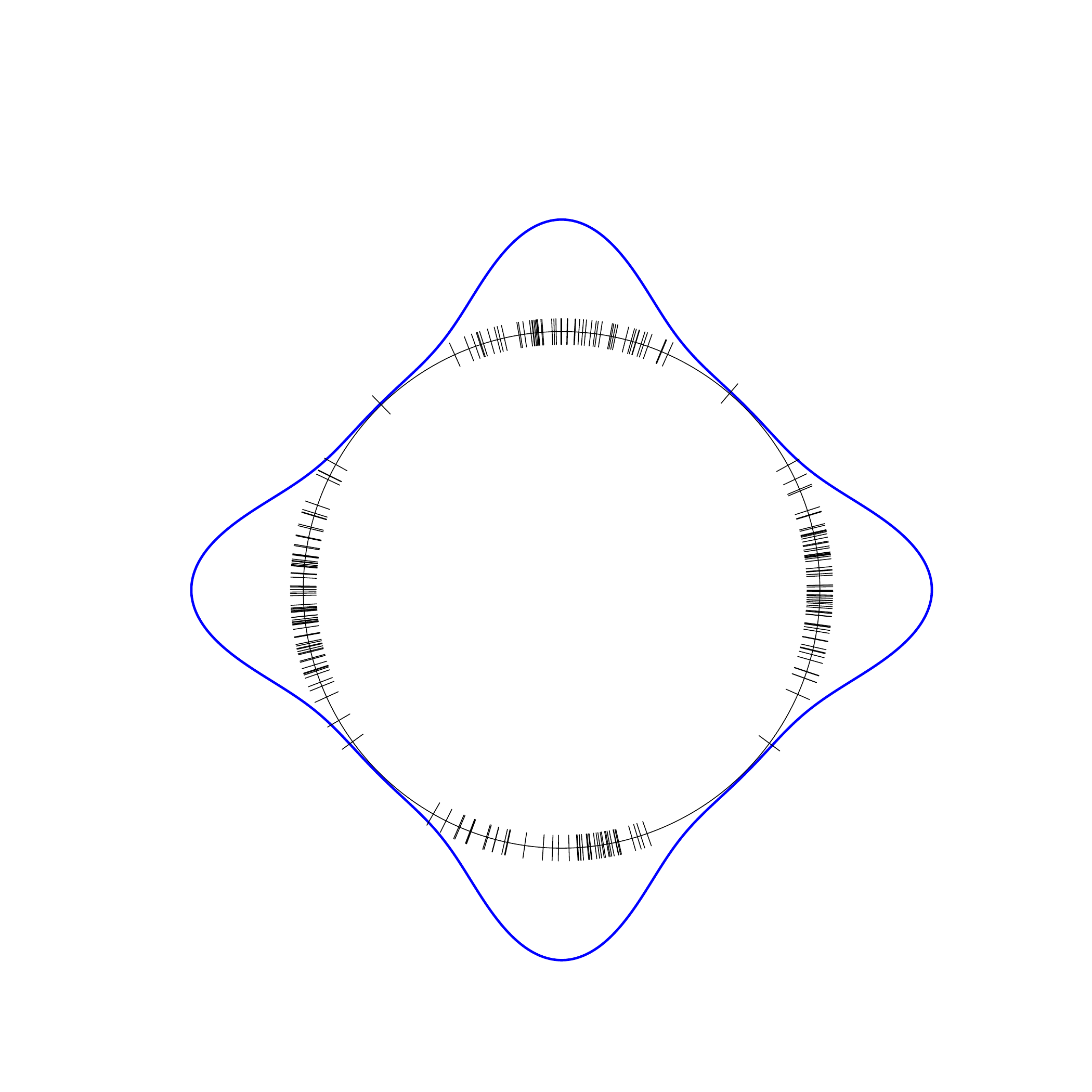}\\
	\includegraphics[width=0.225\textwidth]{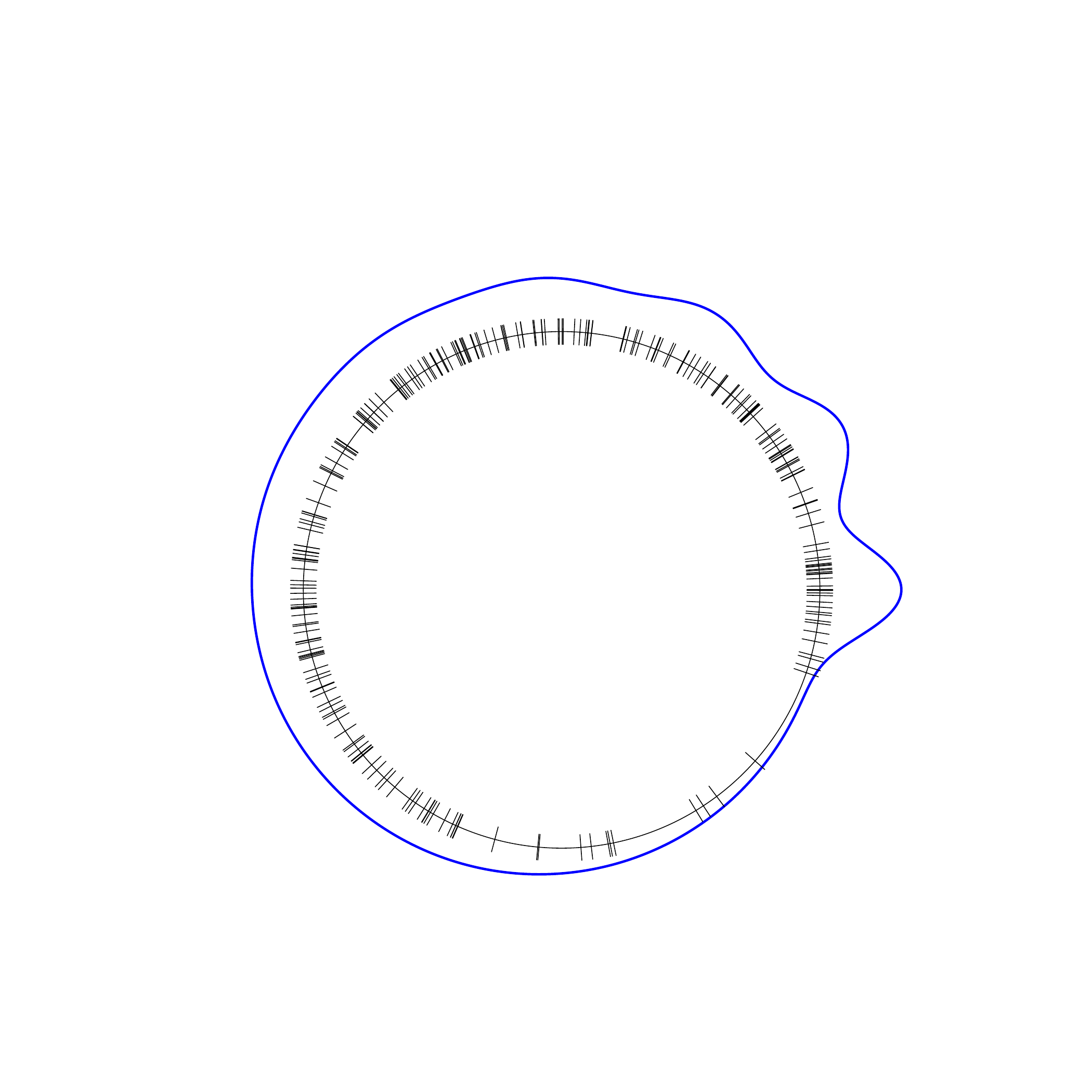}
	\includegraphics[width=0.225\textwidth]{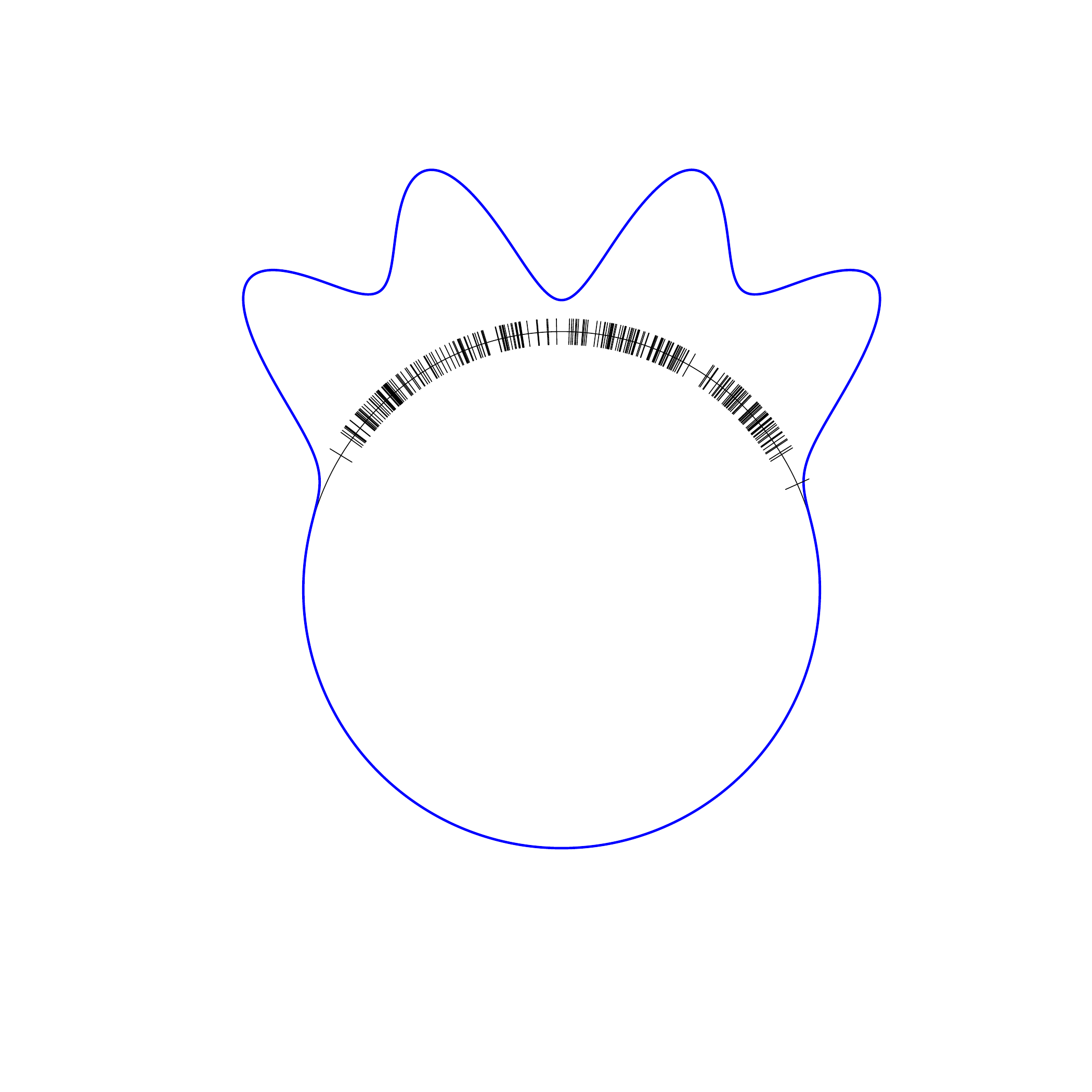}
	\includegraphics[width=0.225\textwidth]{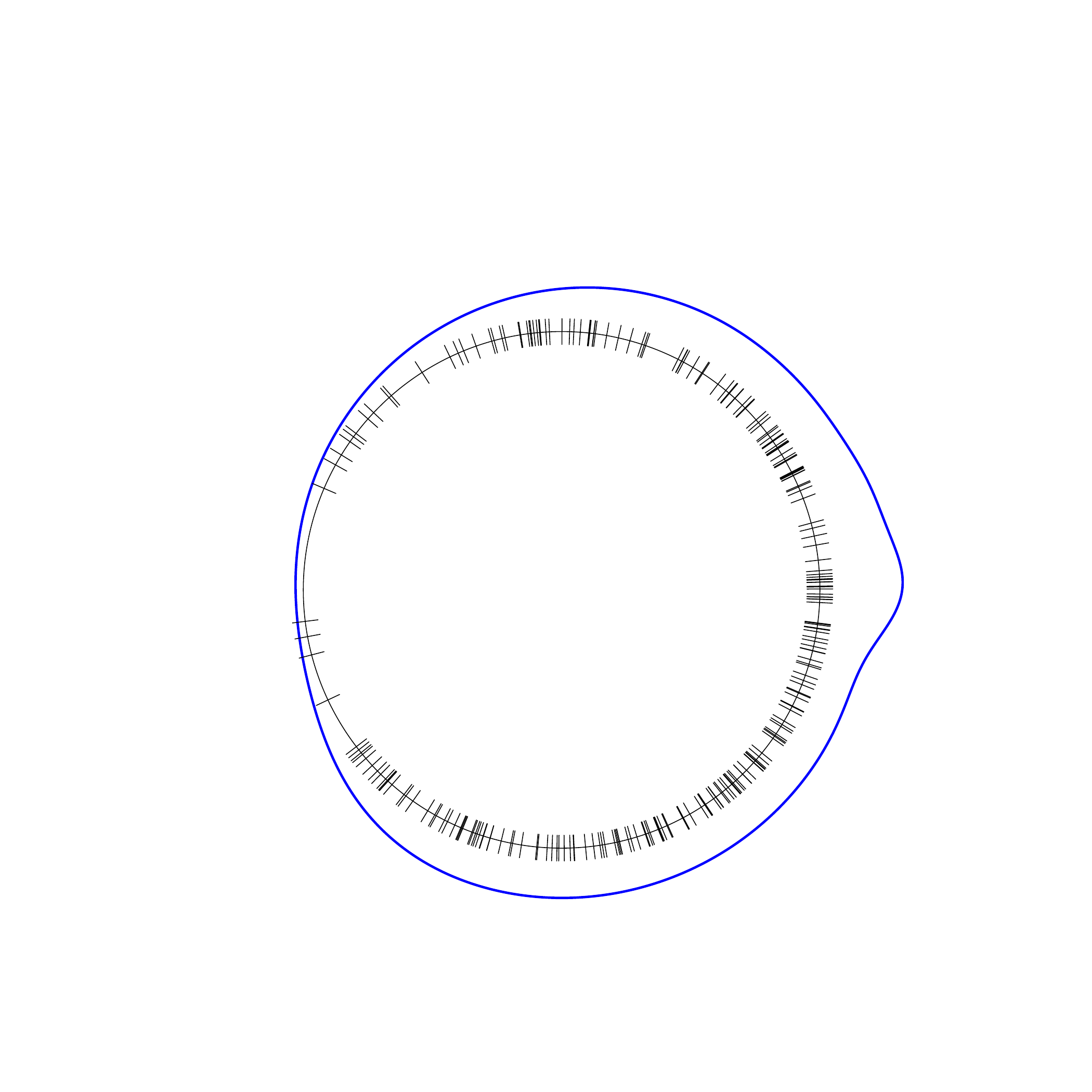}
	\includegraphics[width=0.225\textwidth]{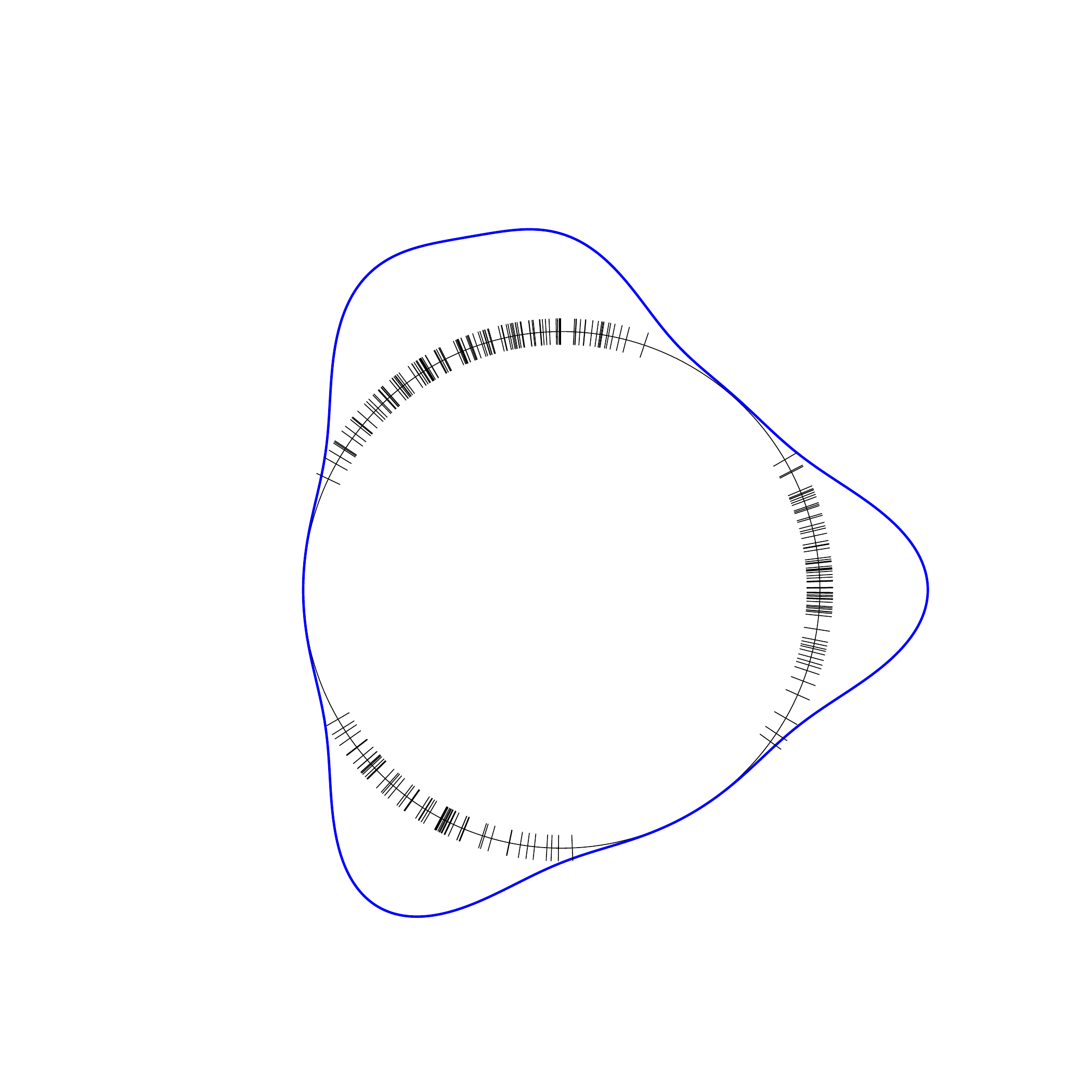}
	\caption{\small Simulation scenarios for the circular case. From left to right and up to down, models M1 to M20. For each model, a sample of size $250$ is drawn. \label{kdebwd:fig:circ}}
\end{figure}

For the circular case, the comparative study has been done for the $20$ models described in Figure \ref{kdebwd:fig:circ} (see Table \ref{kdebwd:tab:models} to see their densities), for the circular selectors $h_{\mathrm{LCV}}$, $h_{\mathrm{LSCV}}$, $h_{\mathrm{TAY}}$, $h_{\mathrm{OLI}}$, $h_{\mathrm{ROT}}$, $h_{\mathrm{AMI}}$ and $h_{\mathrm{EMI}}$ and for the sample sizes $100$, $250$, $500$ and $1000$. Due to space limitations, only the results for sample size $500$ are shown in Table \ref{kdebwd:tab:cir}, and the rest of them are relegated to Appendix \ref{kdebwd:ap:tables}. \\

In addition, to help summarizing the results a ranking similar to Ranking B of \cite{Cao1994} will be constructed. The ranking will be computed according to the following criteria: for each model, the $m$ bandwidth selectors $h_1,\ldots,h_m$ considered are sorted from the best performance (lowest error) to the worst performance (largest error). The best bandwidth receives $m$ points, the second $m-1$ and so on. These points, denoted by $r$, are standardized by $m$ and multiplied by the relative performance of each selector compared with the best one. In other words, the points of the selector $h_k$, if $h_\mathrm{opt}$ is the best one, are $\frac{r_k}{m}\frac{\mathrm{MISE}(h_\mathrm{opt})}{\mathrm{MISE}(h_k)}$. The final score for each selector is the sum of the ranks obtained in all the twenty models (thus, a selector which is the best in all models will have $20$ points). With this ranking, it is easy to group the results in a single and easy to read table. \\

In view of the results, the following conclusions can be extracted. Firstly, $h_\mathrm{ROT}$ performs well in certain unimodal models such as M3 (von Mises) and M6 (skew normal directional), but its performance is very poor with multimodal models like M15 (Watson). In its particular comparison with $h_\mathrm{TAY}$, it can be observed that both selectors share the same order of error, but being $h_\mathrm{ROT}$ better in all the situations except for one: the uniform model (M1). This is due to the ``extra term'' commented in Section \ref{kdebwd:sec:ruleofthumb}: its absence in the denominator makes that $h_\mathrm{TAY}\to\infty$ faster than $h_\mathrm{ROT}$ when the concentration parameter $\kappa\to0$ and, what is a disadvantage for $\kappa>0$, turns out in an advantage for the uniform case. With respect to $h_\mathrm{AMI}$ and $h_\mathrm{EMI}$, although their performance becomes more similar when the sample size increases, something expected, $h_\mathrm{EMI}$ seems to be on average a step ahead from $h_\mathrm{AMI}$, specially for low sample sizes. Among the cross-validated selectors, $h_\mathrm{LCV}$ performs better than $h_\mathrm{LSCV}$, a fact that was previously noted by simulation studies carried out by \cite{Taylor2008} and \cite{Oliveira2012}. Finally, $h_\mathrm{OLI}$ presents the most competitive behaviour among the previous proposals in the literature when the sample size is reasonably large (see Table \nolinebreak[4]\ref{kdebwd:tab:rankcirsph}). \\

The comparison between the circular selectors is summarized in the scores of Table \ref{kdebwd:tab:rankcirsph}. For all the sample sizes considered, $h_\mathrm{EMI}$ is the most competitive selector, followed by $h_\mathrm{AMI}$ for all the sample sizes except $n=100$, where $h_\mathrm{LCV}$ is the second. The effect of the sample effect is also interesting to comment. For $n=100$, $h_\mathrm{LCV}$ and $h_\mathrm{ROT}$ perform surprisingly well, in contrast with $h_\mathrm{OLI}$, which is the second worst selector for this case. When the sample size increases, $h_\mathrm{ROT}$ and $h_\mathrm{TAY}$ have a decreasing performance and $h_\mathrm{OLI}$ stretches differences with $h_\mathrm{AMI}$, showing a similar behaviour. This was something expected as both selectors are based on error criteria that are asymptotically equivalent. The cross-validated selectors show a stable performance for sample sizes larger than\nolinebreak[4] $n=100$.

\pagebreak

\begin{table}[H]
	\centering
	\setlength{\tabcolsep}{5pt}
	\scriptsize
	\begin{tabular}{r@{\hspace{0.275cm}}|r@{\hspace{0.275cm}}|r@{\hspace{0.275cm}}r@{\hspace{0.275cm}}r@{\hspace{0.275cm}}
			r@{\hspace{0.275cm}}r@{\hspace{0.275cm}}r@{\hspace{0.275cm}}
			r@{\hspace{0.275cm}}r@{\hspace{0.275cm}}}\toprule\toprule
		Model &  \multicolumn{1}{c|}{ $h_\mathrm{MISE}$} & \multicolumn{1}{c}{$h_\mathrm{LCV}$} & \multicolumn{1}{c}{$h_\mathrm{LSCV}$} & \multicolumn{1}{c}{$h_\mathrm{TAY}$} & \multicolumn{1}{c}{$h_\mathrm{OLI}$} & \multicolumn{1}{c}{$h_\mathrm{ROT}$} & \multicolumn{1}{c}{$h_\mathrm{AMI}$} & \multicolumn{1}{c}{$h_\mathrm{EMI}$} \\\midrule
		M1 & $0.0000$ & $0.057$ ($0.12$) & $0.059$ ($0.12$) & $\mathbf{0.001}$ ($0.00$) & $0.049$ ($0.10$) & $0.020$ ($0.03$) & $0.020$ ($0.03$) & $0.022$ ($0.03$) \\
		M2 & $0.2298$ & $0.265$ ($0.16$) & $0.297$ ($0.21$) & $0.238$ ($0.15$) & $0.249$ ($0.16$) & $0.235$ ($0.14$) & $0.235$ ($0.14$) & $\mathbf{0.234}$ ($0.15$) \\
		M3 & $0.2815$ & $0.338$ ($0.21$) & $0.356$ ($0.24$) & $0.313$ ($0.20$) & $0.301$ ($0.20$) & $\mathbf{0.294}$ ($0.19$) & $0.298$ ($0.19$) & $0.301$ ($0.19$) \\
		M4 & $0.3429$ & $0.424$ ($0.23$) & $0.413$ ($0.23$) & $0.683$ ($0.29$) & $0.366$ ($0.19$) & $0.534$ ($0.25$) & $\mathbf{0.363}$ ($0.19$) & $0.363$ ($0.19$) \\
		M5 & $0.5938$ & $0.883$ ($0.49$) & $0.708$ ($0.38$) & $2.032$ ($0.61$) & $0.640$ ($0.35$) & $1.723$ ($0.55$) & $\mathbf{0.638}$ ($0.35$) & $0.645$ ($0.36$) \\
		M6 & $0.2789$ & $0.374$ ($0.23$) & $0.354$ ($0.23$) & $0.281$ ($0.14$) & $0.341$ ($0.20$) & $\mathbf{0.281}$ ($0.15$) & $0.323$ ($0.18$) & $0.304$ ($0.17$) \\
		M7 & $0.3013$ & $0.333$ ($0.16$) & $0.353$ ($0.18$) & $6.677$ ($0.07$) & $0.319$ ($0.15$) & $5.533$ ($0.77$) & $\mathbf{0.309}$ ($0.15$) & $0.310$ ($0.15$) \\
		M8 & $0.2408$ & $0.268$ ($0.14$) & $0.298$ ($0.17$) & $0.286$ ($0.12$) & $0.261$ ($0.14$) & $0.254$ ($0.12$) & $0.251$ ($0.12$) & $\mathbf{0.248}$ ($0.12$) \\
		M9 & $0.6208$ & $0.913$ ($0.38$) & $0.720$ ($0.34$) & $1.270$ ($0.31$) & $0.685$ ($0.31$) & $1.152$ ($0.30$) & $0.676$ ($0.31$) & $\mathbf{0.658}$ ($0.30$) \\
		M10 & $0.3180$ & $0.361$ ($0.17$) & $0.390$ ($0.22$) & $0.366$ ($0.16$) & $0.356$ ($0.18$) & $\mathbf{0.341}$ ($0.16$) & $0.347$ ($0.17$) & $0.351$ ($0.17$) \\
		M11 & $0.3056$ & $0.346$ ($0.17$) & $0.372$ ($0.22$) & $0.352$ ($0.17$) & $0.327$ ($0.17$) & $\mathbf{0.326}$ ($0.15$) & $0.328$ ($0.15$) & $0.332$ ($0.15$) \\
		M12 & $0.7322$ & $0.910$ ($0.33$) & $0.822$ ($0.30$) & $1.974$ ($0.30$) & $\mathbf{0.788}$ ($0.27$) & $1.750$ ($0.29$) & $0.854$ ($0.38$) & $0.831$ ($0.39$) \\
		M13 & $0.9349$ & $1.178$ ($0.37$) & $1.011$ ($0.28$) & $4.625$ ($0.33$) & $\mathbf{1.005}$ ($0.31$) & $3.733$ ($0.28$) & $1.095$ ($0.55$) & $1.077$ ($0.55$) \\
		M14 & $0.5106$ & $0.534$ ($0.20$) & $0.570$ ($0.23$) & $12.445$ ($0.53$) & $0.528$ ($0.20$) & $9.085$ ($1.11$) & $0.520$ ($0.19$) & $\mathbf{0.517}$ ($0.19$) \\
		M15 & $0.6101$ & $0.663$ ($0.32$) & $0.709$ ($0.35$) & $44.295$ ($0.39$) & $0.648$ ($0.31$) & $39.961$ ($3.66$) & $0.642$ ($0.31$) & $\mathbf{0.630}$ ($0.30$) \\
		M16 & $0.6006$ & $0.627$ ($0.23$) & $0.664$ ($0.25$) & $14.293$ ($0.00$) & $0.634$ ($0.22$) & $14.231$ ($0.08$) & $0.627$ ($0.22$) & $\mathbf{0.613}$ ($0.22$) \\
		M17 & $0.5891$ & $\mathbf{0.631}$ ($0.17$) & $0.664$ ($0.18$) & $1.280$ ($0.20$) & $0.692$ ($0.15$) & $0.927$ ($0.10$) & $0.743$ ($0.18$) & $0.715$ ($0.17$) \\
		M18 & $1.0646$ & $1.130$ ($0.40$) & $1.116$ ($0.39$) & $4.921$ ($0.24$) & $\mathbf{1.067}$ ($0.38$) & $4.630$ ($0.25$) & $1.104$ ($0.38$) & $1.087$ ($0.38$) \\
		M19 & $0.2718$ & $0.306$ ($0.12$) & $0.323$ ($0.13$) & $0.524$ ($0.15$) & $0.302$ ($0.12$) & $0.341$ ($0.09$) & $0.301$ ($0.12$) & $\mathbf{0.297}$ ($0.10$) \\
		M20 & $0.5550$ & $0.581$ ($0.22$) & $0.618$ ($0.25$) & $13.962$ ($0.91$) & $0.581$ ($0.23$) & $9.299$ ($1.20$) & $0.567$ ($0.22$) & $\mathbf{0.562}$ ($0.21$) \\
		\bottomrule\bottomrule
	\end{tabular}
	\caption{\small Comparative study for the circular case, with sample size $n=500$. Columns of the selector $\bullet$ represent the $\mathrm{MISE}(\bullet)\times100$, with bold type for the minimum of the errors. The standard deviation of the $\mathrm{ISE}\times100$ is given between parentheses.\label{kdebwd:tab:cir}}
\end{table}

\begin{table}[H]
	\centering
	\small
	\begin{tabular}{rr|rrrrrrrrr}\toprule\toprule
		$q$ & $n$ & \multicolumn{1}{c}{$h_\mathrm{LCV}$} & \multicolumn{1}{c}{$h_\mathrm{LSCV}$} & \multicolumn{1}{c}{$h_\mathrm{TAY}$} & \multicolumn{1}{c}{$h_\mathrm{OLI}$}  & \multicolumn{1}{c}{$h_\mathrm{ROT}$} & \multicolumn{1}{c}{$h_\mathrm{AMI}$} & \multicolumn{1}{c}{$h_\mathrm{EMI}$} \\\midrule
		$1$ & $100$ & $11.1494$ & $9.4896$ & $6.5143$ & $6.8864$ & $10.2829$ & $11.1327$ & $\mathbf{14.5329}$\\
		& $250$ & $9.6357$ & $7.6350$ & $5.2053$ & $10.7883$ & $7.9129$ & $13.0558$ & $\mathbf{16.0261}$ \\
		& $500$ & $8.6549$ & $7.7280$ & $4.0933$ & $13.2003$ & $6.8351$ & $14.5268$ & $\mathbf{15.6039}$ \\
		& $1000$ & $9.0128$ & $7.9820$ & $3.7168$ & $14.0234$ & $5.6784$ & $14.7077$ & $\mathbf{15.4358}$ \\\midrule
		$2$ & $100$ & $11.8161$ & $13.4387$ & $\ast$\phantom{$00$} & $\ast$\phantom{$00$} & $6.4424$ & $8.1711$ & $\mathbf{15.3028}$ \\
		& $250$ & $10.2201$ & $12.2789$ & $\ast$\phantom{$00$} & $\ast$\phantom{$00$} & $4.3195$ & $11.2453$ & $\mathbf{17.1272}$ \\
		& $500$ & $8.9001$ & $12.1317$ & $\ast$\phantom{$00$} & $\ast$\phantom{$00$} & $3.3156$ & $13.0011$ & $\mathbf{18.0860}$ \\
		& $1000$ & $8.2036$ & $12.1566$ & $\ast$\phantom{$00$} & $\ast$\phantom{$00$} & $2.9175$ & $13.3693$ & $\mathbf{18.7548}$
		\\
		\bottomrule\bottomrule
	\end{tabular}
	\caption{\small Ranking for the selectors for the circular and spherical cases, for sample sizes $n=100,250,500,1000$. The higher the score in the ranking, the better the performance of the selector. Bold type indicates the best selector.\label{kdebwd:tab:rankcirsph}}
\end{table}

\subsection{Spherical case}
\label{kdebwd:subsec:spherical}

\begin{figure}[H]
	\centering
	\includegraphics[width=0.225\textwidth]{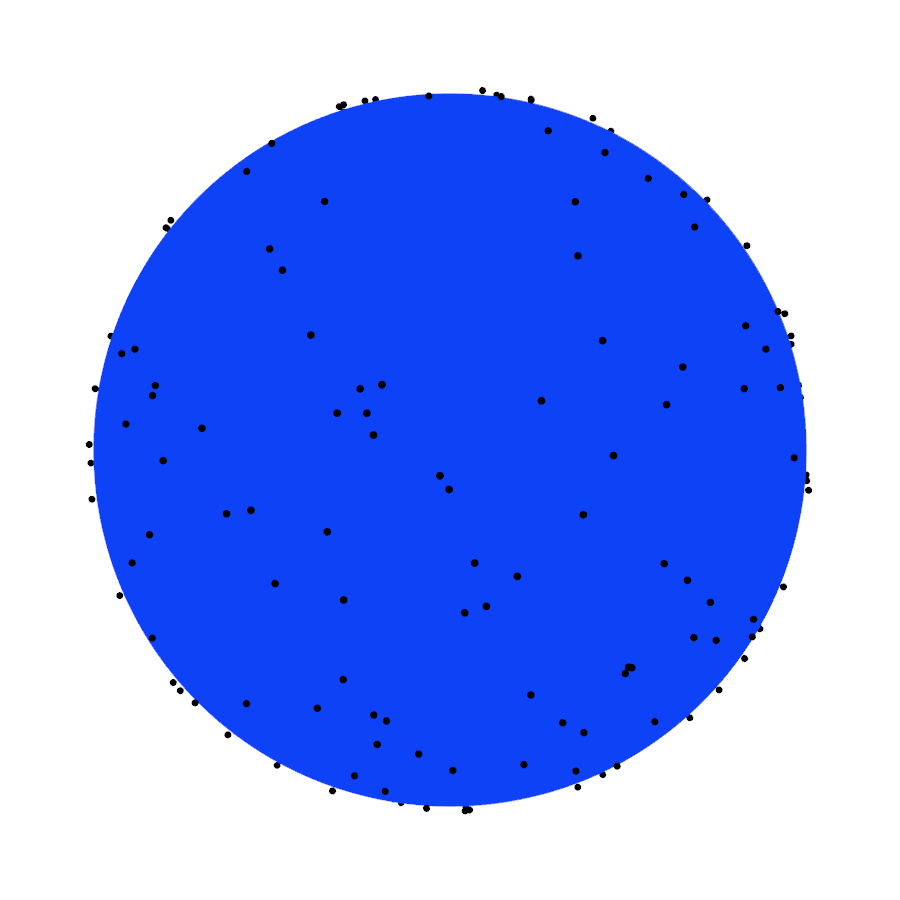}
	\includegraphics[width=0.225\textwidth]{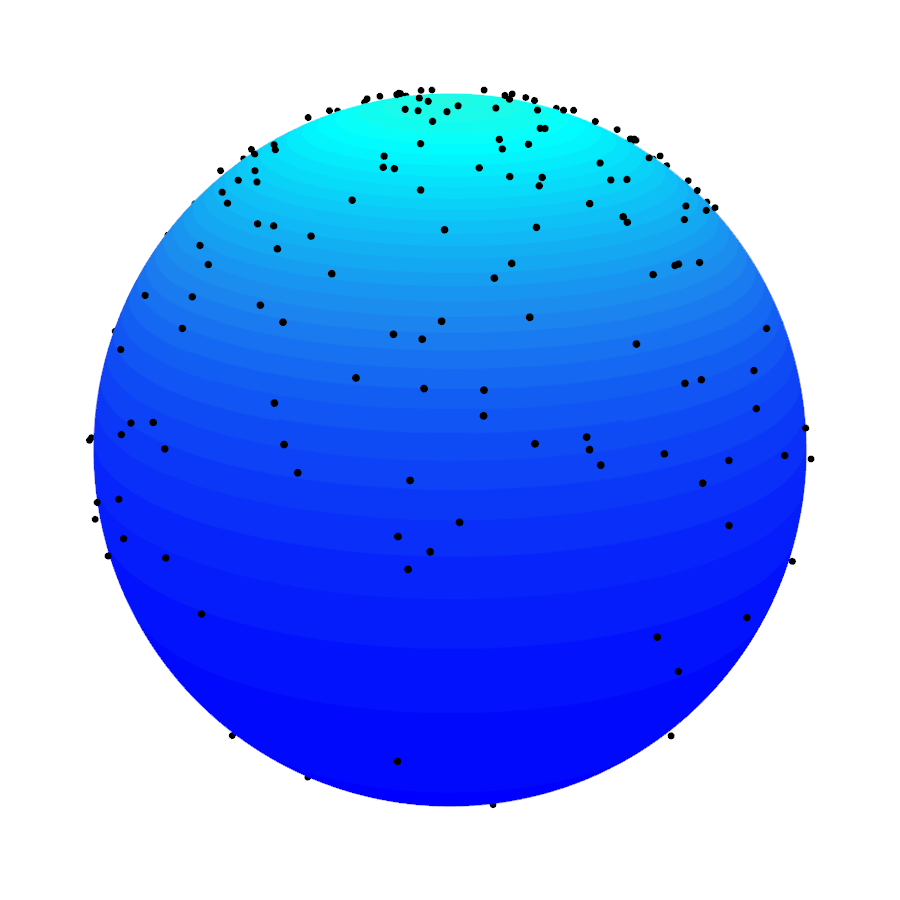}
	\includegraphics[width=0.225\textwidth]{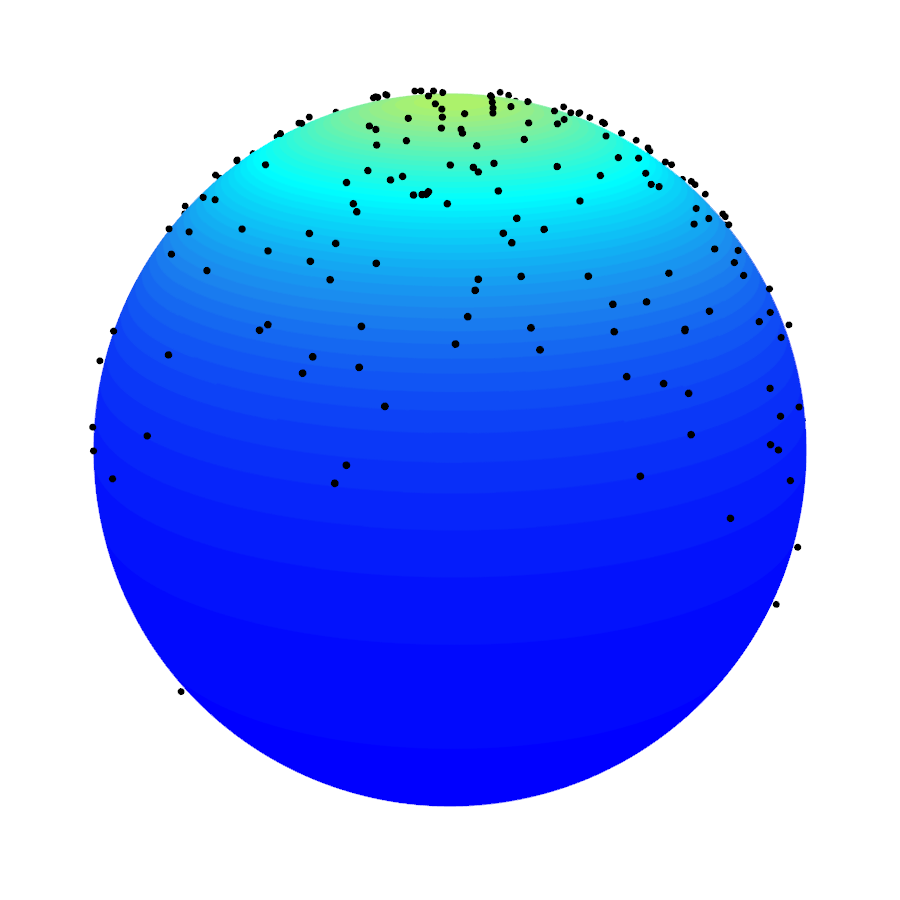}
	\includegraphics[width=0.225\textwidth]{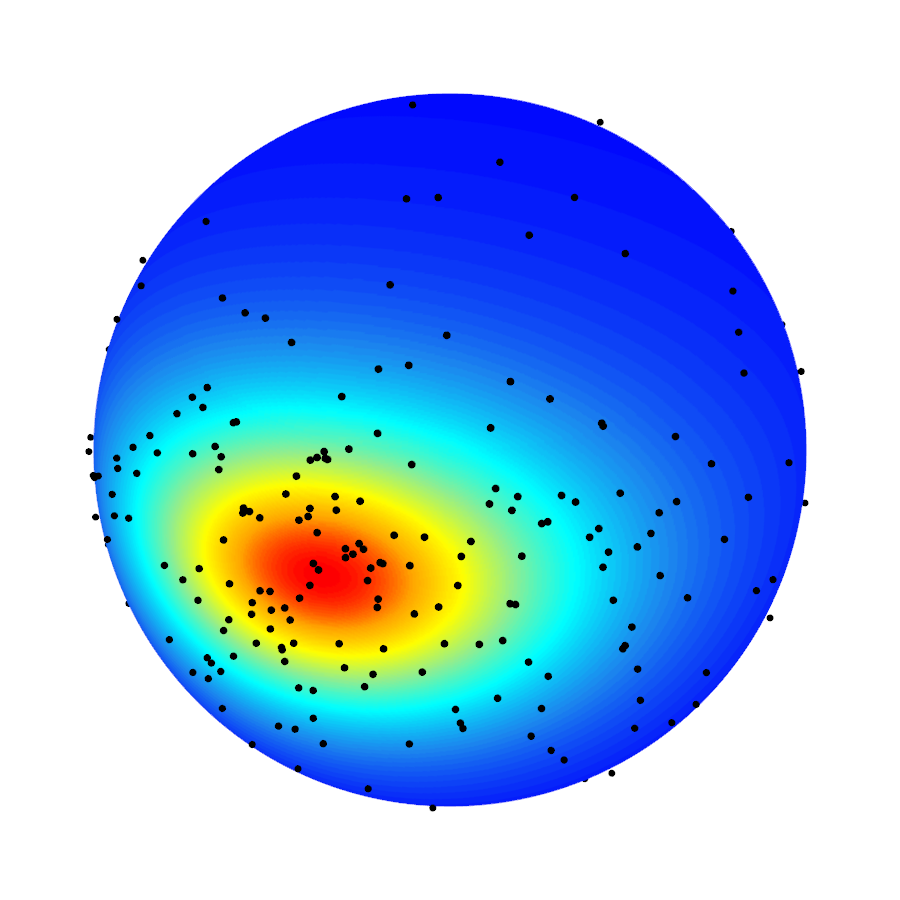}\\
	\includegraphics[width=0.225\textwidth]{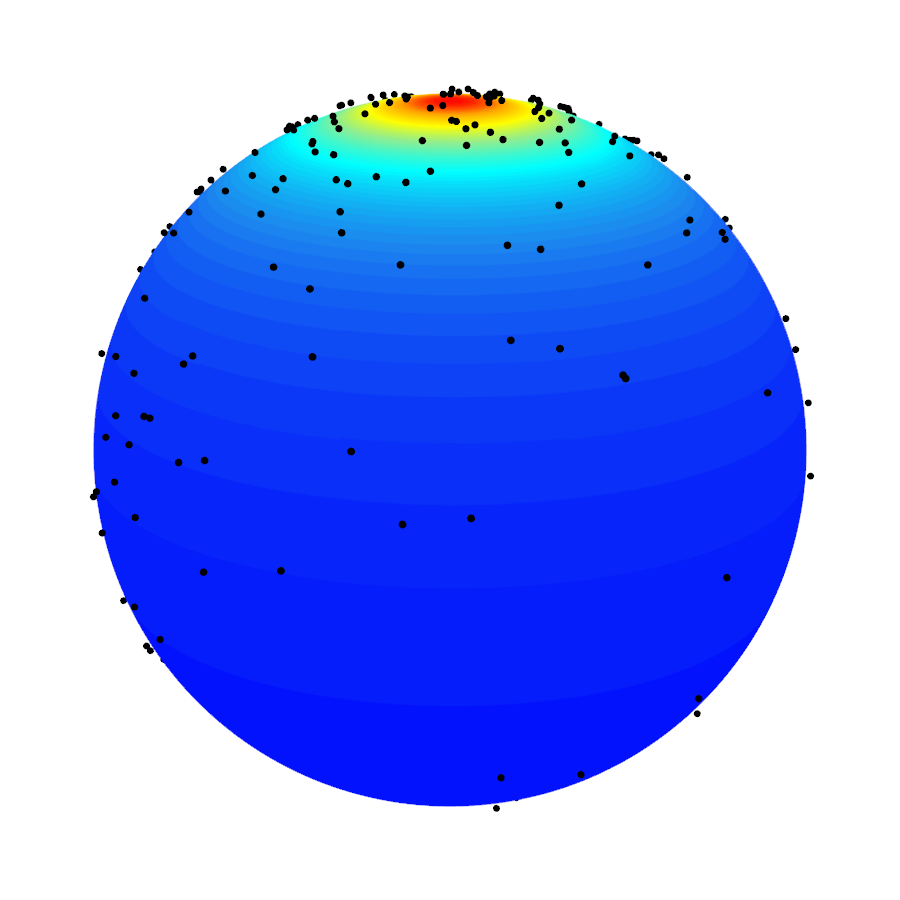}
	\includegraphics[width=0.225\textwidth]{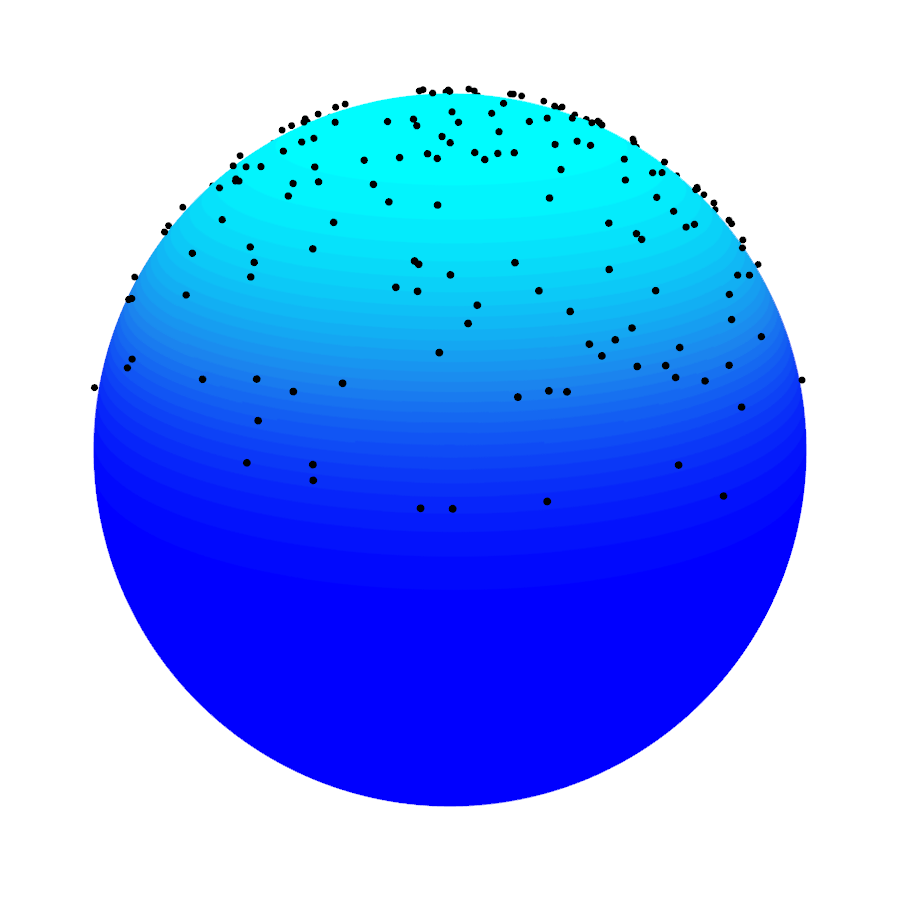}
	\includegraphics[width=0.225\textwidth]{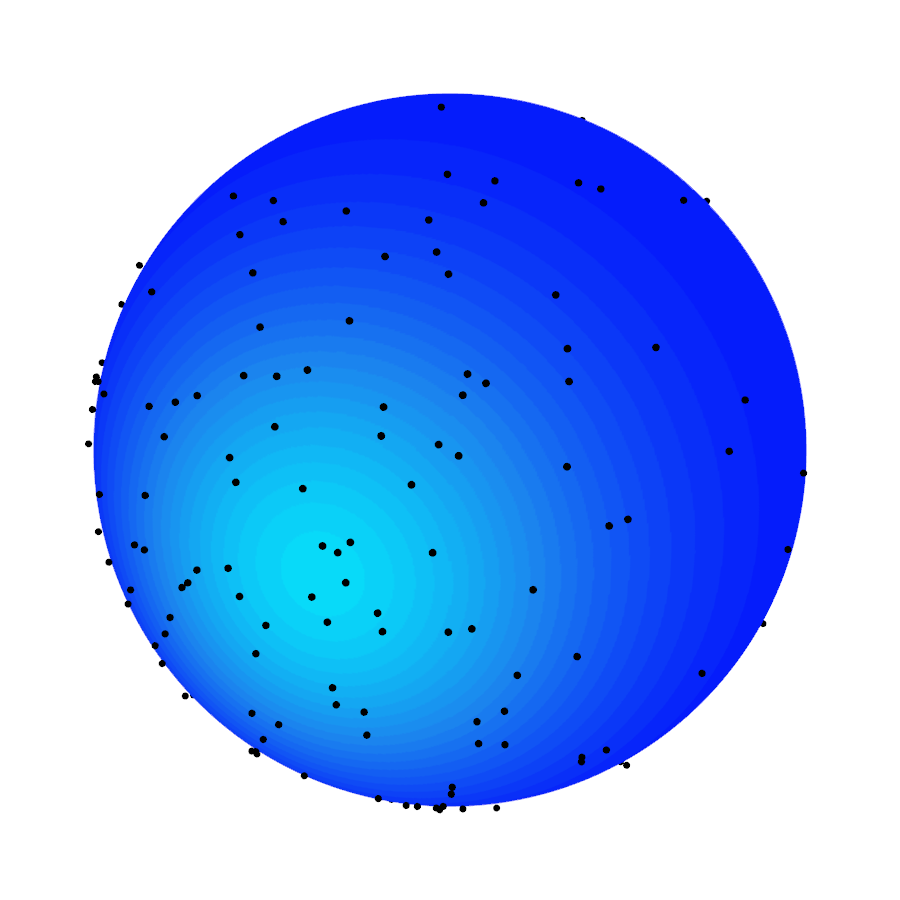}
	\includegraphics[width=0.225\textwidth]{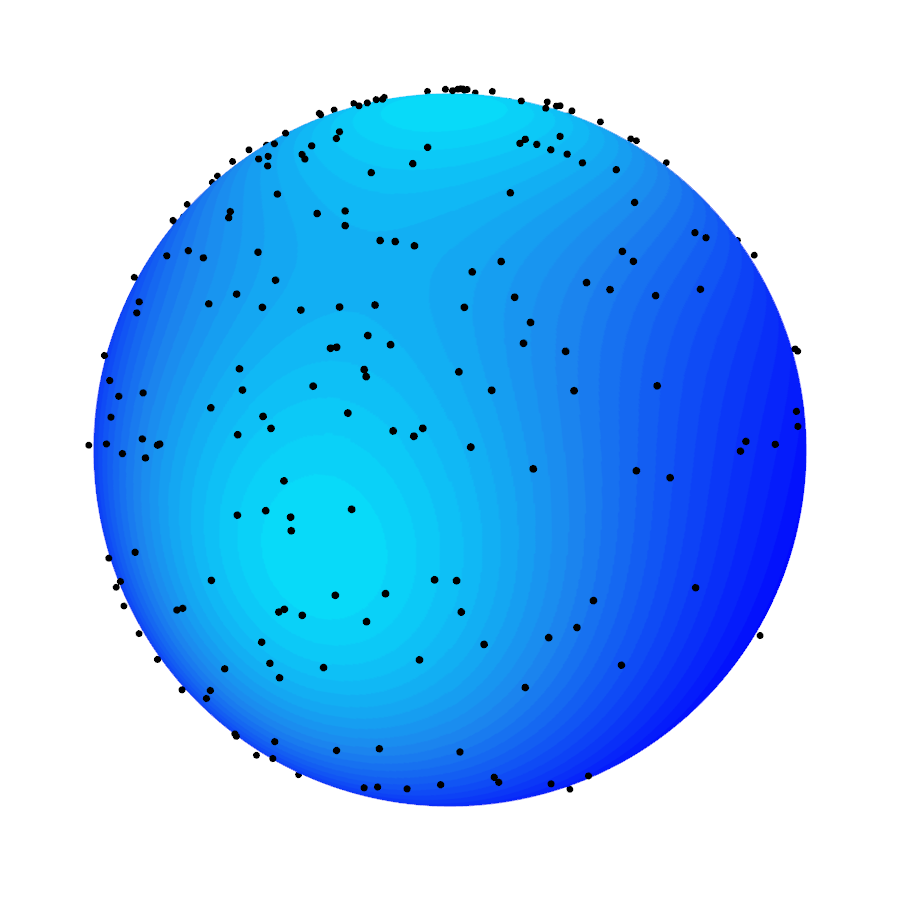}\\
	\includegraphics[width=0.225\textwidth]{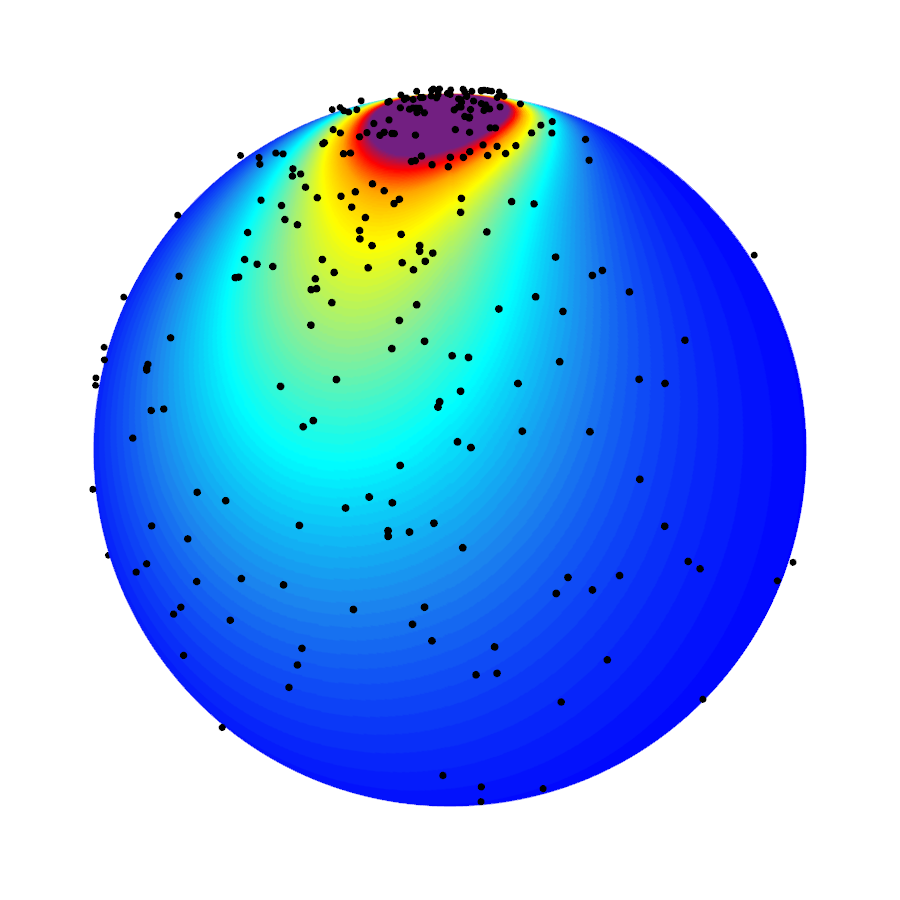}
	\includegraphics[width=0.225\textwidth]{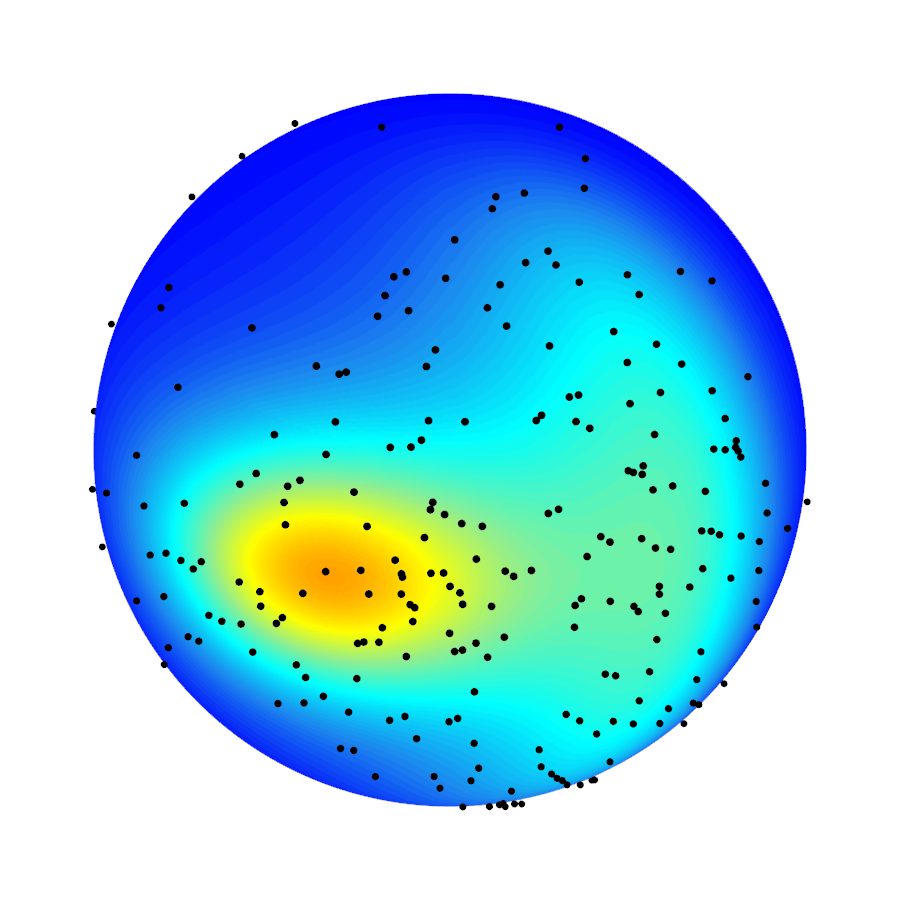}
	\includegraphics[width=0.225\textwidth]{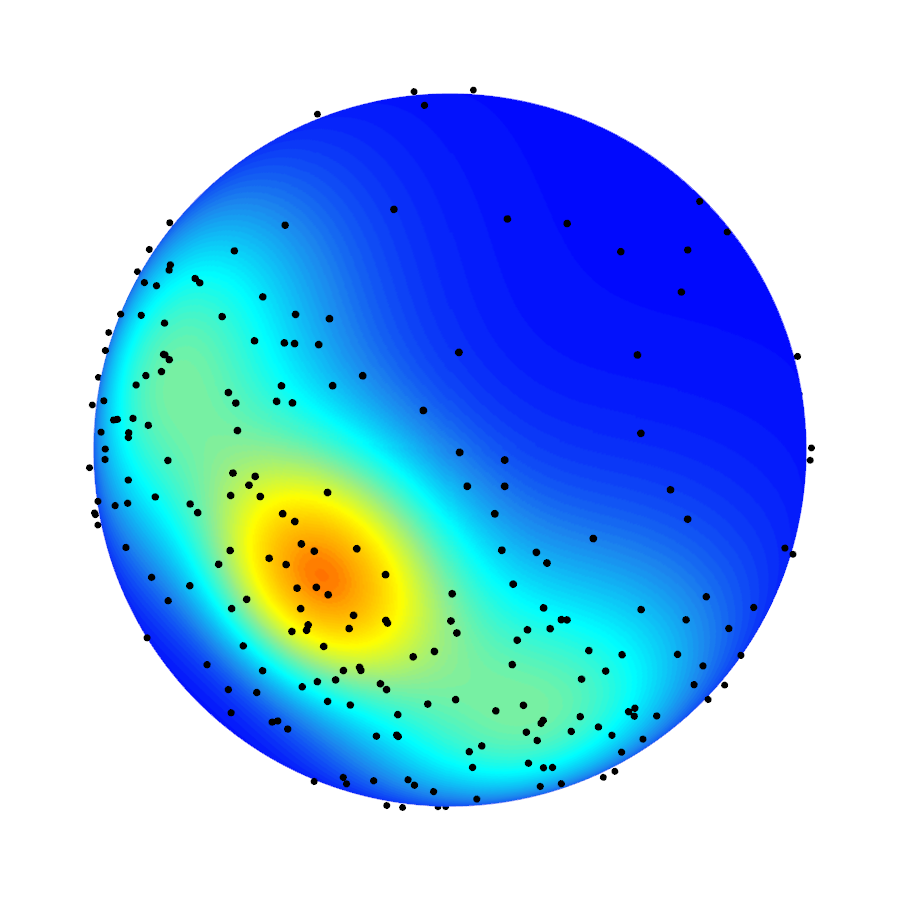}
	\includegraphics[width=0.225\textwidth]{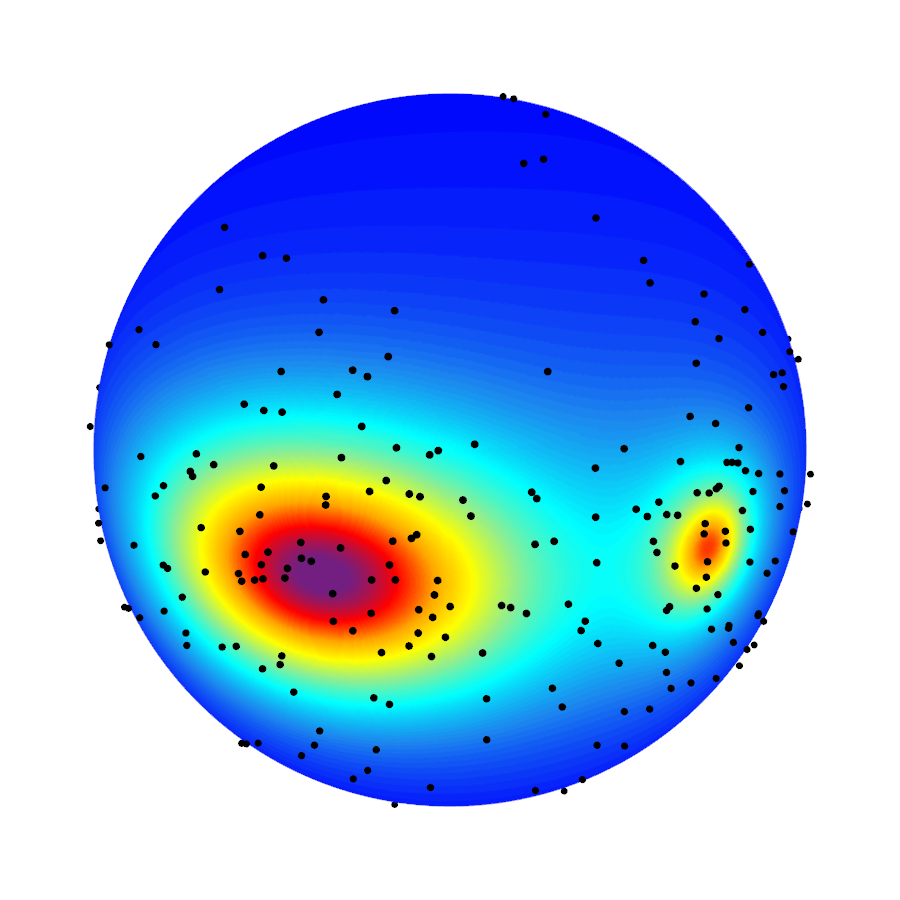}\\
	\includegraphics[width=0.225\textwidth]{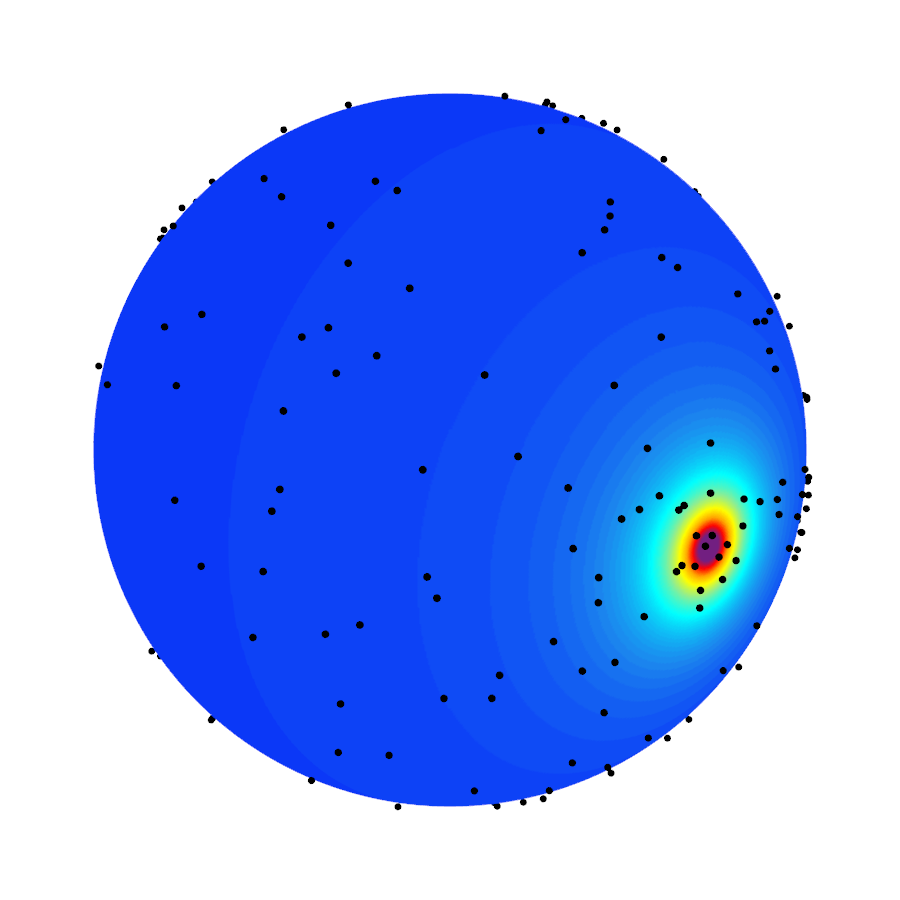}
	\includegraphics[width=0.225\textwidth]{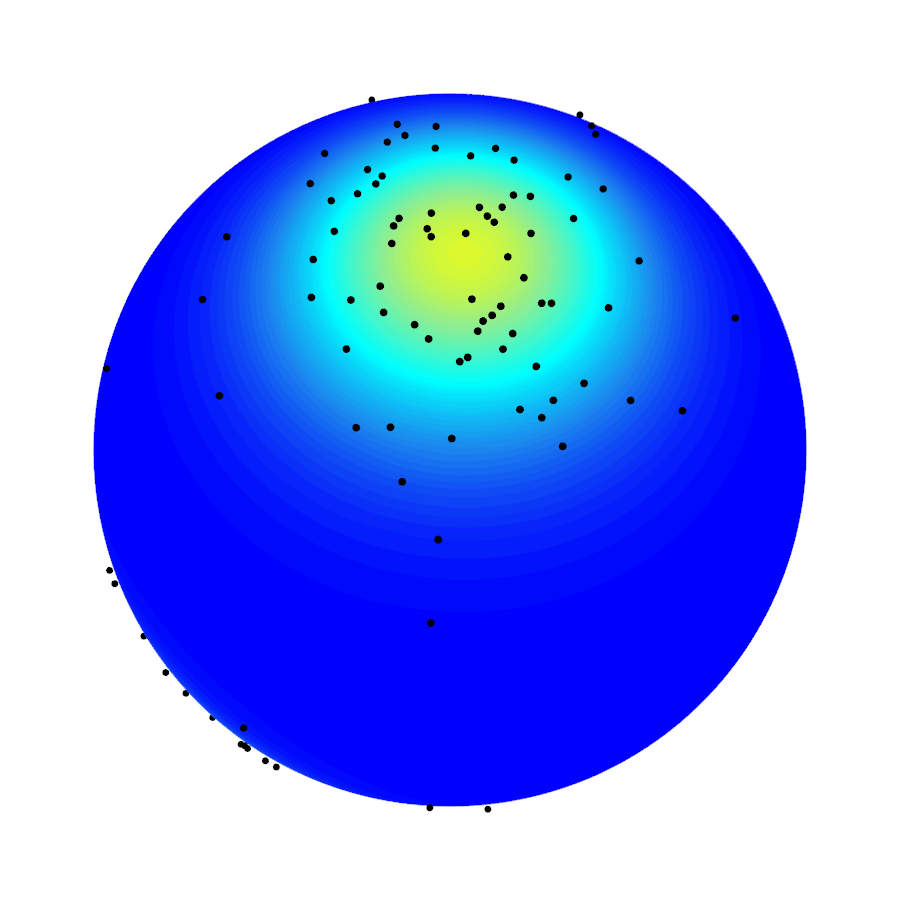}
	\includegraphics[width=0.225\textwidth]{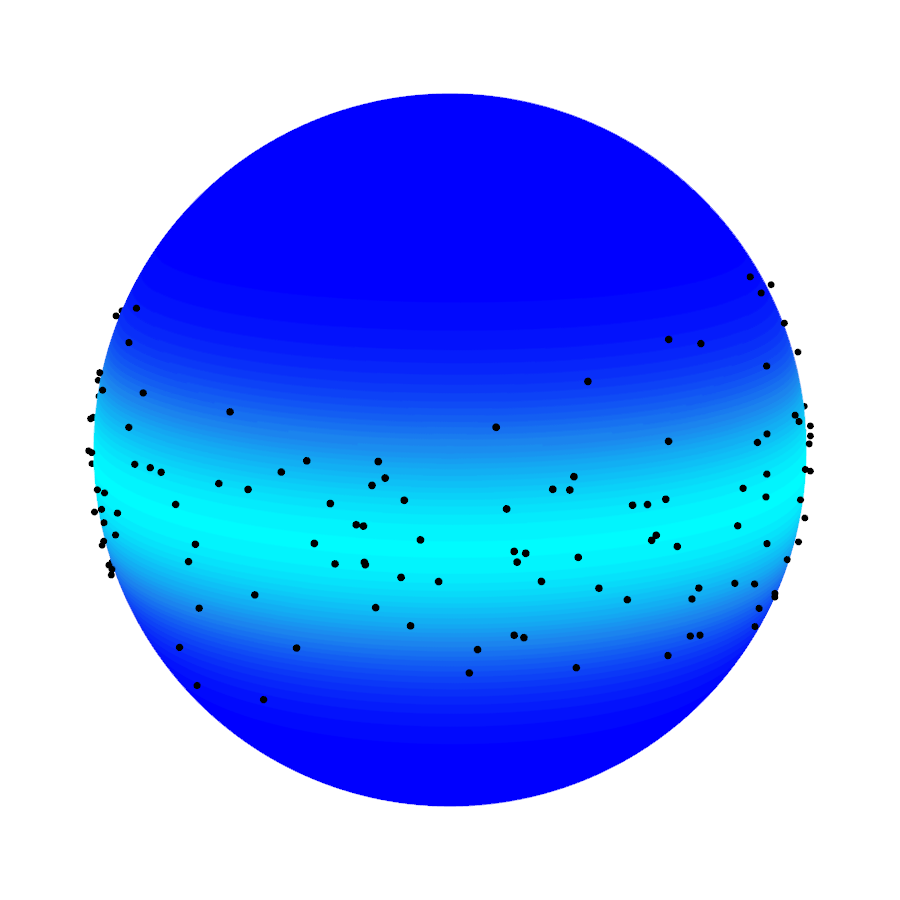}
	\includegraphics[width=0.225\textwidth]{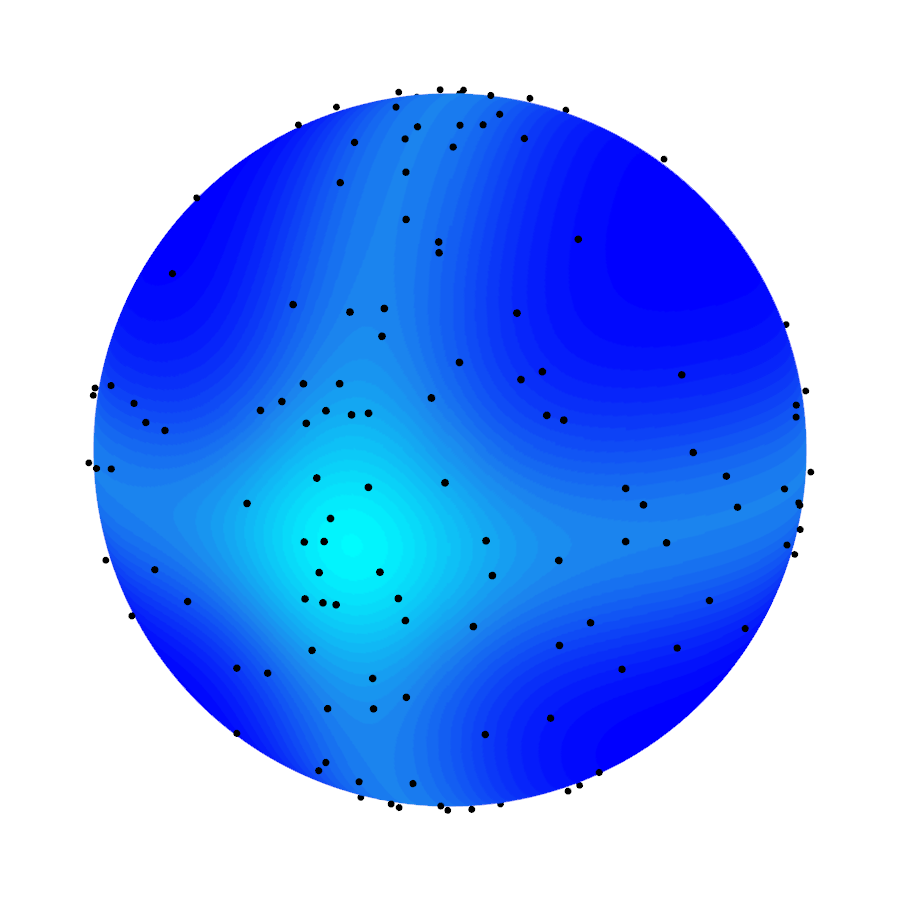}\\
	\includegraphics[width=0.225\textwidth]{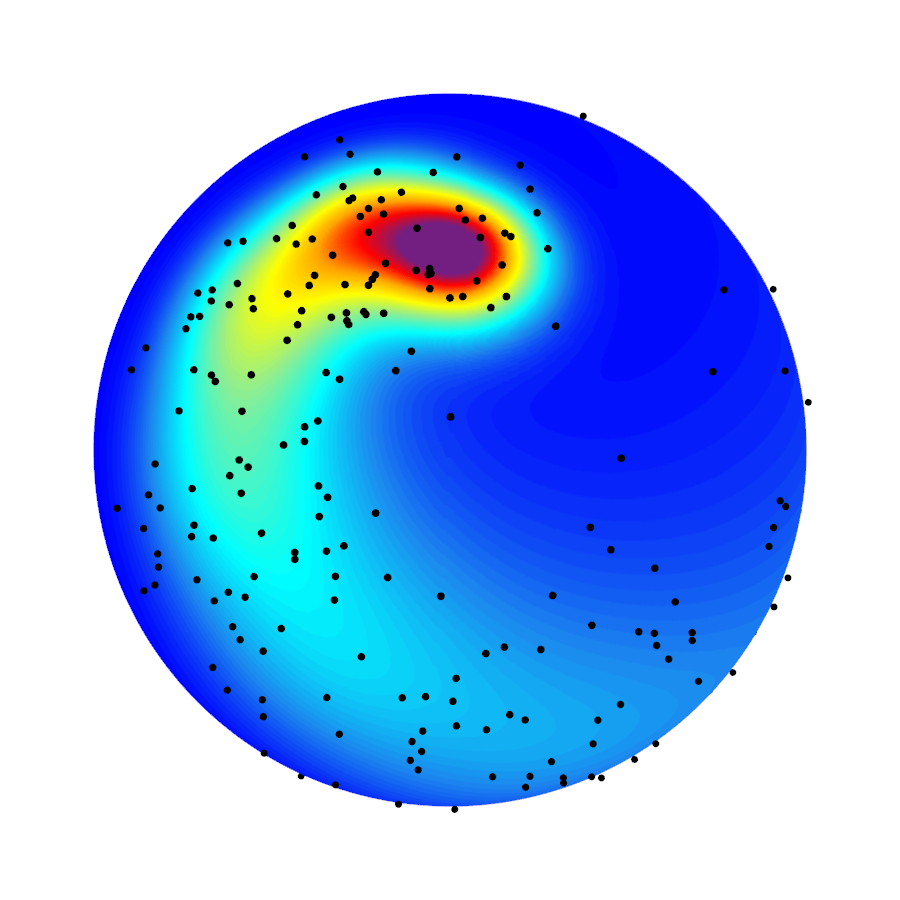}
	\includegraphics[width=0.225\textwidth]{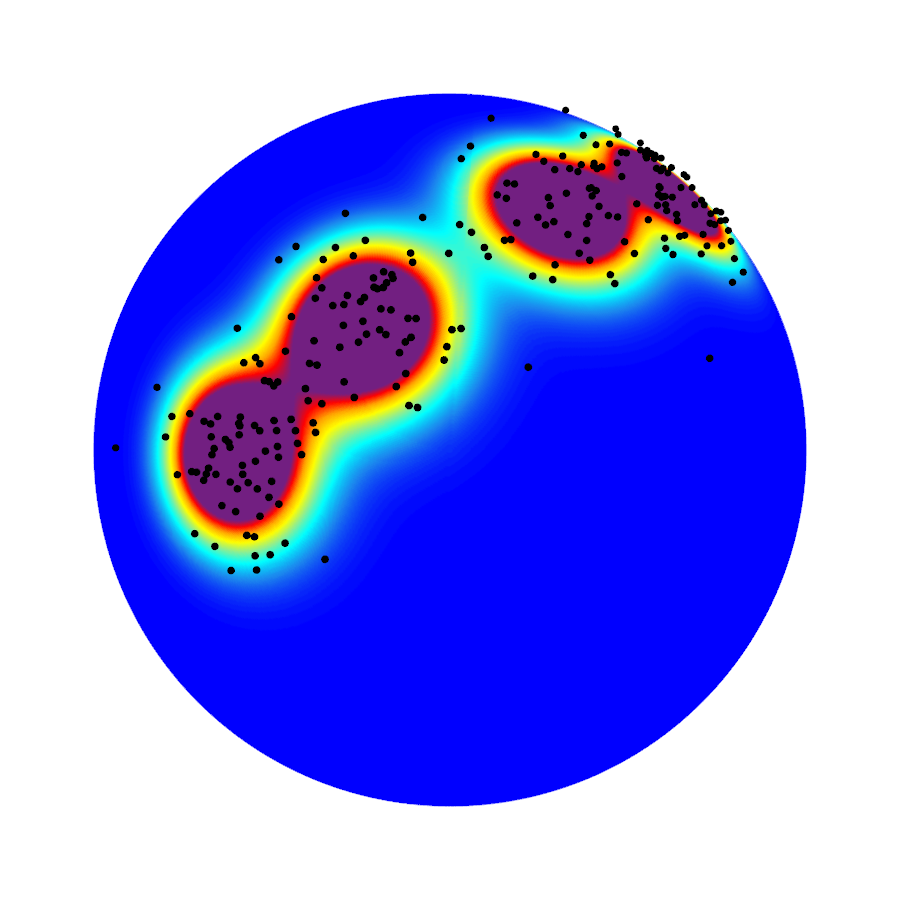}
	\includegraphics[width=0.225\textwidth]{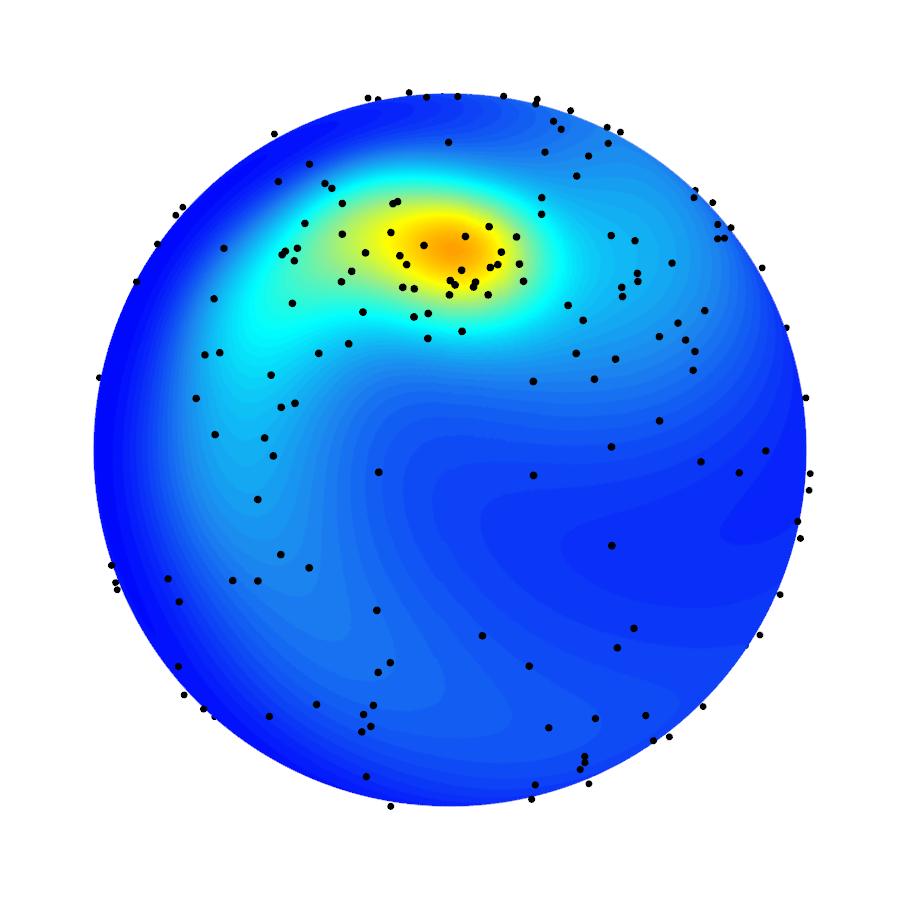}
	\includegraphics[width=0.225\textwidth]{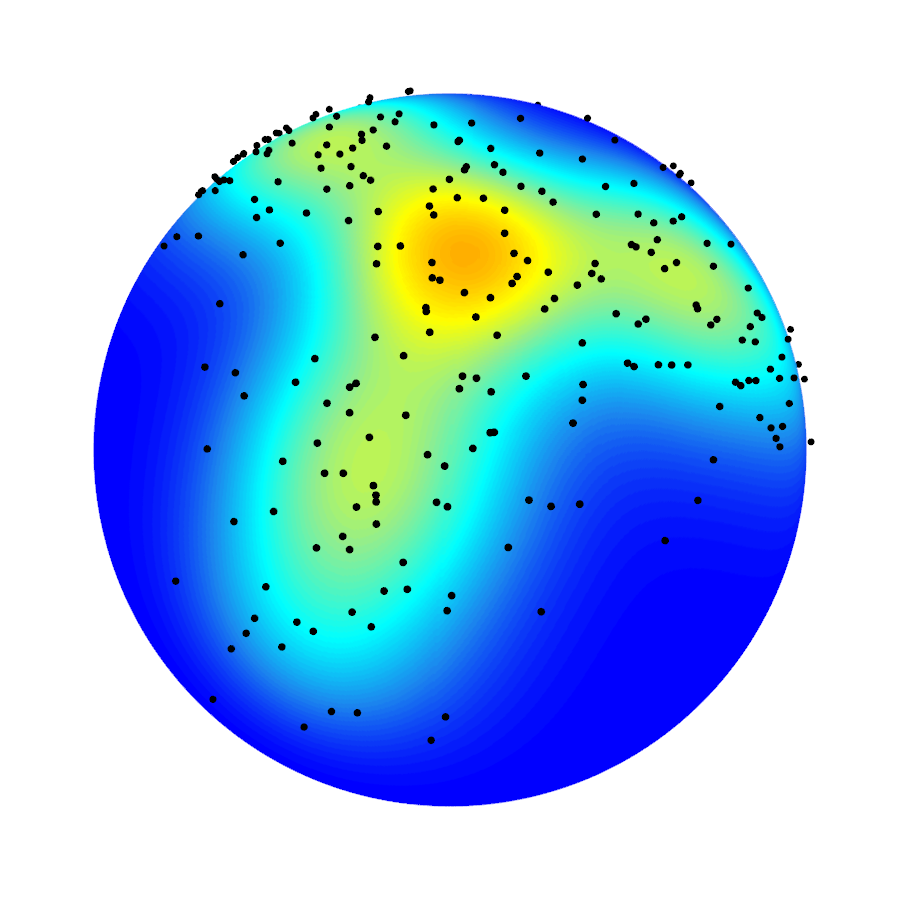}
	\caption{\small Simulation scenarios for the spherical case. From left to right and up to down, models M1 to M20. For each model, a sample of size $250$ is drawn. \label{kdebwd:fig:sph}}
\end{figure}

The comparative study for the spherical case has been done for the directional selectors $h_{\mathrm{LSCV}}$, $h_{\mathrm{ROT}}$, $h_{\mathrm{AMI}}$ and $h_{\mathrm{EMI}}$, in the models given in Figure \ref{kdebwd:fig:sph}. As in the previous case, Table \ref{kdebwd:tab:rankcirsph} contains the scores of the selectors for the different sample sizes, Table \ref{kdebwd:tab:sph} includes the detailed results for $n=500$ and the rest of the sample sizes are shown in Appendix \ref{kdebwd:ap:tables}.\\

In this case the results are even more clear. The $h_\mathrm{EMI}$ selector is by far the best, with an important gap between its competitors for all the sample sizes considered. Further, the effect of computing the exact error instead of the asymptotic one can be appreciated: $h_\mathrm{AMI}$ only is competitive against the cross-validated selectors for sample sizes larger than $n=250$, while $h_\mathrm{EMI}$ remains always the most competitive. In addition, the performance of $h_\mathrm{AMI}$ seems to decrease due to the effect of the dimension in the asymptotic error and does not converge so quick as in the circular case to the performance of $h_\mathrm{EMI}$.

An interesting fact is that $h_\mathrm{LSCV}$ performs better than $h_\mathrm{LCV}$, contrarily to what happens in the circular case. This phenomena is strengthen with higher dimensions, as it can be seen in the next subsection. A possible explanation is the following. For the standard linear case, LCV has been proved to be a bad selector in densities with heavy tails (see \cite{Cao1994}) that are likely to produce outliers. In the circular case, the compact support jointly with periodicity may mitigate this situation, something that does not hold when the dimension increases and the sparsity of the observations is more likely. This makes that among the cross-validated selectors $h_\mathrm{LCV}$ works better for $q=1$ and $h_\mathrm{LSCV}$ for $q>1$.

\begin{table}[H]
\centering
\scriptsize
\begin{tabular}{r|r|rrrrrrr}\toprule\toprule
Model &  \multicolumn{1}{c|}{ $h_\mathrm{MISE}$} & \multicolumn{1}{c}{$h_\mathrm{LCV}$} & \multicolumn{1}{c}{$h_\mathrm{LSCV}$} & \multicolumn{1}{c}{$h_\mathrm{ROT}$} & \multicolumn{1}{c}{$h_\mathrm{AMI}$} & \multicolumn{1}{c}{$h_\mathrm{EMI}$} \\\midrule
M1 & $0.0000$ & $0.023$ ($0.06$) & $0.023$ ($0.06$) & $0.041$ ($0.03$) & $0.017$ ($0.02$) & $\mathbf{0.014}$ ($0.02$) \\
M2 & $0.3058$ & $0.331$ ($0.13$) & $0.340$ ($0.15$) & $0.316$ ($0.13$) & $0.312$ ($0.12$) & $\mathbf{0.310}$ ($0.12$) \\
M3 & $0.4729$ & $0.555$ ($0.22$) & $0.525$ ($0.22$) & $0.542$ ($0.21$) & $\mathbf{0.481}$ ($0.19$) & $0.487$ ($0.19$) \\
M4 & $1.0441$ & $1.501$ ($0.46$) & $1.117$ ($0.34$) & $2.588$ ($0.48$) & $\mathbf{1.088}$ ($0.33$) & $1.093$ ($0.35$) \\
M5 & $0.9621$ & $1.417$ ($0.45$) & $1.024$ ($0.31$) & $2.130$ ($0.40$) & $1.016$ ($0.30$) & $\mathbf{1.003}$ ($0.32$) \\
M6 & $0.4172$ & $0.493$ ($0.15$) & $0.450$ ($0.14$) & $\mathbf{0.427}$ ($0.11$) & $0.496$ ($0.17$) & $0.443$ ($0.13$) \\
M7 & $0.3927$ & $0.417$ ($0.13$) & $0.417$ ($0.12$) & $1.656$ ($0.28$) & $\mathbf{0.399}$ ($0.12$) & $0.400$ ($0.12$) \\
M8 & $0.3380$ & $0.352$ ($0.11$) & $0.370$ ($0.12$) & $0.374$ ($0.11$) & $0.352$ ($0.10$) & $\mathbf{0.343}$ ($0.11$) \\
M9 & $2.6708$ & $5.343$ ($1.26$) & $2.834$ ($0.78$) & $9.710$ ($0.92$) & $2.871$ ($0.65$) & $\mathbf{2.733}$ ($0.72$) \\
M10 & $0.9698$ & $1.230$ ($0.31$) & $\mathbf{1.036}$ ($0.30$) & $1.521$ ($0.29$) & $1.110$ ($0.30$) & $1.101$ ($0.31$) \\
M11 & $1.0349$ & $1.312$ ($0.34$) & $1.097$ ($0.29$) & $2.213$ ($0.35$) & $1.158$ ($0.26$) & $\mathbf{1.067}$ ($0.27$) \\
M12 & $1.5800$ & $2.365$ ($0.59$) & $1.668$ ($0.43$) & $4.123$ ($0.54$) & $\mathbf{1.642}$ ($0.42$) & $1.643$ ($0.44$) \\
M13 & $1.4085$ & $1.674$ ($0.26$) & $1.472$ ($0.23$) & $2.211$ ($0.13$) & $1.729$ ($0.40$) & $\mathbf{1.464}$ ($0.25$) \\
M14 & $1.1299$ & $1.176$ ($0.30$) & $1.182$ ($0.30$) & $8.885$ ($0.77$) & $1.160$ ($0.28$) & $\mathbf{1.137}$ ($0.28$) \\
M15 & $1.1262$ & $\mathbf{1.155}$ ($0.21$) & $1.162$ ($0.21$) & $7.528$ ($0.76$) & $1.302$ ($0.25$) & $1.160$ ($0.21$) \\
M16 & $0.8637$ & $0.890$ ($0.14$) & $\mathbf{0.887}$ ($0.14$) & $3.480$ ($0.22$) & $0.957$ ($0.21$) & $0.889$ ($0.15$) \\
M17 & $1.8989$ & $2.514$ ($0.52$) & $1.971$ ($0.42$) & $6.693$ ($0.45$) & $2.060$ ($0.39$) & $\mathbf{1.950}$ ($0.42$) \\
M18 & $5.0555$ & $5.170$ ($1.08$) & $5.279$ ($1.14$) & $28.468$ ($0.79$) & $5.272$ ($1.06$) & $\mathbf{5.097}$ ($1.08$) \\
M19 & $1.1259$ & $1.262$ ($0.26$) & $\mathbf{1.177}$ ($0.24$) & $2.750$ ($0.24$) & $1.244$ ($0.30$) & $1.199$ ($0.31$) \\
M20 & $1.1810$ & $1.214$ ($0.28$) & $1.250$ ($0.30$) & $2.219$ ($0.28$) & $1.246$ ($0.29$) & $\mathbf{1.195}$ ($0.27$) \\
\bottomrule\bottomrule
\end{tabular}
\caption{\small Comparative study for the spherical case, with sample size $n=500$. Columns of the selector $\bullet$ represent the $\mathrm{MISE}(\bullet)\times100$, with bold type for the minimum of the errors. The standard deviation of the $\mathrm{ISE}\times100$ is given between parentheses.\label{kdebwd:tab:sph}}
\end{table}

\subsection{The effect of dimension}
\label{kdebwd:subsec:dimension}

Finally, the previous selectors are tested in higher dimensions. Table \ref{kdebwd:tab:rankdim} summarizes the information for dimensions $q=3,4,5$ and sample size $n=1000$ (see Table \ref{kdebwd:tab:apdim} in Appendix \ref{kdebwd:ap:tables} for whole results). As it can be seen, $h_\mathrm{EMI}$ continues performing better than its competitors. Also, as previously commented in the spherical case, $h_\mathrm{AMI}$ has a lower performance due to the misfit between AMISE and MISE, which gets worse when the sample size is fixed and the dimension increases. $h_\mathrm{LSCV}$ arises as the second best selector for higher dimensions, outperforming $h_\mathrm{LCV}$, as happened in the spherical\nolinebreak[4] case.

\begin{table}[h]
	\centering
	\small
	\begin{tabular}{r|rrrrrrr}\toprule\toprule
		$q$ & \multicolumn{1}{c}{$h_\mathrm{LCV}$} & \multicolumn{1}{c}{$h_\mathrm{LSCV}$} & \multicolumn{1}{c}{$h_\mathrm{ROT}$} & \multicolumn{1}{c}{$h_\mathrm{AMI}$} & \multicolumn{1}{c}{$h_\mathrm{EMI}$} \\\midrule
		$3$ & $7.4838$ & $15.3405$ & $4.7658$ & $10.6920$ & $\mathbf{17.6956}$ \\
		$4$ & $8.8565$ & $15.4862$ & $5.1370$ &  $9.1871$ & $\mathbf{17.4579}$ \\
		$5$ &$10.3262$ & $15.1815$ & $5.4088$ &  $7.6616$ & $\mathbf{15.5301}$ \\
		\bottomrule\bottomrule
	\end{tabular}
	\caption{\small Ranking for the selectors for dimensions $q=3,4,5$ and sample size $n=1000$. The larger the score in the ranking, the better the performance of the selector. Bold type indicates the best selector. \label{kdebwd:tab:rankdim}}
\end{table}

\section{Data application}
\label{kdebwd:sec:data}

According with the comparative study of the previous section, the $h_\mathrm{EMI}$ selector poses in average the best performance of all the considered selectors. In this section it will be applied to estimate the density of two real datasets.

\subsection{Wind direction}
\label{kdebwd:subsec:wind}

\begin{figure}[h]
\centering
\includegraphics[scale=0.375]{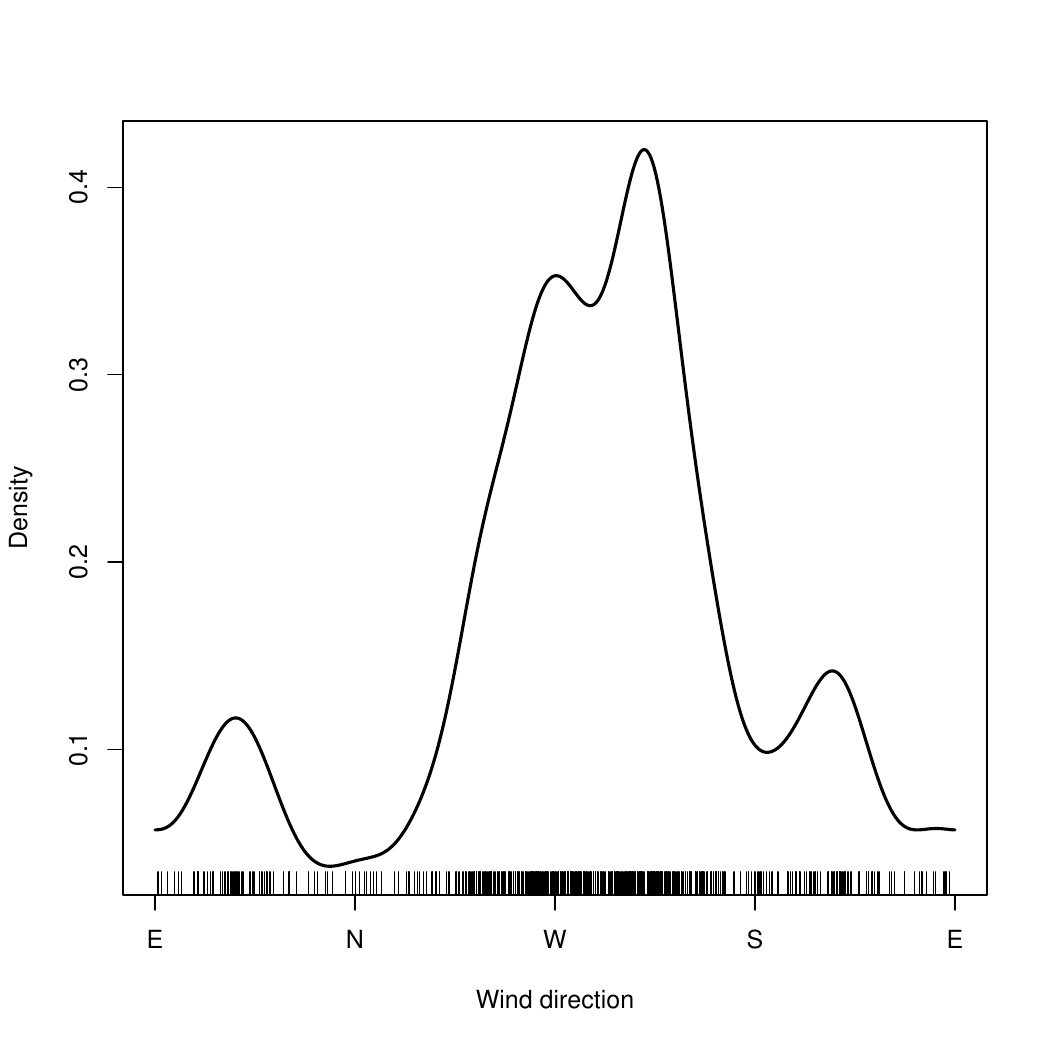}\includegraphics[scale=0.375]{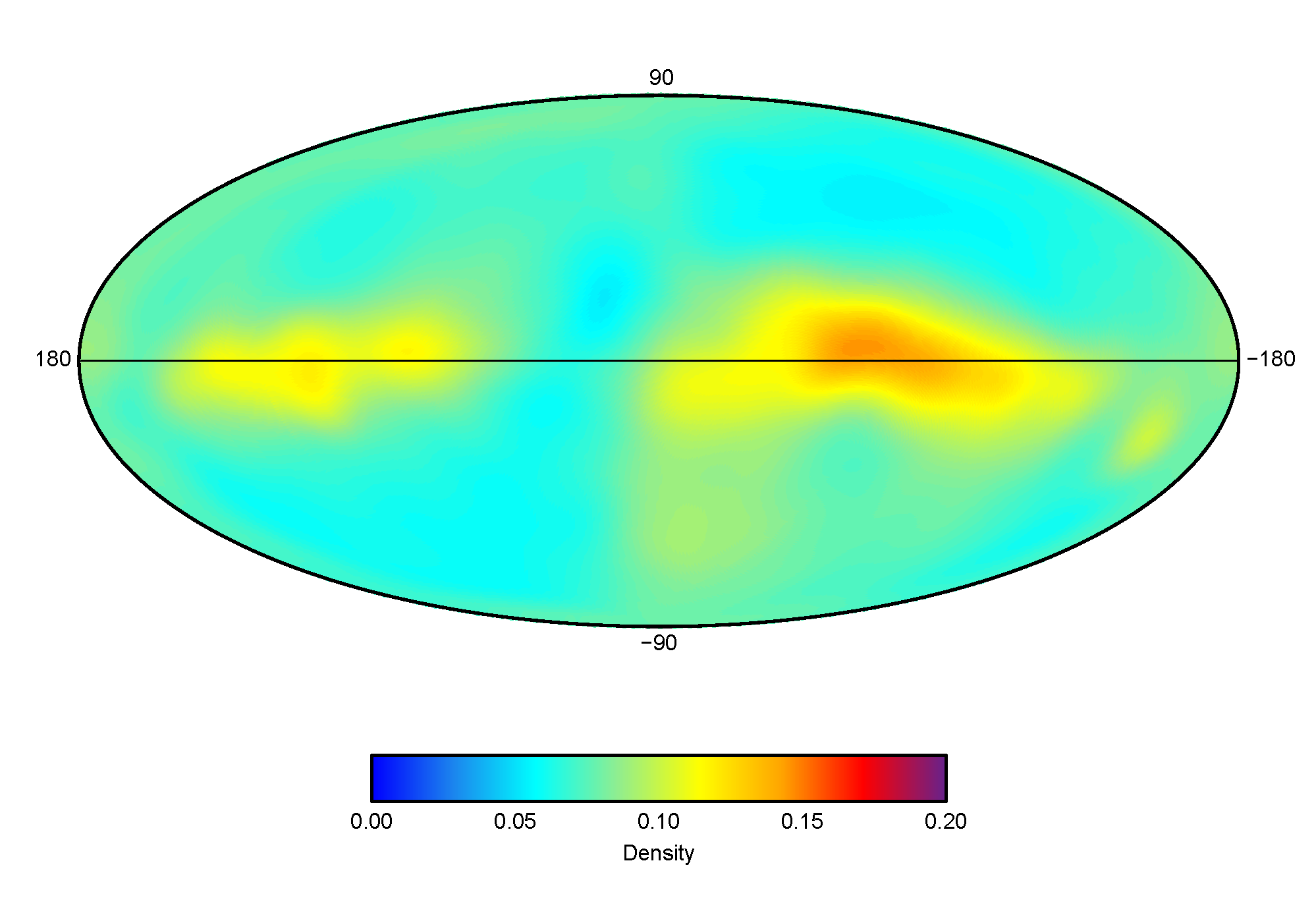}
\caption{\small Left: density of the wind direction in the meteorological station of A Mourela. Right: density of the stars collected in the Hipparcos catalogue, represented in galactic coordinates and Aitoff projection. \label{kdebwd:fig:data}}
\end{figure}

Wind direction is a typical example of circular data. The data of this illustration was recorded in the meteorological station of A Mourela ($7^\circ$ $51$' $21.91$'' W, $43^\circ$ $25$' $52.35$'' N), located near the coal power plant of As Pontes, in the northwest of Spain. The wind direction has a big impact on the dispersion of the pollutants from the coal power plant and a reliable estimation of its unknown density is useful for a further study of the pollutants transportation. The wind direction was measured minutely at the top of a pole of $80$ metres during the month of June, 2012. In order to mitigate serial dependence, the data has been hourly averaged by computing the circular mean, resulting in a sample of size $673$. The resulting bandwidth is $h_\mathrm{EMI}=0.1896$, obtained from the data-driven mixture of $3$ von Mises. Left plot of Figure \ref{kdebwd:fig:data} represents the estimated density, which shows a clear predominance of the winds from the west and three main modes. Running time, measured in a $3.5$ GHz core, is $1.21$ seconds ($0.89$ for the mixtures fitting and $0.32$ for bandwidth optimization).

\subsection{Position of stars}
\label{kdebwd:subsec:stars}

A challenging field where spherical data is present is astronomy. Usually, the position of stars is referred to the position that occupy in the celestial sphere, \textit{i.e.}, the location in the earth surface that arises as the intersection with the imaginary line that joins the centre of the earth with the star. A massive enumeration of near stars is given in the Hipparcos catalogue \citep{Hipparcos}, that collects the findings of the Hipparcos mission carried out by the European Space Agency in 1989--1993. An improved version of the original dataset, available from \cite{VanLeeuwen2007}, contains a corrected collection of the position of the stars on the celestial sphere as well as other star variables. \\

For many years, most of the statistical tools used to describe this kind of data were histograms adapted to the spherical case, where the choice of the bin width was done manually (see page 328 of \cite{Hipparcos}). In this illustration, a smooth estimation of the spherical density is given using the optimal smoothing of the $h_\mathrm{EMI}$ selector. Using the $117955$ star positions from the dataset of \cite{VanLeeuwen2007}, the underlying density is approximated with $12$ components automatically obtained, resulting the bandwidth $h_\mathrm{EMI}=0.1064$. Note that the analysis of such a large dataset by cross-validatory techniques would demand an enormous amount of computing time and memory resources, whereas the running time for $h_\mathrm{EMI}$ is reasonable, with $256.01$ seconds ($247.34$ for the mixtures fitting and $8.67$ for bandwidth optimization). The right plot of Figure \ref{kdebwd:fig:data} shows the density of the position of the measured stars. This plot is given in the Aitoff projection (see \cite{Hipparcos}) and in galactic coordinates, which means that the equator represents the position of the galactic rotation plane. The higher concentrations of stars are located around two spots, that represent the Orion's arm (left) and the Gould's Belt (right) of our galaxy.

\section{Conclusions}
\label{kdebwd:sec:conclusions}

Three new bandwidth selectors for directional data are proposed. The rule of thumb extends and improves significantly the previous proposal of \cite{Taylor2008}, but also fails estimating densities with multimodality. On the other hand, the selectors based on mixtures are competitive with the previous proposals in the literature, being the EMI selector the most competitive on average among all, for different sample sizes and dimensions, but specially for low or moderate sample sizes. The performance of AMI selector is one step behind EMI, a difference that is reduced when sample size increases.\\

In the comparison study, new rotationally symmetric models have been introduced and  other interesting conclusions have been obtained. First, LCV is also a competitive selector for the circular case and outperforms LSCV, something that was known in the literature of circular data. However, this situation is reversed for the spherical case and higher dimensions, where LSCV is competitive and performs better than LCV.\\

The final conclusion of this paper is simple: the EMI bandwidth selector presents a reliable choice for kernel density estimation with directional data and its performance is at least as competitive as the existing proposals until the moment.

\section*{Acknowledgements}
The author gratefully acknowledges the comments and guidance of professors Rosa M. Crujeiras and Wenceslao Gonz\'alez-Manteiga. The work of the author has been supported by FPU grant AP2010-0957 from the Spanish Ministry of Education. Support of Project MTM2008-03010, from the Spanish Ministry of Science and Innovation, Project 10MDS207015PR from Direcci\'on Xeral de I+D, Xunta de Galicia and IAP network StUDyS, from Belgian Science Policy, are acknowledged. The author acknowledges the helpful comments by the Editor and an Associate \nopagebreak[4]Editor.

\appendix

\section{Proofs}
\label{kdebwd:ap:proofs}

\begin{proof}[Proof of Proposition \ref{kdebwd:prop:rot}]

By simple differentiation, the operator (\ref{kdebwd:Psi_dir}) in a von Mises density $\mathrm{vM}(\bmu,\kappa)$\nolinebreak[4] is
\begin{align*}
\Psi\lrp{f_{\mathrm{vM}}(\cdot;\bmu,\kappa),\bx}=\kappa C_q(\kappa)e^{\kappa\bx^T\bmu}\lrp{-\bx^T\bmu+\kappa q^{-1}\lrp{1-(\bx^T\bmu)^2}}.
\end{align*}
Then, by the change of variables of (\ref{kdebwd:change}),
\begin{align*}
R\lrp{\Psi(f_{\mathrm{vM}}(\cdot;\bmu,\kappa),\cdot)}&=\Iq{\Psi(f_{\mathrm{vM}}(\cdot;\bmu,\kappa),\bx)^2}{\bx}\\
&=\kappa^2C_q(\kappa)^2\int_{\Om{q-1}}\int_{-1}^1 e^{2\kappa t}\lrp{-t+\frac{\kappa}{q}(1-t^2)}^2(1-t^2)^{\frac{q}{2}-1}\,dt\,\om{q-1}(d\bxi)\\
&=\frac{\kappa^{q+1}}{2^{q}\pi^{\frac{q}{2}+1}\mathcal{I}_{\frac{q-1}{2}}(\kappa)^2\Gamma\lrp{\frac{q}{2}}}\int_{-1}^1 e^{2\kappa t}\lrp{-t+\frac{\kappa}{q}(1-t^2)}^2(1-t^2)^{\frac{q}{2}-1}\,dt.
\end{align*}
The integral can divided into three terms expanding the square. After two integrations by parts, the sum of the first two is
\begin{align*}
\int_{-1}^1 e^{2\kappa t}(1-t^2)^{\frac{q}{2}-1}t^2\,dt-\frac{2\kappa}{q}\int_{-1}^1 e^{2\kappa t}(1-t^2)^{\frac{q}{2}}t\,dt&=\frac{1}{q}\int_{-1}^1 e^{2\kappa t}(1-t^2)^{\frac{q}{2}}\,dt.
\end{align*}
This integral and the last term follows immediately by the integral form of the modified Bessel function, yielding
\begin{align*}
R\lrp{\Psi(f_{\mathrm{vM}}(\cdot;\bmu,\kappa),\cdot)}=\frac{\kappa^{\frac{q+1}{2}}}{2^{q+2}\pi^\frac{q+1}{2}\mathcal{I}_{\frac{q-1}{2}}(\kappa)^2q}\lrc{2q\mathcal{I}_{\frac{q+1}{2}}(2\kappa)+(2+q)\kappa\mathcal{I}_{\frac{q+3}{2}}(2\kappa)}.
\end{align*}
The particular case $q=2$ follows by using $\Ical_{-\frac{1}{2}}(z)=\sqrt{\frac{2}{\pi z}}\sinh(z)$, $\Ical_{\frac{1}{2}}(z)=\sqrt{\frac{2}{\pi z}}\cosh(z)$ and relations $\Ical_{\nu-1}(z)=\Ical_{\nu+1}(z)+\frac{2\nu}{z}\Ical_\nu(z)$ and $\Ical_{\nu+1}(z)=\Ical_{\nu-1}(z)-\frac{2\nu}{z}\Ical_\nu(z)$. Also, for the von Mises kernel $L(r)=e^{-r}$, it is easy to see that
\begin{align*}
\lambda_{q}(L)=(2\pi)^\frac{q}{2},\quad b_q(L)=\frac{q}{2}\quad\text{and}\quad d_q(L)=2^{-\frac{q}{2}}.
\end{align*}
\end{proof}

\section{Models for the simulation study}
\label{kdebwd:ap:models}

Table \ref{kdebwd:tab:models} collects the densities of the different models used in the simulation study. Apart from the notations introduced in Section \ref{kdebwd:sec:comparative} for the families of directional densities, the following terminology is used. First, the vector $\bO$ represent a vector with $q$ zeros. Second, functions $\rho_1$ and $\rho_2$ give the polar and spherical parametrization of a vector from a single and a pair of angles, respectively:
\begin{align*}
\rho_1(\theta)=(\cos(\theta),\sin(\theta)),\,\rho_2(\theta,\phi)=(\cos(\theta)\sin(\phi),\sin(\theta)\sin(\phi),\cos(\phi)),\quad\theta\in[0,2\pi),\,\phi\in[0,\pi).
\end{align*}
Thirdly, the notation $\#i$ for an index $i$ varying in the ordered set $S$ aims to represent the position of $i$ in $S$. Finally, the matrix $\bSigma_1$ is such that the first three elements of $\mathrm{diag}(\bSigma_1)$ are $\frac{1}{2},\,\frac{1}{4},\,\frac{1}{8}$ and the rest of them are $1$. Matrix $\bSigma_2$ are just like matrix $\bSigma_1$ but with the diagonal reversed.

\section{Extended tables for the simulation study}
\label{kdebwd:ap:tables}
Tables \ref{kdebwd:tab:apcir} and \ref{kdebwd:tab:apsph} show the results for sample sizes $100$, $250$ and $1000$ for the circular and spherical cases, respectively. The case $n=500$ is collected in Tables \ref{kdebwd:tab:cir} and \ref{kdebwd:tab:sph}. Finally, Table \nopagebreak[4]\ref{kdebwd:tab:apdim} contains the simulation results for sample size $n=1000$ and dimensions $q=3,4,5$.
\newpage
\vspace*{\fill}
\begin{table}[H]
	\scriptsize
	\centering
	\setlength{\tabcolsep}{3pt}
	\begin{tabular}{cp{3.8cm}l}
		\toprule\toprule
		Model & Description & Density \\[0.05cm]\midrule
		M1 & Uniform (Unif.) & $\om{q}^{-1}$\\[0.05cm]
		M2 & Von Mises (vM) & $\mathrm{vM}((\bO,1),2)$ \\[0.05cm]
		M3 & Projected normal (PN), \phantom{aa} rotationally symmetric & $\mathrm{PN}\lrp{(\bO,1),\frac{1}{2}I_{q+1}}$ \\[0.05cm]
		M4 & Projected normal, non \phantom{aaa} rotationally symmetric & $\mathrm{PN}((1,\bO),2\bSigma_1)$ \\[0.05cm]
		M5 & Directional Cauchy (DC)& $\mathrm{DC}((\bO,1),10)$ \\[0.05cm]
		M6 & Skew normal directional & $\mathrm{SND}\lrp{(\bO,1),\frac{1}{2},\frac{1}{2},5}$\\[0.05cm]
		M7 & Watson & $\mathrm{W}((1,\bO),2)$\\[0.05cm]\midrule
		M8 & Mixture of two 90$^\circ$ vM & $\frac{1}{2}\mathrm{vM}((\bO,1),3)+\frac{1}{2}\mathrm{vM}((1,\bO),3)$\\[0.05cm]
		M9 & Skewed mixture of vM (8 vM) &  \\[0.05cm]
		& $q=1$ & $\frac{1}{8}\mathrm{vM}\lrp{(0,1),\lrp{\frac{5}{3}}^8}+\frac{1}{8}\sum\limits_{i\in\lrb{1,2,3,4,6,8,9}} \mathrm{vM}\lrp{\rho_1\lrp{\frac{i\pi}{20}},\lrp{\frac{5}{3}}^{\#i}}$\\[0.05cm]
		& $q>1$ & $\frac{1}{8}\mathrm{vM}\lrp{(\bOd,1),\lrp{\frac{5}{3}}^8}+\frac{1}{8}\sum\limits_{i\in\lrb{1,2,3,4,6,8,9}} \mathrm{vM}\lrp{\lrp{\bOd,\rho_2\lrp{0,\frac{(10-i)\pi}{20}}},\lrp{\frac{5}{3}}^{\#i}}$\\[0.05cm]
		M10 & Mixture of two PN & $\frac{1}{2}\mathrm{PN}\lrp{(1,\bO),\bSigma_1}+\frac{1}{2}\mathrm{PN}\big(\big(\frac{\sqrt{2}}{2},\frac{\sqrt{2}}{2},\mathbf{0}_{q-1}\big),\bSigma_2\big)$\\[0.05cm]
		M11 & Bandage (5 vM) & \\[0.05cm]
		& $q=1$ & $\frac{2}{10}\mathrm{vM}\lrp{\rho_1\lrp{0},20}+\frac{2}{10}\sum\limits_{i\in\lrb{-1,1}} \mathrm{vM}\lrp{\rho_1\lrp{\frac{i\pi}{6}},10}$\\[0.05cm]
		& & $+\frac{1}{10}\sum\limits_{i\in\lrb{-1,1}}\lrc{\mathrm{vM}\lrp{\rho_1\lrp{\frac{i\pi}{4}},5}+\mathrm{vM}\lrp{\rho_1\lrp{\frac{i\pi}{2}},1}}$\\[0.05cm]
		& $q>1$ &$\frac{2}{10}\mathrm{vM}\lrp{\lrp{\rho_2\lrp{0,\frac{\pi}{2}},\bOd},20}+\frac{2}{10}\sum\limits_{i\in\lrb{-1,1}}\mathrm{vM}\lrp{\lrp{\rho_2\lrp{\frac{i\pi}{6},\frac{(4+i)\pi}{8}},\bOd},10}$\\[0.05cm]
		& &$+\frac{2}{10}\sum\limits_{i\in\lrb{-1,1}}\Big[\mathrm{vM}\lrp{\lrp{\rho_2\lrp{\frac{i\pi}{4},\frac{i\pi}{3}},\bOd},5}+\mathrm{vM}\lrp{\lrp{\rho_2\lrp{\frac{i\pi}{2},\frac{i\pi}{3}},\bOd},1}\Big]$\\[0.05cm]
		M12 & Mixture of PN and DC & $\frac{3}{4}\mathrm{PN}((1,\bO),\bSigma_1)+\frac{1}{4}\mathrm{DC}\big(\big(\frac{1}{2},\frac{\sqrt{3}}{2},\mathbf{0}_{q-1}\big),50\big)$\\[0.05cm]
		M13 & Mixture of Unif. and DC & $\frac{4}{5}\om{q}^{-1}+\frac{1}{5}\mathrm{DC}\big(\big(\frac{1}{2},\frac{\sqrt{3}}{2},\mathbf{0}_{q-1}\big),100\big)$\\[0.05cm]\midrule
		M14 & Trimodal (3 vM)& $\frac{1}{3} \mathrm{vM}((\bO,1),10)+\frac{1}{3}\mathrm{vM}\lrp{\lrp{\bOd,\rho_1\lrp{\frac{5\pi}{4}}},10}+\frac{1}{3}\mathrm{vM}\lrp{\lrp{\bOd,\rho_1\lrp{\frac{7\pi}{4}}},10}$\\[0.05cm]
		M15 & Small circle & $\mathrm{SC}((\bO,1),10)$\\[0.05cm]
		M16 & Double small circle & $\frac{1}{2}\mathrm{SC}((\bO,1),10)+\frac{1}{2}\mathrm{SC}((1,\bO),10)$\\[0.05cm]
		M17 & Spiral (10 vM) & \\[0.05cm]
		& $q=1$ & $\frac{1}{10}\sum_{i=0}^9 \mathrm{vM}\lrp{\rho_1\lrp{\frac{3\pi i}{18}},\lrp{\frac{3}{2}}^{10-i}}$\\[0.05cm]
		& $q>1$ & $\frac{1}{10}\sum_{i=0}^9 \mathrm{vM}\lrp{\lrp{\rho_2\lrp{\frac{3\pi i}{18},\frac{3\pi i}{36}},\bOd},\lrp{\frac{3}{2}}^{10-i}}$\\[0.05cm]\midrule
		M18 & Claw (4 vM) & $\frac{1}{4}\sum_{i=0}^1\Big[ \mathrm{vM}\lrp{\lrp{\bOd,\rho_1\lrp{\frac{(2i+1)\pi}{4}},50}}+\mathrm{vM}\lrp{\lrp{\bOd,\rho_1\lrp{\frac{(i+2)\pi}{5}}},50}\Big]$\\[0.05cm]
		M19 & Double spiral (20 vM)& \\[0.05cm]
		& $q=1$ & $\frac{1}{20}\sum_{i=0}^9\lrc{ \mathrm{vM}\lrp{\rho_1\lrp{\frac{3\pi i}{18}},\lrp{\frac{3}{2}}^{10-i}}+\mathrm{vM}\lrp{\rho_1\lrp{-\frac{3\pi i}{18}},10}}$\\[0.05cm]
		& $q>1$ & $\frac{1}{20}\sum_{i=0}^9\Big[ \mathrm{vM}\lrp{\lrp{\rho_2\lrp{\frac{3\pi i}{18},\frac{3\pi i}{36}},\bOd},\lrp{\frac{3}{2}}^{10-i}}$\\[0.05cm]
		& & $+\mathrm{vM}\lrp{\lrp{\rho_2\lrp{\frac{3\pi i}{18},-\frac{3\pi i}{36}},\bOd},10}\Big]$\\[0.05cm]
		M20 & Windmill (4 vM)& \\[0.05cm]
		& $q=1$ & $\frac{2}{11}\mathrm{vM}\lrp{(0,1),20}+\frac{1}{11}\sum_{i=1}^3\mathrm{vM}\lrp{\rho_1\lrp{\frac{2i\pi}{3}},15}$\\[0.05cm]
		& $q>1$ & $\frac{2}{11}\mathrm{vM}\lrp{(\bO,1),20}+\frac{1}{11}\sum_{i=1}^3\sum\limits_{j\in\lrb{3,5,6}}\mathrm{vM}\lrp{\lrp{\rho_2\lrp{\frac{2i\pi}{3},\frac{\pi}{j}},\bOd},15}$\\[0.05cm]
		\bottomrule\bottomrule
	\end{tabular}
	\caption{\small Directional densities considered in the simulation study.\label{kdebwd:tab:models}}
\end{table}
\vspace*{\fill}
\newpage

\begin{table}[H]
\centering
\scriptsize
\setlength{\tabcolsep}{2pt}
\begin{tabular}{@{\hspace{0.2cm}}r@{\hspace{0.27cm}}|r@{\hspace{0.27cm}}|r@{\hspace{0.27cm}}r@{\hspace{0.27cm}}r@{\hspace{0.27cm}}
r@{\hspace{0.27cm}}r@{\hspace{0.27cm}}r@{\hspace{0.27cm}}
r@{\hspace{0.27cm}}r}\toprule\toprule
Model &  \multicolumn{1}{c|}{ $h_\mathrm{MISE}$} & \multicolumn{1}{c}{$h_\mathrm{LCV}$} & \multicolumn{1}{c}{$h_\mathrm{LSCV}$} & \multicolumn{1}{c}{$h_\mathrm{TAY}$} & \multicolumn{1}{c}{$h_\mathrm{OLI}$} & \multicolumn{1}{c}{$h_\mathrm{ROT}$} & \multicolumn{1}{c}{$h_\mathrm{AMI}$} & \multicolumn{1}{c}{$h_\mathrm{EMI}$} \\\midrule
M1 & $0.0000$ & $0.286$ ($0.56$) & $0.293$ ($0.60$) & $\mathbf{0.019}$ ($0.06$) & $0.680$ ($1.03$) & $0.109$ ($0.16$) & $0.115$ ($0.20$) & $0.120$ ($0.20$) \\
M2 & $0.7525$ & $0.938$ ($0.62$) & $1.140$ ($1.02$) & $0.803$ ($0.57$) & $1.229$ ($1.15$) & $\mathbf{0.790}$ ($0.55$) & $0.812$ ($0.67$) & $0.802$ ($0.62$) \\
M3 & $0.8828$ & $1.152$ ($0.72$) & $1.282$ ($1.09$) & $0.985$ ($0.66$) & $1.345$ ($1.24$) & $\mathbf{0.936}$ ($0.62$) & $0.953$ ($0.64$) & $0.962$ ($0.65$) \\
M4 & $1.1173$ & $\mathbf{1.492}$ ($0.90$) & $1.526$ ($1.05$) & $1.890$ ($1.02$) & $1.558$ ($1.18$) & $1.542$ ($0.86$) & $1.585$ ($1.00$) & $1.551$ ($0.94$) \\
M5 & $1.9219$ & $3.375$ ($1.71$) & $2.547$ ($1.61$) & $4.536$ ($1.79$) & $\mathbf{2.357}$ ($1.38$) & $3.982$ ($1.61$) & $2.415$ ($1.54$) & $2.384$ ($1.57$) \\
M6 & $0.8810$ & $1.264$ ($0.95$) & $1.228$ ($1.03$) & $\mathbf{0.907}$ ($0.53$) & $1.654$ ($1.37$) & $0.918$ ($0.55$) & $1.203$ ($0.90$) & $1.073$ ($0.74$) \\
M7 & $0.9914$ & $1.145$ ($0.64$) & $1.248$ ($0.79$) & $6.584$ ($0.20$) & $1.317$ ($0.96$) & $5.571$ ($0.74$) & $1.091$ ($0.67$) & $\mathbf{1.064}$ ($0.61$) \\
M8 & $0.7534$ & $0.885$ ($0.55$) & $1.046$ ($0.84$) & $0.823$ ($0.40$) & $1.255$ ($1.12$) & $\mathbf{0.775}$ ($0.43$) & $0.935$ ($0.59$) & $0.859$ ($0.50$) \\
M9 & $1.8991$ & $2.727$ ($1.19$) & $2.386$ ($1.35$) & $2.743$ ($0.89$) & $2.516$ ($1.40$) & $2.574$ ($0.88$) & $2.401$ ($1.22$) & $\mathbf{2.245}$ ($1.13$) \\
M10 & $0.9551$ & $1.154$ ($0.62$) & $1.355$ ($1.16$) & $1.029$ ($0.54$) & $1.510$ ($1.22$) & $\mathbf{0.995}$ ($0.54$) & $1.060$ ($0.80$) & $1.048$ ($0.68$) \\
M11 & $0.8626$ & $1.078$ ($0.68$) & $1.200$ ($0.95$) & $0.976$ ($0.63$) & $1.363$ ($1.24$) & $\mathbf{0.913}$ ($0.58$) & $0.932$ ($0.61$) & $0.933$ ($0.60$) \\
M12 & $2.1940$ & $2.995$ ($1.00$) & $\mathbf{2.731}$ ($1.33$) & $3.461$ ($0.79$) & $2.967$ ($1.42$) & $3.217$ ($0.77$) & $3.278$ ($1.02$) & $3.265$ ($0.93$) \\
M13 & $2.6714$ & $3.305$ ($1.01$) & $\mathbf{3.142}$ ($1.03$) & $5.012$ ($0.46$) & $3.405$ ($1.06$) & $4.486$ ($0.51$) & $3.879$ ($1.10$) & $3.611$ ($1.18$) \\
M14 & $1.7224$ & $1.872$ ($0.84$) & $2.028$ ($1.04$) & $12.790$ ($0.52$) & $2.076$ ($1.14$) & $11.392$ ($1.12$) & $1.853$ ($0.86$) & $\mathbf{1.789}$ ($0.82$) \\
M15 & $2.3079$ & $2.608$ ($1.50$) & $2.892$ ($1.74$) & $43.759$ ($1.41$) & $2.701$ ($1.64$) & $39.639$ ($3.94$) & $2.483$ ($1.36$) & $\mathbf{2.408}$ ($1.33$) \\
M16 & $2.2045$ & $2.354$ ($0.98$) & $2.585$ ($1.23$) & $14.312$ ($0.05$) & $2.620$ ($1.21$) & $14.303$ ($0.03$) & $2.490$ ($1.04$) & $\mathbf{2.325}$ ($0.97$) \\
M17 & $1.2089$ & $1.448$ ($0.56$) & $1.488$ ($0.65$) & $1.740$ ($0.41$) & $1.953$ ($1.11$) & $\mathbf{1.372}$ ($0.32$) & $1.523$ ($0.61$) & $1.413$ ($0.41$) \\
M18 & $3.5569$ & $\mathbf{3.929}$ ($1.56$) & $4.108$ ($1.58$) & $7.254$ ($0.72$) & $4.082$ ($1.61$) & $7.028$ ($0.74$) & $4.360$ ($1.47$) & $4.380$ ($1.44$) \\
M19 & $0.6717$ & $0.875$ ($0.53$) & $0.914$ ($0.56$) & $1.114$ ($0.50$) & $1.320$ ($1.09$) & $\mathbf{0.766}$ ($0.32$) & $0.887$ ($0.55$) & $0.813$ ($0.41$) \\
M20 & $1.8640$ & $2.034$ ($0.93$) & $2.203$ ($1.16$) & $14.834$ ($0.89$) & $2.360$ ($1.28$) & $12.778$ ($1.56$) & $2.046$ ($0.93$) & $\mathbf{1.943}$ ($0.86$) \\
\midrule
M1 & $0.0000$ & $0.114$ ($0.22$) & $0.117$ ($0.23$) & $\mathbf{0.004}$ ($0.01$) & $0.137$ ($0.27$) & $0.040$ ($0.07$) & $0.040$ ($0.07$) & $0.043$ ($0.07$) \\
M2 & $0.3760$ & $0.456$ ($0.29$) & $0.513$ ($0.41$) & $0.393$ ($0.27$) & $0.444$ ($0.35$) & $0.387$ ($0.26$) & $0.387$ ($0.26$) & $\mathbf{0.387}$ ($0.26$) \\
M3 & $0.4492$ & $0.557$ ($0.36$) & $0.617$ ($0.48$) & $0.495$ ($0.32$) & $0.531$ ($0.40$) & $\mathbf{0.468}$ ($0.30$) & $0.478$ ($0.32$) & $0.483$ ($0.32$) \\
M4 & $0.5716$ & $0.738$ ($0.43$) & $0.718$ ($0.46$) & $1.066$ ($0.52$) & $\mathbf{0.648}$ ($0.38$) & $0.850$ ($0.44$) & $0.671$ ($0.42$) & $0.667$ ($0.42$) \\
M5 & $1.0223$ & $1.594$ ($0.88$) & $1.294$ ($0.73$) & $2.961$ ($1.01$) & $1.139$ ($0.63$) & $2.550$ ($0.92$) & $\mathbf{1.125}$ ($0.63$) & $1.133$ ($0.65$) \\
M6 & $0.4702$ & $0.658$ ($0.43$) & $0.626$ ($0.44$) & $\mathbf{0.476}$ ($0.25$) & $0.686$ ($0.48$) & $0.478$ ($0.26$) & $0.586$ ($0.36$) & $0.536$ ($0.31$) \\
M7 & $0.5067$ & $0.573$ ($0.29$) & $0.632$ ($0.37$) & $6.641$ ($0.12$) & $0.573$ ($0.32$) & $5.496$ ($0.79$) & $0.529$ ($0.28$) & $\mathbf{0.527}$ ($0.26$) \\
M8 & $0.4070$ & $0.459$ ($0.25$) & $0.533$ ($0.38$) & $0.462$ ($0.21$) & $0.492$ ($0.34$) & $\mathbf{0.421}$ ($0.22$) & $0.449$ ($0.27$) & $0.432$ ($0.25$) \\
M9 & $1.0059$ & $1.521$ ($0.61$) & $1.228$ ($0.60$) & $1.766$ ($0.47$) & $1.185$ ($0.56$) & $1.624$ ($0.46$) & $1.146$ ($0.54$) & $\mathbf{1.096}$ ($0.50$) \\
M10 & $0.5141$ & $0.600$ ($0.31$) & $0.659$ ($0.43$) & $0.568$ ($0.28$) & $0.644$ ($0.39$) & $\mathbf{0.539}$ ($0.28$) & $0.551$ ($0.31$) & $0.555$ ($0.29$) \\
M11 & $0.4758$ & $0.562$ ($0.31$) & $0.607$ ($0.36$) & $0.532$ ($0.29$) & $0.550$ ($0.34$) & $\mathbf{0.498}$ ($0.27$) & $0.506$ ($0.27$) & $0.511$ ($0.27$) \\
M12 & $1.1958$ & $1.549$ ($0.55$) & $1.388$ ($0.56$) & $2.501$ ($0.44$) & $\mathbf{1.366}$ ($0.53$) & $2.261$ ($0.43$) & $1.781$ ($0.72$) & $1.735$ ($0.78$) \\
M13 & $1.5100$ & $1.912$ ($0.61$) & $\mathbf{1.692}$ ($0.52$) & $4.813$ ($0.40$) & $1.740$ ($0.55$) & $4.055$ ($0.38$) & $1.981$ ($0.93$) & $1.896$ ($0.94$) \\
M14 & $0.8561$ & $0.908$ ($0.37$) & $0.981$ ($0.45$) & $12.619$ ($0.55$) & $0.925$ ($0.43$) & $10.234$ ($1.22$) & $0.886$ ($0.37$) & $\mathbf{0.874}$ ($0.36$) \\
M15 & $1.0708$ & $1.190$ ($0.63$) & $1.297$ ($0.69$) & $44.200$ ($0.60$) & $1.184$ ($0.64$) & $40.171$ ($3.55$) & $1.139$ ($0.59$) & $\mathbf{1.113}$ ($0.58$) \\
M16 & $1.0426$ & $1.088$ ($0.42$) & $1.156$ ($0.47$) & $14.296$ ($0.01$) & $1.132$ ($0.43$) & $14.246$ ($0.06$) & $1.107$ ($0.41$) & $\mathbf{1.071}$ ($0.40$) \\
M17 & $0.9007$ & $\mathbf{0.973}$ ($0.25$) & $1.017$ ($0.29$) & $1.474$ ($0.29$) & $1.046$ ($0.34$) & $1.078$ ($0.16$) & $1.067$ ($0.27$) & $1.005$ ($0.20$) \\
M18 & $1.8266$ & $1.974$ ($0.78$) & $1.989$ ($0.74$) & $5.904$ ($0.36$) & $\mathbf{1.878}$ ($0.72$) & $5.641$ ($0.36$) & $2.015$ ($0.80$) & $1.960$ ($0.80$) \\
M19 & $0.4184$ & $0.494$ ($0.24$) & $0.529$ ($0.30$) & $0.729$ ($0.27$) & $0.522$ ($0.28$) & $0.482$ ($0.16$) & $0.503$ ($0.23$) & $\mathbf{0.476}$ ($0.19$) \\
M20 & $0.9527$ & $1.009$ ($0.41$) & $1.075$ ($0.46$) & $14.453$ ($0.88$) & $1.042$ ($0.44$) & $11.014$ ($1.41$) & $0.988$ ($0.40$) & $\mathbf{0.969}$ ($0.39$) \\
\midrule
M1 & $0.0000$ & $0.032$ ($0.06$) & $0.031$ ($0.06$) & $\mathbf{0.000}$ ($0.00$) & $0.023$ ($0.05$) & $0.010$ ($0.01$) & $0.010$ ($0.01$) & $0.011$ ($0.02$) \\
M2 & $0.1386$ & $0.157$ ($0.09$) & $0.176$ ($0.13$) & $0.142$ ($0.09$) & $0.144$ ($0.09$) & $0.140$ ($0.08$) & $0.140$ ($0.08$) & $\mathbf{0.140}$ ($0.08$) \\
M3 & $0.1664$ & $0.190$ ($0.12$) & $0.201$ ($0.13$) & $0.183$ ($0.11$) & $\mathbf{0.172}$ ($0.10$) & $0.173$ ($0.11$) & $0.176$ ($0.11$) & $0.177$ ($0.11$) \\
M4 & $0.2051$ & $0.247$ ($0.13$) & $0.240$ ($0.13$) & $0.432$ ($0.18$) & $0.215$ ($0.11$) & $0.331$ ($0.15$) & $\mathbf{0.214}$ ($0.11$) & $0.214$ ($0.11$) \\
M5 & $0.3545$ & $0.502$ ($0.27$) & $0.405$ ($0.20$) & $1.389$ ($0.37$) & $\mathbf{0.373}$ ($0.19$) & $1.160$ ($0.34$) & $0.374$ ($0.19$) & $0.378$ ($0.20$) \\
M6 & $0.1718$ & $0.219$ ($0.13$) & $0.212$ ($0.13$) & $0.173$ ($0.09$) & $0.194$ ($0.10$) & $\mathbf{0.172}$ ($0.09$) & $0.193$ ($0.10$) & $0.184$ ($0.10$) \\
M7 & $0.1793$ & $0.195$ ($0.09$) & $0.205$ ($0.10$) & $6.694$ ($0.04$) & $0.186$ ($0.08$) & $5.545$ ($0.78$) & $\mathbf{0.183}$ ($0.08$) & $0.184$ ($0.08$) \\
M8 & $0.1472$ & $0.160$ ($0.08$) & $0.175$ ($0.09$) & $0.180$ ($0.08$) & $0.153$ ($0.08$) & $0.157$ ($0.07$) & $0.151$ ($0.07$) & $\mathbf{0.150}$ ($0.07$) \\
M9 & $0.3762$ & $0.500$ ($0.21$) & $0.417$ ($0.18$) & $0.897$ ($0.20$) & $0.400$ ($0.17$) & $0.802$ ($0.20$) & $0.402$ ($0.17$) & $\mathbf{0.392}$ ($0.17$) \\
M10 & $0.1916$ & $0.213$ ($0.10$) & $0.232$ ($0.13$) & $0.227$ ($0.10$) & $\mathbf{0.205}$ ($0.10$) & $0.209$ ($0.10$) & $0.213$ ($0.10$) & $0.215$ ($0.10$) \\
M11 & $0.1926$ & $0.215$ ($0.10$) & $0.229$ ($0.12$) & $0.232$ ($0.10$) & $\mathbf{0.202}$ ($0.09$) & $0.212$ ($0.09$) & $0.208$ ($0.09$) & $0.210$ ($0.09$) \\
M12 & $0.4395$ & $0.534$ ($0.18$) & $0.481$ ($0.16$) & $1.507$ ($0.20$) & $\mathbf{0.459}$ ($0.15$) & $1.313$ ($0.19$) & $0.465$ ($0.16$) & $0.463$ ($0.16$) \\
M13 & $0.5716$ & $0.718$ ($0.23$) & $0.602$ ($0.17$) & $4.362$ ($0.27$) & $\mathbf{0.591}$ ($0.19$) & $3.406$ ($0.21$) & $0.609$ ($0.21$) & $0.616$ ($0.22$) \\
M14 & $0.2951$ & $0.306$ ($0.11$) & $0.320$ ($0.13$) & $12.094$ ($0.55$) & $0.299$ ($0.11$) & $7.599$ ($0.87$) & $0.298$ ($0.11$) & $\mathbf{0.297}$ ($0.11$) \\
M15 & $0.3526$ & $0.384$ ($0.19$) & $0.395$ ($0.19$) & $44.368$ ($0.25$) & $0.376$ ($0.18$) & $40.070$ ($3.76$) & $0.373$ ($0.18$) & $\mathbf{0.367}$ ($0.17$) \\
M16 & $0.3408$ & $0.355$ ($0.12$) & $0.372$ ($0.13$) & $14.293$ ($0.00$) & $0.358$ ($0.12$) & $14.217$ ($0.10$) & $0.356$ ($0.12$) & $\mathbf{0.349}$ ($0.12$) \\
M17 & $0.3638$ & $\mathbf{0.385}$ ($0.10$) & $0.393$ ($0.11$) & $1.111$ ($0.12$) & $0.486$ ($0.12$) & $0.815$ ($0.05$) & $0.468$ ($0.13$) & $0.466$ ($0.15$) \\
M18 & $0.6353$ & $0.670$ ($0.23$) & $0.651$ ($0.22$) & $3.963$ ($0.17$) & $0.688$ ($0.23$) & $3.665$ ($0.18$) & $0.649$ ($0.22$) & $\mathbf{0.644}$ ($0.22$) \\
M19 & $0.1769$ & $0.194$ ($0.07$) & $0.201$ ($0.08$) & $0.383$ ($0.09$) & $0.189$ ($0.06$) & $0.248$ ($0.06$) & $\mathbf{0.188}$ ($0.06$) & $0.193$ ($0.06$) \\
M20 & $0.3286$ & $0.339$ ($0.12$) & $0.359$ ($0.14$) & $13.143$ ($0.92$) & $0.335$ ($0.12$) & $7.403$ ($0.87$) & $0.333$ ($0.12$) & $\mathbf{0.331}$ ($0.12$) \\
\bottomrule\bottomrule
\end{tabular}
\caption{\small Comparative study for the circular case, with up to down blocks corresponding to sample sizes $100$, $250$ and $1000$, respectively. Columns of the selector $\bullet$ represent the $\mathrm{MISE}(\bullet)\times100$, with bold type for the minimum of the errors. The standard deviation of the $\mathrm{ISE}\times100$ is given between parentheses. \label{kdebwd:tab:apcir}}
\end{table}

\begin{table}[H]
\centering
\scriptsize
\begin{tabular}{r|r|rrrrrrr}\toprule\toprule
Model &  \multicolumn{1}{c|}{ $h_\mathrm{MISE}$} & \multicolumn{1}{c}{$h_\mathrm{LCV}$} & \multicolumn{1}{c}{$h_\mathrm{LSCV}$} & \multicolumn{1}{c}{$h_\mathrm{ROT}$} & \multicolumn{1}{c}{$h_\mathrm{AMI}$} & \multicolumn{1}{c}{$h_\mathrm{EMI}$} \\\midrule
M1 & $0.0000$ & $\mathbf{0.074}$ ($0.23$) & $0.121$ ($0.32$) & $0.149$ ($0.12$) & $0.145$ ($0.72$) & $0.083$ ($0.20$) \\
M2 & $0.8133$ & $0.920$ ($0.45$) & $1.047$ ($0.73$) & $\mathbf{0.838}$ ($0.41$) & $0.889$ ($0.73$) & $0.851$ ($0.47$) \\
M3 & $1.2481$ & $1.465$ ($0.63$) & $1.527$ ($0.81$) & $1.361$ ($0.61$) & $1.334$ ($0.89$) & $\mathbf{1.309}$ ($0.62$) \\
M4 & $2.6742$ & $3.777$ ($1.31$) & $3.087$ ($1.35$) & $4.867$ ($1.24$) & $3.161$ ($1.40$) & $\mathbf{3.022}$ ($1.30$) \\
M5 & $2.3676$ & $3.260$ ($1.10$) & $\mathbf{2.760}$ ($1.09$) & $3.770$ ($0.96$) & $3.353$ ($1.57$) & $2.992$ ($1.22$) \\
M6 & $1.0781$ & $1.362$ ($0.61$) & $1.295$ ($0.67$) & $\mathbf{1.092}$ ($0.38$) & $1.304$ ($0.90$) & $1.158$ ($0.47$) \\
M7 & $1.0165$ & $1.179$ ($0.58$) & $1.176$ ($0.47$) & $2.199$ ($0.33$) & $1.186$ ($0.92$) & $\mathbf{1.078}$ ($0.43$) \\
M8 & $0.8731$ & $0.958$ ($0.40$) & $1.048$ ($0.53$) & $\mathbf{0.921}$ ($0.34$) & $1.056$ ($0.64$) & $0.947$ ($0.41$) \\
M9 & $6.7507$ & $11.184$ ($2.90$) & $7.647$ ($2.67$) & $14.962$ ($1.99$) & $8.056$ ($2.53$) & $\mathbf{7.186}$ ($2.14$) \\
M10 & $2.3766$ & $2.851$ ($0.79$) & $2.751$ ($1.13$) & $2.946$ ($0.73$) & $\mathbf{2.653}$ ($1.20$) & $2.660$ ($0.81$) \\
M11 & $2.5706$ & $3.266$ ($0.87$) & $\mathbf{2.926}$ ($1.13$) & $4.029$ ($0.83$) & $3.588$ ($1.88$) & $3.263$ ($1.08$) \\
M12 & $3.8591$ & $5.629$ ($1.56$) & $\mathbf{4.312}$ ($1.49$) & $6.991$ ($1.30$) & $4.996$ ($2.14$) & $4.559$ ($1.76$) \\
M13 & $2.4304$ & $2.689$ ($0.37$) & $2.681$ ($0.53$) & $\mathbf{2.629}$ ($0.23$) & $3.691$ ($2.32$) & $2.769$ ($0.49$) \\
M14 & $3.0506$ & $3.200$ ($0.90$) & $3.364$ ($1.17$) & $12.687$ ($1.31$) & $3.376$ ($0.95$) & $\mathbf{3.128}$ ($0.88$) \\
M15 & $2.8638$ & $\mathbf{3.031}$ ($0.70$) & $3.083$ ($0.81$) & $8.674$ ($0.64$) & $4.156$ ($1.91$) & $3.125$ ($0.80$) \\
M16 & $2.1417$ & $\mathbf{2.263}$ ($0.49$) & $2.300$ ($0.56$) & $3.928$ ($0.20$) & $3.033$ ($1.35$) & $2.911$ ($0.94$) \\
M17 & $4.6150$ & $5.730$ ($1.24$) & $\mathbf{5.030}$ ($1.35$) & $9.316$ ($0.93$) & $5.560$ ($2.16$) & $5.047$ ($1.32$) \\
M18 & $13.2289$ & $\mathbf{13.572}$ ($3.49$) & $14.586$ ($4.42$) & $40.641$ ($1.60$) & $15.735$ ($4.23$) & $13.717$ ($3.54$) \\
M19 & $2.5921$ & $2.907$ ($0.65$) & $\mathbf{2.861}$ ($0.76$) & $3.883$ ($0.48$) & $4.067$ ($1.67$) & $3.570$ ($0.96$) \\
M20 & $3.0018$ & $\mathbf{3.174}$ ($0.89$) & $3.373$ ($1.09$) & $4.261$ ($0.73$) & $3.850$ ($1.56$) & $3.316$ ($0.88$) \\
\midrule
M1 & $0.0000$ & $0.046$ ($0.11$) & $0.047$ ($0.11$) & $0.070$ ($0.05$) & $0.038$ ($0.12$) & $\mathbf{0.027}$ ($0.04$) \\
M2 & $0.4615$ & $0.508$ ($0.23$) & $0.536$ ($0.28$) & $0.477$ ($0.22$) & $0.475$ ($0.20$) & $\mathbf{0.471}$ ($0.21$) \\
M3 & $0.7147$ & $0.844$ ($0.36$) & $0.817$ ($0.40$) & $0.808$ ($0.34$) & $\mathbf{0.729}$ ($0.31$) & $0.739$ ($0.32$) \\
M4 & $1.6108$ & $2.271$ ($0.71$) & $1.753$ ($0.59$) & $3.462$ ($0.73$) & $\mathbf{1.704}$ ($0.55$) & $1.710$ ($0.58$) \\
M5 & $1.4470$ & $2.070$ ($0.70$) & $1.587$ ($0.55$) & $2.776$ ($0.62$) & $1.642$ ($0.60$) & $\mathbf{1.555}$ ($0.57$) \\
M6 & $0.6433$ & $0.775$ ($0.27$) & $0.721$ ($0.27$) & $\mathbf{0.655}$ ($0.20$) & $0.733$ ($0.31$) & $0.677$ ($0.22$) \\
M7 & $0.5942$ & $0.634$ ($0.20$) & $0.643$ ($0.20$) & $1.871$ ($0.31$) & $0.613$ ($0.19$) & $\mathbf{0.607}$ ($0.19$) \\
M8 & $0.5106$ & $0.538$ ($0.18$) & $0.573$ ($0.22$) & $0.552$ ($0.18$) & $0.556$ ($0.22$) & $\mathbf{0.524}$ ($0.17$) \\
M9 & $3.9353$ & $7.525$ ($1.76$) & $4.242$ ($1.20$) & $11.752$ ($1.27$) & $4.392$ ($1.06$) & $\mathbf{4.087}$ ($1.09$) \\
M10 & $1.4223$ & $1.783$ ($0.47$) & $\mathbf{1.567}$ ($0.49$) & $2.016$ ($0.42$) & $1.647$ ($0.58$) & $1.667$ ($0.44$) \\
M11 & $1.5512$ & $1.951$ ($0.52$) & $1.687$ ($0.49$) & $2.888$ ($0.53$) & $1.858$ ($0.65$) & $\mathbf{1.661}$ ($0.47$) \\
M12 & $2.3194$ & $3.469$ ($0.92$) & $2.516$ ($0.76$) & $5.174$ ($0.82$) & $2.497$ ($0.70$) & $\mathbf{2.421}$ ($0.71$) \\
M13 & $1.8303$ & $2.070$ ($0.32$) & $\mathbf{1.953}$ ($0.33$) & $2.383$ ($0.17$) & $2.673$ ($0.91$) & $2.033$ ($0.41$) \\
M14 & $1.7407$ & $1.811$ ($0.47$) & $1.856$ ($0.51$) & $10.627$ ($1.13$) & $1.824$ ($0.46$) & $\mathbf{1.761}$ ($0.46$) \\
M15 & $1.6979$ & $\mathbf{1.759}$ ($0.34$) & $1.773$ ($0.36$) & $8.061$ ($0.73$) & $2.111$ ($0.55$) & $1.785$ ($0.35$) \\
M16 & $1.2929$ & $\mathbf{1.338}$ ($0.24$) & $1.344$ ($0.24$) & $3.675$ ($0.21$) & $1.677$ ($0.43$) & $1.704$ ($0.41$) \\
M17 & $2.7758$ & $3.619$ ($0.77$) & $2.927$ ($0.66$) & $7.752$ ($0.60$) & $3.122$ ($0.71$) & $\mathbf{2.901}$ ($0.65$) \\
M18 & $7.7070$ & $7.889$ ($1.78$) & $8.149$ ($1.92$) & $33.585$ ($1.08$) & $8.272$ ($1.77$) & $\mathbf{7.803}$ ($1.74$) \\
M19 & $1.6228$ & $1.801$ ($0.37$) & $\mathbf{1.720}$ ($0.35$) & $3.211$ ($0.33$) & $2.136$ ($0.76$) & $1.951$ ($0.75$) \\
M20 & $1.7820$ & $1.845$ ($0.48$) & $1.924$ ($0.54$) & $2.961$ ($0.43$) & $1.996$ ($0.57$) & $\mathbf{1.828}$ ($0.48$) \\
\midrule
M1 & $0.0000$ & $0.013$ ($0.03$) & $0.013$ ($0.03$ & $0.023$ ($0.02$) & $0.008$ ($0.01$) & $\mathbf{0.007}$ ($0.01$) \\
M2 & $0.2002$ & $0.214$ ($0.08$) & $0.217$ ($0.08$ & $0.207$ ($0.08$) & $0.203$ ($0.07$) & $\mathbf{0.202}$ ($0.07$) \\
M3 & $0.3069$ & $0.356$ ($0.13$) & $0.328$ ($0.12$ & $0.357$ ($0.12$) & $\mathbf{0.311}$ ($0.11$) & $0.315$ ($0.11$) \\
M4 & $0.6790$ & $0.983$ ($0.30$) & $0.714$ ($0.21$ & $1.912$ ($0.32$) & $0.704$ ($0.20$) & $\mathbf{0.699}$ ($0.21$) \\
M5 & $0.6403$ & $0.960$ ($0.28$) & $0.671$ ($0.19$ & $1.622$ ($0.26$) & $\mathbf{0.658}$ ($0.18$) & $0.661$ ($0.20$) \\
M6 & $0.2789$ & $0.324$ ($0.09$) & $0.297$ ($0.09$ & $\mathbf{0.288}$ ($0.07$) & $0.327$ ($0.09$) & $0.295$ ($0.08$) \\
M7 & $0.2559$ & $0.270$ ($0.08$) & $0.268$ ($0.07$ & $1.452$ ($0.28$) & $\mathbf{0.258}$ ($0.07$) & $0.261$ ($0.07$) \\
M8 & $0.2196$ & $0.227$ ($0.07$) & $0.234$ ($0.07$ & $0.249$ ($0.07$) & $0.225$ ($0.06$) & $\mathbf{0.221}$ ($0.07$) \\
M9 & $1.7392$ & $3.617$ ($0.77$) & $1.811$ ($0.48$) & $7.790$ ($0.65$) & $1.827$ ($0.42$) & $\mathbf{1.768}$ ($0.46$) \\
M10 & $0.6306$ & $0.794$ ($0.20$) & $0.666$ ($0.18$) & $1.104$ ($0.19$) & $0.661$ ($0.17$) & $\mathbf{0.655}$ ($0.18$) \\
M11 & $0.6922$ & $0.869$ ($0.22$) & $0.722$ ($0.18$) & $1.687$ ($0.23$) & $0.733$ ($0.15$) & $\mathbf{0.704}$ ($0.17$) \\
M12 & $1.0539$ & $1.576$ ($0.36$) & $1.093$ ($0.24$) & $3.211$ ($0.37$) & $1.094$ ($0.25$) & $\mathbf{1.093}$ ($0.26$) \\
M13 & $1.0348$ & $1.302$ ($0.19$) & $1.069$ ($0.15$) & $2.045$ ($0.10$) & $1.136$ ($0.16$) & $\mathbf{1.061}$ ($0.16$) \\
M14 & $0.7395$ & $0.773$ ($0.18$) & $0.764$ ($0.18$) & $7.232$ ($0.51$) & $0.751$ ($0.16$) & $\mathbf{0.742}$ ($0.17$) \\
M15 & $0.7442$ & $\mathbf{0.756}$ ($0.12$) & $0.762$ ($0.13$) & $7.040$ ($0.77$) & $0.817$ ($0.13$) & $0.757$ ($0.12$) \\
M16 & $0.5674$ & $0.581$ ($0.08$) & $0.578$ ($0.08$) & $3.290$ ($0.24$) & $0.618$ ($0.10$) & $\mathbf{0.573}$ ($0.08$) \\
M17 & $1.2668$ & $1.696$ ($0.34$) & $1.304$ ($0.26$) & $5.657$ ($0.32$) & $1.350$ ($0.23$) & $\mathbf{1.289}$ ($0.26$) \\
M18 & $3.2966$ & $3.396$ ($0.67$) & $3.381$ ($0.67$) & $23.727$ ($0.59$) & $3.373$ ($0.64$) & $\mathbf{3.311}$ ($0.65$) \\
M19 & $0.7614$ & $0.851$ ($0.16$) & $\mathbf{0.784}$ ($0.14$) & $2.308$ ($0.17$) & $0.809$ ($0.14$) & $0.788$ ($0.16$) \\
M20 & $0.7675$ & $0.782$ ($0.17$) & $0.800$ ($0.18$) & $1.614$ ($0.19$) & $0.787$ ($0.17$) & $\mathbf{0.772}$ ($0.17$) \\
\bottomrule\bottomrule
\end{tabular}
\caption{\small Comparative study for the spherical case, with up to down blocks corresponding to sample sizes $100$, $250$ and $1000$, respectively. Columns of the selector $\bullet$ represent the $\mathrm{MISE}(\bullet)\times100$, with bold type for the minimum of the errors. The standard deviation of the $\mathrm{ISE}\times100$ is given between parentheses. \label{kdebwd:tab:apsph}}
\end{table}

\begin{table}[H]
\centering
\scriptsize
\begin{tabular}{r|r|rrrrrrr}\toprule\toprule
Model &  \multicolumn{1}{c|}{ $h_\mathrm{MISE}$} & \multicolumn{1}{c}{$h_\mathrm{LCV}$} & \multicolumn{1}{c}{$h_\mathrm{LSCV}$} & \multicolumn{1}{c}{$h_\mathrm{ROT}$} & \multicolumn{1}{c}{$h_\mathrm{AMI}$} & \multicolumn{1}{c}{$h_\mathrm{EMI}$} \\\midrule
M1 & $0.0000$ & $0.006$ ($0.02$) & $0.006$ ($0.02$) & $0.008$ ($0.01$) & $0.008$ ($0.01$) & $\mathbf{0.005}$ ($0.01$) \\
M2 & $0.2201$ & $0.230$ ($0.06$) & $0.229$ ($0.06$) & $0.224$ ($0.05$) & $0.224$ ($0.05$) & $\mathbf{0.221}$ ($0.06$) \\
M3 & $0.4536$ & $0.528$ ($0.13$) & $0.470$ ($0.12$) & $\mathbf{0.456}$ ($0.11$) & $0.456$ ($0.11$) & $0.462$ ($0.12$) \\
M4 & $1.2583$ & $1.568$ ($0.23$) & $\mathbf{1.281}$ ($0.19$) & $2.350$ ($0.24$) & $1.291$ ($0.18$) & $1.288$ ($0.20$) \\
M5 & $0.6392$ & $0.846$ ($0.18$) & $0.657$ ($0.14$) & $0.909$ ($0.16$) & $0.667$ ($0.12$) & $\mathbf{0.651}$ ($0.14$) \\
M6 & $0.3575$ & $0.401$ ($0.07$) & $0.370$ ($0.06$) & $0.387$ ($0.06$) & $0.397$ ($0.08$) & $\mathbf{0.370}$ ($0.06$) \\
M7 & $0.2808$ & $0.294$ ($0.06$) & $0.288$ ($0.06$) & $1.425$ ($0.18$) & $\mathbf{0.283}$ ($0.06$) & $0.284$ ($0.06$) \\
M8 & $0.2623$ & $0.269$ ($0.06$) & $0.271$ ($0.06$) & $0.266$ ($0.06$) & $0.271$ ($0.05$) & $\mathbf{0.264}$ ($0.06$) \\
M9 & $6.9786$ & $17.556$ ($2.08$) & $7.139$ ($1.49$) & $19.477$ ($1.75$) & $7.303$ ($1.29$) & $\mathbf{7.064}$ ($1.44$) \\
M10 & $1.2743$ & $1.567$ ($0.25$) & $\mathbf{1.302}$ ($0.20$) & $1.903$ ($0.26$) & $1.308$ ($0.21$) & $1.303$ ($0.22$) \\
M11 & $1.4724$ & $2.117$ ($0.34$) & $1.506$ ($0.26$) & $2.465$ ($0.34$) & $1.612$ ($0.22$) & $\mathbf{1.485}$ ($0.26$) \\
M12 & $1.4808$ & $1.922$ ($0.26$) & $\mathbf{1.506}$ ($0.20$) & $2.500$ ($0.26$) & $1.522$ ($0.20$) & $1.512$ ($0.22$) \\
M13 & $0.4761$ & $0.509$ ($0.06$) & $\mathbf{0.492}$ ($0.05$) & $0.594$ ($0.04$) & $0.666$ ($0.22$) & $0.505$ ($0.07$) \\
M14 & $1.4327$ & $1.543$ ($0.25$) & $1.459$ ($0.24$) & $10.893$ ($0.94$) & $1.466$ ($0.22$) & $\mathbf{1.436}$ ($0.23$) \\
M15 & $0.8662$ & $0.883$ ($0.08$) & $0.879$ ($0.08$) & $4.351$ ($0.35$) & $0.993$ ($0.10$) & $\mathbf{0.877}$ ($0.08$) \\
M16 & $0.5830$ & $0.592$ ($0.06$) & $\mathbf{0.591}$ ($0.06$) & $1.895$ ($0.12$) & $0.636$ ($0.07$) & $0.706$ ($0.08$) \\
M17 & $5.2376$ & $10.032$ ($1.16$) & $5.373$ ($1.01$) & $19.474$ ($0.94$) & $5.506$ ($0.85$) & $\mathbf{5.320}$ ($0.92$) \\
M18 & $13.3175$ & $13.954$ ($2.18$) & $13.529$ ($2.13$) & $43.085$ ($2.21$) & $13.713$ ($2.09$) & $\mathbf{13.357}$ ($2.11$) \\
M19 & $2.6298$ & $3.743$ ($0.47$) & $\mathbf{2.699}$ ($0.48$) & $7.433$ ($0.43$) & $2.758$ ($0.45$) & $2.701$ ($0.48$) \\
M20 & $2.1853$ & $2.276$ ($0.33$) & $2.222$ ($0.32$) & $4.823$ ($0.32$) & $2.268$ ($0.29$) & $\mathbf{2.193}$ ($0.31$) \\
\midrule
M1 & $0.0000$ & $\mathbf{0.004}$ ($0.01$) & $0.004$ ($0.01$) & $0.008$ ($0.01$) & $0.008$ ($0.01$) & $0.004$ ($0.00$) \\
M2 & $0.2356$ & $0.243$ ($0.05$) & $0.243$ ($0.05$) & $0.243$ ($0.05$) & $0.243$ ($0.05$) & $\mathbf{0.237}$ ($0.05$) \\
M3 & $0.6224$ & $0.705$ ($0.14$) & $0.637$ ($0.12$) & $0.629$ ($0.11$) & $0.629$ ($0.11$) & $\mathbf{0.629}$ ($0.12$) \\
M4 & $1.8521$ & $2.023$ ($0.17$) & $\mathbf{1.870}$ ($0.16$) & $2.899$ ($0.18$) & $1.927$ ($0.17$) & $1.886$ ($0.16$) \\
M5 & $0.4269$ & $0.488$ ($0.09$) & $0.437$ ($0.07$) & $0.479$ ($0.08$) & $0.495$ ($0.08$) & $\mathbf{0.436}$ ($0.08$) \\
M6 & $0.4448$ & $0.486$ ($0.06$) & $\mathbf{0.455}$ ($0.06$) & $0.507$ ($0.05$) & $0.507$ ($0.05$) & $0.463$ ($0.05$) \\
M7 & $0.2765$ & $0.287$ ($0.05$) & $0.282$ ($0.05$) & $0.819$ ($0.09$) & $0.284$ ($0.04$) & $\mathbf{0.278}$ ($0.05$) \\
M8 & $0.2953$ & $0.301$ ($0.05$) & $0.303$ ($0.05$) & $\mathbf{0.296}$ ($0.05$) & $0.311$ ($0.05$) & $0.297$ ($0.05$) \\
M9 & $9.9498$ & $19.713$ ($1.88$) & $11.588$ ($1.53$) & $19.502$ ($1.66$) & $14.847$ ($4.08$) & $\mathbf{10.004}$ ($1.43$) \\
M10 & $2.3370$ & $2.629$ ($0.23$) & $\mathbf{2.364}$ ($0.21$) & $3.338$ ($0.24$) & $2.521$ ($0.31$) & $2.369$ ($0.22$) \\
M11 & $2.9424$ & $4.518$ ($0.57$) & $2.993$ ($0.43$) & $4.557$ ($0.54$) & $3.358$ ($0.38$) & $\mathbf{2.963}$ ($0.43$) \\
M12 & $1.9610$ & $2.290$ ($0.21$) & $\mathbf{1.980}$ ($0.17$) & $2.769$ ($0.21$) & $2.071$ ($0.21$) & $1.996$ ($0.19$) \\
M13 & $0.1732$ & $0.180$ ($0.02$) & $\mathbf{0.180}$ ($0.02$) & $0.180$ ($0.02$) & $0.207$ ($0.12$) & $0.187$ ($0.02$) \\
M14 & $2.2517$ & $2.393$ ($0.28$) & $2.278$ ($0.26$) & $11.171$ ($0.81$) & $2.391$ ($0.25$) & $\mathbf{2.255}$ ($0.26$) \\
M15 & $0.8263$ & $0.842$ ($0.05$) & $0.836$ ($0.05$) & $2.384$ ($0.16$) & $0.883$ ($0.07$) & $\mathbf{0.834}$ ($0.05$) \\
M16 & $0.5189$ & $\mathbf{0.525}$ ($0.04$) & $0.526$ ($0.04$) & $1.069$ ($0.06$) & $0.527$ ($0.06$) & $0.568$ ($0.06$) \\
M17 & $18.2152$ & $38.755$ ($2.66$) & $18.815$ ($3.26$) & $55.977$ ($1.91$) & $19.331$ ($3.03$) & $\mathbf{18.403}$ ($2.94$) \\
M18 & $74.2135$ & $87.710$ ($16.21$) & $77.381$ ($16.68$) & $215.090$ ($11.43$) & $75.489$ ($16.94$) & $\mathbf{74.833}$ ($16.58$) \\
M19 & $7.8653$ & $12.311$ ($0.97$) & $8.093$ ($1.26$) & $18.594$ ($0.72$) & $8.325$ ($1.23$) & $\mathbf{8.061}$ ($1.22$) \\
M20 & $4.1058$ & $4.235$ ($0.44$) & $4.152$ ($0.44$) & $7.416$ ($0.45$) & $4.461$ ($0.42$) & $\mathbf{4.115}$ ($0.43$) \\
\midrule
M1 & $0.0000$ & $\mathbf{0.001}$ ($0.00$) & $0.004$ ($0.01$) & $0.010$ ($0.01$) & $0.010$ ($0.01$) & $0.004$ ($0.00$) \\
M2 & $0.2539$ & $0.260$ ($0.05$) & $0.260$ ($0.05$) & $0.266$ ($0.04$) & $0.266$ ($0.04$) & $\mathbf{0.255}$ ($0.05$) \\
M3 & $0.8986$ & $1.004$ ($0.17$) & $0.912$ ($0.15$) & $0.915$ ($0.14$) & $0.915$ ($0.14$) & $\mathbf{0.905}$ ($0.15$) \\
M4 & $2.4674$ & $2.522$ ($0.14$) & $\mathbf{2.484}$ ($0.14$) & $3.340$ ($0.15$) & $2.620$ ($0.18$) & $2.521$ ($0.15$) \\
M5 & $0.3296$ & $0.356$ ($0.06$) & $\mathbf{0.336}$ ($0.05$) & $0.339$ ($0.05$) & $0.379$ ($0.07$) & $0.352$ ($0.06$) \\
M6 & $0.5562$ & $0.592$ ($0.07$) & $\mathbf{0.565}$ ($0.06$) & $0.666$ ($0.05$) & $0.666$ ($0.05$) & $0.586$ ($0.06$) \\
M7 & $0.2715$ & $0.280$ ($0.04$) & $0.276$ ($0.04$) & $0.551$ ($0.05$) & $0.287$ ($0.03$) & $\mathbf{0.273}$ ($0.04$) \\
M8 & $0.3342$ & $0.341$ ($0.05$) & $0.341$ ($0.05$) & $0.336$ ($0.05$) & $0.358$ ($0.04$) & $\mathbf{0.335}$ ($0.05$) \\
M9 & $12.4415$ & $23.477$ ($2.05$) & $20.881$ ($3.64$) & $21.104$ ($1.89$) & $31.948$ ($19.84$) & $\mathbf{12.539}$ ($1.57$) \\
M10 & $2.6473$ & $2.775$ ($0.17$) & $\mathbf{2.669}$ ($0.16$) & $3.108$ ($0.18$) & $2.800$ ($0.28$) & $2.964$ ($0.28$) \\
M11 & $6.3810$ & $10.020$ ($0.94$) & $6.454$ ($0.87$) & $9.398$ ($0.92$) & $7.191$ ($0.88$) & $\mathbf{6.430}$ ($0.82$) \\
M12 & $2.4791$ & $2.700$ ($0.18$) & $\mathbf{2.496}$ ($0.16$) & $3.072$ ($0.18$) & $2.580$ ($0.18$) & $2.569$ ($0.18$) \\
M13 & $0.0704$ & $0.085$ ($0.01$) & $0.076$ ($0.01$) & $\mathbf{0.071}$ ($0.01$) & $0.071$ ($0.01$) & $0.072$ ($0.01$) \\
M14 & $3.1635$ & $3.300$ ($0.33$) & $3.205$ ($0.31$) & $11.378$ ($0.73$) & $3.576$ ($0.30$) & $\mathbf{3.167}$ ($0.31$) \\
M15 & $0.8038$ & $0.816$ ($0.04$) & $\mathbf{0.812}$ ($0.04$) & $1.625$ ($0.09$) & $0.879$ ($0.22$) & $0.926$ ($0.28$) \\
M16 & $0.4712$ & $\mathbf{0.476}$ ($0.03$) & $0.477$ ($0.03$) & $0.738$ ($0.04$) & $0.683$ ($0.11$) & $0.775$ ($0.15$) \\
M17 & $20.2938$ & $37.587$ ($2.73$) & $22.014$ ($3.55$) & $52.149$ ($2.25$) & $31.485$ ($15.35$) & $\mathbf{20.508}$ ($3.34$) \\
M18 & $74.4378$ & $74.868$ ($13.64$) & $83.389$ ($16.68$) & $133.737$ ($9.48$) & $108.546$ ($36.67$) & $\mathbf{74.639}$ ($13.26$) \\
M19 & $8.3636$ & $11.984$ ($0.98$) & $8.832$ ($1.14$) & $17.672$ ($0.81$) & $12.430$ ($5.23$) & $\mathbf{8.494}$ ($1.08$) \\
M20 & $7.4037$ & $7.701$ ($0.73$) & $7.467$ ($0.70$) & $11.502$ ($0.68$) & $8.450$ ($0.82$) & $\mathbf{7.415}$ ($0.71$) \\
\bottomrule\bottomrule
\end{tabular}
\caption{\small Comparative study for higher dimensions with sample size $n=1000$: up to down blocks correspond to dimensions $q=3,4,5$. Columns of the selector $\bullet$ represent the $\mathrm{MISE}(\bullet)\times100$, with bold type for the minimum of the errors. The standard deviation of the $\mathrm{ISE}\times100$ is given between parentheses. \label{kdebwd:tab:apdim}}
\end{table}

\end{document}